\documentclass[Afour,sageh,times]{sagej}

\usepackage{moreverb,url}
\usepackage{graphicx}
\usepackage[T1]{fontenc}
\usepackage{url}
\usepackage{float}
\usepackage{mathtools}
\usepackage{amsmath}
\usepackage{amssymb}
\usepackage{nameref} 
\usepackage{multirow}
\usepackage{tikz}
\usetikzlibrary{arrows.meta, positioning}
\usepackage[colorlinks,bookmarksopen,bookmarksnumbered,citecolor=red,urlcolor=red]{hyperref}

\newcommand\BibTeX{{\rmfamily B\kern-.05em \textsc{i\kern-.025em b}\kern-.08em
T\kern-.1667em\lower.7ex\hbox{E}\kern-.125emX}}

\def\volumeyear{2016}
\usepackage{siunitx}
\usepackage{lmodern}
\usepackage{relsize}
\usepackage{fix-cm}
\usepackage{xr-hyper}
\makeatletter
\newcommand*{\addFileDependency}[1]{
\typeout{(#1)}
\@addtofilelist{#1}

\IfFileExists{#1}{}{\typeout{No file #1.}}
}\makeatother

\begin{document}

\runninghead{Karami et al.}

\title{Comparative study of Bayesian and Frequentist methods for epidemic forecasting: Insights from simulated and historical data}

\author{Hamed Karami\affilnum{1}, Ruiyan Luo\affilnum{2}, Pejman Sanaei\affilnum{1}, and Gerardo Chowell\affilnum{2}}

\affiliation{\affilnum{1}Department of Mathematics and Statistics, Georgia State University, Atlanta, GA 30303, USA\\
\affilnum{2}Department of Population Health Sciences, School of Public Health, Georgia State University, Atlanta, GA, USA
}

\corrauth{Gerardo Chowell, Department of Population Health Sciences, School of Public Health, Georgia State University, Atlanta, GA, USA}

\email{gchowell@gsu.edu}

\begin{abstract}
Accurate epidemic forecasting is critical for effective public health interventions. This study compares Bayesian and Frequentist estimation frameworks within deterministic compartmental epidemic models, focusing on nonlinear least squares (NLS) optimization versus Bayesian inference using a normal likelihood and MCMC sampling via \texttt{Stan}. Rather than evaluating all methodological variants, we compare forecasting performance under a shared modeling structure and error assumption. Our findings apply to specific implementations of the two approaches, not broad generalizations.

We assess performance on both simulated datasets (with $\mathcal{R}_0$ values of 2 and 1.5) and historical datasets, including the 1918 influenza pandemic, the 1896–97 Bombay plague, and the COVID-19 pandemic. Evaluation metrics include Mean Absolute Error (MAE), Root Mean Squared Error (RMSE), Weighted Interval Score (WIS), and 95\% prediction interval coverage.

Forecasting performance depends on epidemic phase and dataset characteristics, with no method consistently outperforming across all contexts. The Frequentist method performs well at the peak in simulated data and in the post-peak phase of real outbreaks but tends to be less accurate during the pre-peak phase. In contrast, Bayesian methods, particularly those with uniform priors, offer better predictive accuracy early in the epidemic. Bayesian approaches also typically provide stronger uncertainty quantification, especially valuable when data are sparse or noisy. Frequentist methods, however, often yield more accurate point forecasts, with lower MAE, RMSE, and WIS in many scenarios, though their interval estimates may be less robust.

We also discuss how prior choice influences Bayesian forecasts and examine how extending forecasting horizons affects convergence and model efficiency. These findings offer practical guidance for researchers and decision-makers in choosing estimation strategies tailored to the epidemic phase and data quality, ultimately supporting more effective public health interventions and resource planning.
\end{abstract}


\maketitle

\section{Introduction}\label{sec:intro}
Epidemic forecasting has become increasingly critical for predicting the spread of infectious diseases, enabling timely and effective public health interventions. Accurate forecasts are essential for guiding resource allocation, informing policy decisions, and mitigating the impact of epidemics and pandemics. Model-based forecasts have been pivotal in managing various health crises in the past decade. During the COVID-19 pandemic, forecasts played a significant role in informing resource allocation and social distancing policies \cite{dixon2022comparison,cheng2023real,lutz2019applying,shearer2020infectious,rahimi2023review,shinde2020forecasting,bertozzi2020challenges,ioannidis2022forecasting,chowell2022ensemble,chowell2022sub}. The US CDC's FluSight Challenge leveraged models to optimize influenza vaccine distribution and public health messaging \cite{reich2019collaborative,mcgowan2019collaborative,biggerstaff2014estimates}. Similarly, during the West African and DRC Ebola outbreaks, models were instrumental in predicting the spread and evaluating the effectiveness of interventions, thereby supporting international response efforts \cite{chowell2017perspectives,funk2019assessing,meltzer2014estimating,chretien2015mathematical,chowell2014west,roosa2020multi}. More recently, forecasting models for mpox have been used to predict its spread and assess containment measures \cite{bleichrodt2023real,bleichrodt2023retrospective,charniga2024nowcasting,chowell2024growthpredict}. These examples underscore the need to understand not only the modeling structures but also the estimation methodologies underlying forecasts.\\

Bayesian and Frequentist estimation methods are the two predominant paradigms for calibrating compartmental models to epidemic data. Bayesian estimation methods have increasingly been applied for calibrating epidemic models based on ordinary differential equations (ODEs) by incorporating prior knowledge and updating parameter estimates as new data becomes available \cite{greenland2009bayesian,mckinley2014simulation,kypraios2017tutorial,girolami2008bayesian,grinsztajn2021bayesian,bouman2024bayesian,gelman2020bayesian,belasso2023bayesian}. These methods apply Bayes' theorem to combine prior distributions of parameters with the likelihood of observed data, producing posterior distributions that explicitly incorporate uncertainty in parameter information and expert knowledge into the modeling process. Bayesian methods typically define a likelihood function, such as the probability of observed data given the model parameters, and specify prior distributions for these parameters. The posterior distributions of the parameters are usually approximated using Markov Chain Monte Carlo (MCMC) algorithms.  This provides a comprehensive measure of parameter uncertainty, and allows for constructing credible intervals that indicate parameter values consistent with the observed data. The flexibility and robustness of Bayesian methods make them particularly useful in emerging epidemics where data may be sparse or noisy. Tools like Stan facilitate the implementation of Bayesian estimation and forecasting, allowing for rigorous uncertainty quantification and model validation \cite{martin2020computing,dunson2001commentary,harel2018multiple,grinsztajn2006bayesian,annis2017bayesian,kelter2020bayesian,sennhenn2018bayesian,sorensen2015bayesian,monnahan2017faster,burkner2017brms}.\\

Frequentist estimation methods typically involve calibrating epidemic models based ODEs by optimizing a likelihood function to estimate model parameters that best fit observed data \cite{gneiting2008probabilistic,mwambi2011frequentist,chowell2017fitting,chowell2020real}. These methods often minimize an objective function, such as the sum of squared differences between observed and predicted values, using algorithms like gradient descent or the Levenberg-Marquardt algorithm. Frequentist approaches generally assume specific distributions for measurement errors, such as Gaussian or Poisson distributions, to incorporate observation error structures. Bootstrapping techniques are used to quantify parameter uncertainty by generating multiple resampled datasets and re-estimating model parameters, allowing for the construction of confidence and prediction intervals \cite{banks2014modeling,pruitt2024role,transtrum2012optimal,huang2024nonlinear}. Once calibrated, the ODE models can forecast epidemic trajectories. The QuantDiffForecast toolbox can be used to fit models to data and generate predictions with quantified uncertainty \cite{chowell2024parameter}.\\

Parameter estimation for epidemic models as ordinary differential equations using Bayesian and Frequentist has evolved through multiple methodological approaches, each addressing different computational and statistical challenges. Frequentist methods primarily rely on nonlinear least squares techniques, where researchers fit model predictions to observed epidemic data such as case counts and mortality rates~\cite{bates1988nonlinear,cao2012penalized,ramsay2017dynamic,seber2003nonlinear,luo2024estimation}. This approach involves using numerical solvers to approximate differential equation solutions and then optimizing parameter values to minimize differences between model predictions and observed data. While effective, this method can become computationally demanding when dealing with noisy datasets or complex nonlinear epidemic dynamics. To improve efficiency, researchers have developed two-stage procedures that first smooth noisy data before parameter estimation~\cite{varah1982spline,liang2008parameter}, and more sophisticated approaches that simultaneously optimize both data fitting and adherence to the underlying differential equations~\cite{ramsay2007parameter}. Recent advances have also incorporated machine learning techniques, including neural networks, into parameter estimation frameworks~\cite{rudi2022parameter}. Bayesian methods offer an alternative approach that can better navigate complex parameter spaces and avoid getting trapped in suboptimal solutions~\cite{gelman1996physiological}. These methods typically use sampling algorithms like Metropolis-Hastings to explore parameter distributions, though this can be computationally intensive since each parameter proposal requires solving the differential equation system~\cite{huang2006hierarchical}. Bayesian frameworks have also embraced more flexible approaches, including Gaussian process methods that can handle incomplete data and hierarchical models that account for multiple sources of uncertainty~\cite{huang2020bayesian}. Modern implementations often combine these concepts with advanced sampling techniques to improve computational efficiency. In our study, we compare these two paradigms by implementing nonlinear least squares optimization for the Frequentist approach and Bayesian inference with normal likelihood using MCMC sampling, allowing us to evaluate their relative strengths in epidemic forecasting contexts.\\

In this paper, we compare the forecasting performance of Frequentist and Bayesian estimation methods in the context of epidemic forecasting using compartmental models based on ODEs, with both simulated and real epidemic data. Although modern Bayesian models have become mainstream---owing to their ability to incorporate prior information and handle incompletely observed data---the practical differences between Bayesian and Frequentist approaches can remain substantial, especially in scenarios with limited data or under model misspecification. Rather than attempting an exhaustive comparison of all methodological variants, we use a fixed model structure with a normal likelihood and assess performance differences that arise purely from the estimation approach. Specifically, we implement nonlinear least squares estimation for the Frequentist method and Bayesian inference via MCMC sampling in Stan.  We evaluate these methods using simulated datasets with different reproductive numbers ($\mathcal{R}_0 = 2$ and $\mathcal{R}_0 = 1.5$) as well as historical epidemic data including the 1918 influenza pandemic, the 1896--97 Bombay plague, and the COVID-19 pandemic. By leveraging these diverse datasets, we aim to evaluate how each estimation framework performs in short-term and long-term forecasting based on four performance metrics: the mean absolute error (MAE), the root mean squared error (RMSE), the coverage of the 95\% prediction interval, and the weighted interval score (WIS).

While acknowledging that convergence between Bayesian and Frequentist methods can be expected as data accumulate, our study focuses explicitly on the early to mid-stages of epidemics—where prior assumptions and methodological choices may significantly influence forecasting performance. The remainder of the paper is structured as follows: we first present the models examined in this article (Section~\nameref{sec:models}). Then, we present the methodology and introduce Bayesian and the Frequentist approaches (Section~\nameref{sec:method}). We then go through all the case studies examined in this article in detail (Section~\nameref{sec:data}). Next, to assess the forecasting performance of the methods, we present the performance metrics (Section~\nameref{sec:metrics}). We continue by presenting results for both simulated and real-world datasets (Section~\nameref{sec:result}). Finally, we conclude the article by discussing the advantages and disadvantages of each method (Section~\nameref{sec:discussion}).

\section{Models}
\label{sec:models}

In this article, we employ three compartmental epidemic models to capture different dynamics of disease spread: 1) the SEIR model, which incorporates a reporting proportion without accounting for disease-induced deaths, 2) the SEIRD model, which includes both a reporting proportion and disease-induced deaths, and 3) the SEIUR model, which explicitly tracks both reported and unreported infected cases.

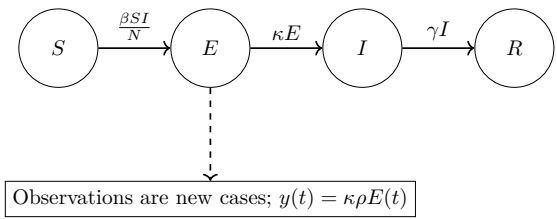
\begin{figure}
    \centering
    \scalebox{0.8}{ 
    \begin{tikzpicture}[node distance=2.5cm, auto]
        \node [draw, circle, minimum size=1.3cm] (S) {$S$};
        \node [draw, circle, minimum size=1.3cm, right of=S] (E) {$E$};
        \node [draw, circle, minimum size=1.3cm, right of=E] (I) {$I$};
        \node [draw, circle, minimum size=1.3cm, right of=I] (R) {$R$};
        
        \node [draw, rectangle, below of=E, node distance=2.5cm] (obs) {Observations are new cases; $y(t) = \kappa\rho E(t)$};
        
        \draw [->, thick] (S) -- node[midway, above] {$\frac{\beta SI}{N}$} (E);
        \draw [->, thick] (E) -- node[midway, above] {$\kappa E$} (I);
        \draw [->, thick] (I) -- node[midway, above] {$\gamma I$} (R);
        \draw [->, dashed, thick] (E) -- (obs);
    \end{tikzpicture}
    }
    \caption{\footnotesize Compartmental diagram of the SEIR model with underreporting. Circles show the epidemiological compartments for the different states of the system. Solid arrows indicate the transitions between compartments. The dashed arrow indicate the source of the observations, which are the newly reported infected individuals.}
    \label{fig:seir_model}
\end{figure}

\noindent \textbf{SEIR model}. The SEIR model categorizes the population into four epidemiological states: $S$ (susceptible), $E$ (exposed), $I$ (Infectious), and $R$ (recovered), with an additional state, $C$, representing the cumulative number of infected individuals. This model is governed by the following system of ordinary differential equations:
\begin{equation} \label{eq:SEIR}
\begin{gathered}
\frac{dS}{dt} = -\beta \frac{IS}{N}, \quad \frac{dE}{dt} = \beta \frac{IS}{N} - \kappa E, \quad \frac{dI}{dt} = \kappa E - \gamma I,\\\quad \frac{dR}{dt} = \gamma I,\quad \frac{dC}{dt} = \kappa \rho E.
\end{gathered}
\end{equation}
Here, $\beta > 0$ is the transmission rate, $\kappa > 0$ the incubation rate, $\gamma > 0$ the recovery rate, $\rho \in \left[0, 1\right]$ the reporting proportion, and $N$ the total population size, which is assumed to be known. The initial conditions are defined as $\left(S_0, E_0, I_0, R_0, C_0\right) = \left(N - \text{cases}\left[0\right], 0, \text{cases}\left[0\right], 0, \text{cases}\left[0\right]\right)$, where $\text{cases}\left[0\right]$.
 denotes the initial number of reported cases. The model is structured to generate observations as a time-series of new reported cases $\frac{dC}{dt}$. Figure~\ref{fig:seir_model} shows the compartmental diagram of this model with the source of the observations.

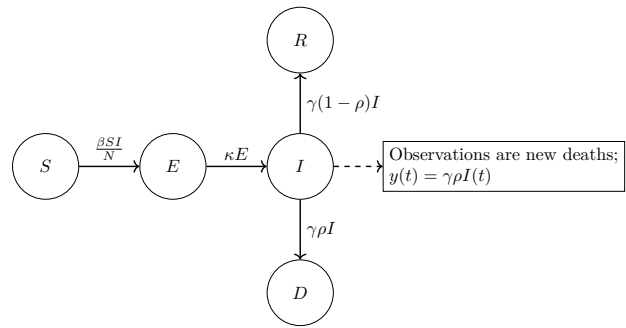
\begin{figure}
    \centering
    \scalebox{0.67}{
    \begin{tikzpicture}[node distance=2.5cm, auto]
        \node [draw, circle, minimum size=1.3cm] (S) {$S$};
        \node [draw, circle, minimum size=1.3cm, right of=S] (E) {$E$};
        \node [draw, circle, minimum size=1.3cm, right of=E] (I) {$I$};
        \node [draw, circle, minimum size=1.3cm, above of=I, node distance=2.5cm] (R) {$R$};
        \node [draw, circle, minimum size=1.3cm, below of=I, node distance=2.5cm] (D) {$D$};
        
        \node [draw, rectangle, right of=I, node distance=4cm, align=left] (obs) {Observations are new deaths; \\ $y(t) = \gamma\rho I(t)$};
        
        \draw [->, thick] (S) -- node[midway, above] {$\frac{\beta SI}{N}$} (E);
        \draw [->, thick] (E) -- node[midway, above] {$\kappa E$} (I);
        \draw [->, thick] (I) -- node[midway, right] {$\gamma (1-\rho) I$} (R);
        \draw [->, thick] (I) -- node[midway, right] {$\gamma \rho I$} (D);
        \draw [->, dashed, thick] (I) -- (obs);
    \end{tikzpicture}}
    \caption{\footnotesize Compartmental diagram of the SEIRD model with underreporting. Circles show the epidemiological compartments for the different states of the system. Solid arrows indicate the transitions between compartments. The dashed arrow indicate the source of the observations, which are the daily deaths.}
    \label{fig:seird_model}
\end{figure}

\begin{figure}
    \centering
    \scalebox{0.5}{
    \begin{tikzpicture}[node distance=3cm, auto]
        \node [draw, circle, minimum size=1.3cm] (S) {$S$};
        \node [draw, circle, minimum size=1.3cm, right of=S] (E) {$E$};
        \node [draw, circle, minimum size=1.3cm, above right=of E] (I) {$I$};
        \node [draw, circle, minimum size=1.3cm, below right=of E] (U) {$U$};
        \node [draw, circle, minimum size=1.3cm, right of=E, node distance=12cm] (R) {$R$};
        
        \node [draw, rectangle, right of=E, node distance=6cm] (obs) {Observations are new cases; $y(t) = \kappa\rho E(t)$};

        \draw [->, thick] (S) -- node[midway, above] {$\frac{\beta (I+U)S}{N}$} (E);
        \draw [->, thick] (E) -- node[midway, sloped, above] {$\kappa \rho E$} (I);
        \draw [->, thick] (E) -- node[midway, sloped, below] {$\kappa (1-\rho) E$} (U);
        \draw [->, thick] (I) -- node[midway, sloped, above] {$\gamma I$} (R);
        \draw [->, thick] (U) -- node[midway, sloped, below] {$\gamma U$} (R);
        \draw [->, dashed, thick] (E) -- (obs);
    \end{tikzpicture}
    }
    \caption{\footnotesize Compartmental diagram of the SEIUR model with underreporting. Circles show the epidemiological compartments for the different states of the system. Solid arrows indicate the transitions between compartments. The dashed arrow indicate the source of the observations, which are the newly reported infected individuals.}
    \label{fig:seiur_model}
\end{figure}
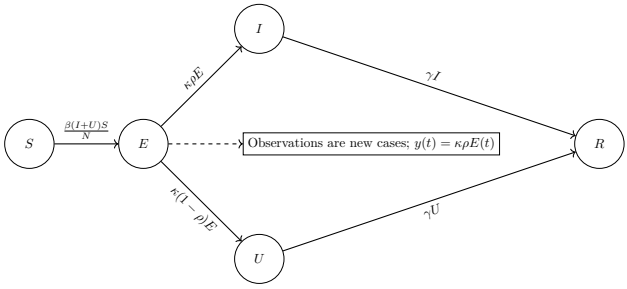

\noindent \textbf{SEIRD model}. The SEIRD model also keeps track of the number of disease-induced deaths and is given by:
\begin{equation}
\label{eq:SEIRD}
\begin{gathered}
\frac{dS}{dt} = -\beta \frac{IS}{N}, \quad 
\frac{dE}{dt} = \beta \frac{IS}{N} - \kappa E, \quad
\frac{dI}{dt} = \kappa E - \gamma I,\quad\\
\frac{dR}{dt} = \gamma (1-\rho) I,\quad
\frac{dD}{dt} = \gamma \rho I,
\end{gathered}
\end{equation}
$\beta$, $\kappa$, and $N$ are defined as above whereas $\gamma$ captures the rate from infection to recovery ($R$) or death ($D$) and $\rho$ denotes the proportion of deaths out of the total cases.  We define the initial conditions for the model as follows: $\left(S_0, E_0, I_0, R_0, D_0\right) = \left(N - \text{cases}\left[0\right], 0, \text{cases}\left[0\right], 0, 0\right)$. The observations correspond to the number of new disease-induced deaths given by $\frac{dD}{dt}$.  A compartmental diagram with the observation operator in this model is presented in Figure~\ref{fig:seird_model}. \\
\noindent \textbf{SEIUR model}. The SEIUR model keeps track of the number of reported and unreported infected cases as
\begin{equation}
\begin{gathered}
\label{eq:SEIUR}
\frac{dS}{dt} = -\beta \frac{(I+U)S}{N}, \quad
\frac{dE}{dt} = \beta \frac{(I+U)S}{N} - \kappa E, \quad\\
\frac{dI}{dt} = \kappa \rho E - \gamma I,
\frac{dU}{dt} = \kappa (1-\rho) E - \gamma U,\quad\\
\frac{dR}{dt} = \gamma (I+U), \quad
\frac{dC}{dt} = \kappa \rho E,
\end{gathered}
\end{equation}
$\beta$, $\kappa$, and $N$ are defined as above whereas $\gamma$ captures the rate from infection (reported and unreported) to recovery $\left(R\right)$ and $\rho$ denotes the reporting proportion. Also,  we are dividing the infectious people into two groups, the reported infectious people $I$, and the unreported infectious people $U$. We adopt the following initial conditions for the model: $\left(S_0, E_0, I_0, U_0, R_0, C_0\right) = \left(N - \text{cases}\left[0\right], 0, \text{cases}\left[0\right], 0, 0, \text{cases}\left[0\right]\right)$. The observations correspond to the curve of reported cases given by $\frac{dC}{dt}$. 
It is notable that if the initial conditions are known, the parameters in all the models explained above, ($\beta$,$\kappa$,$\gamma$,$\rho$), are structurally identifiable as shown in~\cite{chowell2023structural}. Moreover, note that in all these model, the basic reproduction number $\mathcal{R}_0$, is calculated using the formula: $\mathcal{R}_0 = \dfrac{\beta}{\gamma}$. Figure~\ref{fig:seiur_model} represents the compartmental diagram and the observation operator in this model. 

\section{Estimation Methods}\label{sec:method}

In this section, we detail the Bayesian and Frequentist methods for parameter estimation and forecasting. Let  $\mathbf{Y}=(y_{t_1},\cdots, y_{t_n})$ denote the observed data from which the parameters $\boldsymbol{\theta}$ are to be estimated. Here we have $\boldsymbol{\theta}=\left(\beta, \gamma, \kappa, \rho\right)$ for the SEIR and SEIRD models, and $\boldsymbol{\theta}=\left(\beta,\rho\right)$ for the SEIUR model. 

\subsection{Bayesian Inference}
Bayesian inference leverages both prior knowledge and observed data to derive the posterior distribution of the model parameters~\cite{van2021bayesian}. This approach is particularly powerful in scenarios where prior knowledge exists or when data is sparse or noisy, allowing for more robust uncertainty quantification. In Bayes' rule
\begin{equation}
    \label{eq:bayes}
    p\left(\theta \vert \mathbf{Y} \right) \propto p\left(\boldsymbol{\theta}\right)p\left(\mathbf{Y}\vert \theta\right),
\end{equation}
$p\left(\boldsymbol{\theta}\right)$ denotes the prior distribution of parameters, $p\left(\mathbf{Y}\vert \boldsymbol{\theta}\right)$ is the likelihood, and $p\left(\boldsymbol{\theta} \vert Y \right)$ denotes the posterior distribution of the parameters $\boldsymbol{\theta}$. 

\noindent \textbf{The likelihood}. Assuming the observation errors independently have normal distributions,  for the SEIR (and SEIUR) model, the likelihood model is
\begin{align}
    \label{eq:seirLd}
    y_{t_j}|\boldsymbol{\theta} &\sim N\left(\kappa\rho E_\theta\left(t_j\right), \sigma^2\right) \nonumber \\
    &\quad \text{independently for } j=1, \ldots, n,
\end{align}

where $E_\mathbf{\theta}\left(t\right)$ denotes the $E$ component of the solution to the SEIR (and SEIUR) model given parameters $\boldsymbol{\theta}$. For the SEIRD model, the likelihood is given by 
\begin{align}
    \label{eq:seirdLd}
    y_{t_j}|\boldsymbol{\theta} &\sim N\left(\gamma\rho I_\theta\left(t_j\right), \sigma^2\right) \nonumber \\
    &\quad \text{independently for } j=1, \ldots, n,
\end{align}
where $I_\mathbf{\theta}(t)$ denotes the $I$ component of the solution to the SEIRD model given parameters $\mathbf{\theta}$.

\noindent \textbf{Prior distributions}. Prior distributions encapsulate existing knowledge or beliefs about the parameters before observing the data. For simulated data, we utilize prior distributions centered around the true parameter values with varying degrees of variance, where smaller variances indicate stronger informative priors. For real data, we reference existing studies and apply both informative and uniform priors, allowing for comparative analysis across different datasets.

\noindent \textbf{Computer code}. In prior work, researchers developed a Bayesian toolbox (\texttt{BayesianFitForecast}) for disease transmission ODE modeling, designed to guide researchers through building, fitting, and diagnosing these models, model performance evaluation with several metrics (MAE, RMSE, WIS, and coverage of the $95\%$ prediction interval) and to provide options for conducting forecasts with quantified uncertainty~\cite{karami2024bayesianfitforecast}. 
In this toolbox, the user does not need to deal with Stan directly and after setting the option file, the Stan file will be automatically generated. Besides the features mentioned above, the user accesses an estimation of parameters, the trace plot, the convergence, and the histogram of the parameters.\\
\noindent \textbf{Iterations and convergence statistics.} Large enough number of iterations were conducted,  with half of these iterations used as burn-in to allow the Markov chain to reach a stable state. We use the convergence diagnostic Rhat values, which compare the between- and within-chain estimates for model parameters, to check the convergence of the Markov chains.  Posterior distributions and forecasts are summarized using posterior medians and credible intervals, providing a comprehensive view of the parameter estimates and predictive uncertainties.

\subsection{Frequentist Methods}
Unlike Bayesian inference, which incorporates prior distributions, Frequentist estimation relies solely on the observed data to estimate model parameters. This method operates under the assumption that the true parameter values are fixed but unknown, and they are inferred by optimizing a likelihood function to best align the model with the observed data. The parameters can be estimated using sample data, and additional statistical procedures are employed to quantify the uncertainty of these estimates. Model parameters can be estimated using nonlinear least squares fitting or maximum likelihood estimation (MLE)~\cite{roosa2019comparative}, with specific assumptions about the observation error in the data. In our analyses, we assume a normal error structure, consistent with the Bayesian estimation approach. Afterwards, the parameters are estimated by the nonlinear least squares method, i.e., 
\begin{equation}
    \hat{\boldsymbol{\theta}} = \arg\min_{\boldsymbol{\theta}} \sum_{j=1}^{n} \left( y_{t_j} - \mu_j \right)^2
\end{equation}
where $\mu _j = \kappa\rho E_\theta\left(t_j\right)$ for the SEIR (and SEIUR) model and $\gamma\rho I_\theta\left(t_j\right)$ for the SEIRD model.\\

\noindent \textbf{Uncertainty quantification}. To quantify parameter uncertainty, a parametric bootstrapping approach is employed~\cite{chowell2017fitting}. This method involves generating new datasets by resampling data and then estimating parameters for each resampled dataset. The steps are as follows:

\begin{enumerate}
    \item \textbf{Generate Bootstrap Samples:} Create $B$ bootstrap samples $\{y_{t_1}^b, \ldots, y_{t_n}^b\}_{b=1}^B$, where $y_{t_j}^b\sim N\left(\hat{\mu}_j, \hat{\sigma}^2\right)$ and $\hat{\mu}_j$ are estimates of $\mu _j$ by plugging in the non-linear least square estimates: $\hat{\mu}_j$ is equal to $ \hat{\kappa}\hat{\rho} E_{\hat{\theta}}\left(t_j\right)$ in the SEIR and SEIUR models and equal to $\hat{\kappa}\hat{\rho} I_{\hat{\theta}}\left(t_j\right)$ in the SEIRD model. 
    \item \textbf{Parameter Estimation:} For each bootstrap sample $\{y_{t_1}^b, \ldots, y_{t_n}^b\}$, re-estimate the model parameters and denote as $\hat{\theta}^b$, $b=1, \ldots,B$.
    \item \textbf{Construct Confidence Intervals:} Use the quantiles of $\{\hat{\theta}^b\}_{b=1}^B$ to construct the confidence intervals (CIs) for parameters. For example, the 95\% CI can be constructed from the 2.5th and 97.5th percentiles of the bootstrap estimates.
\end{enumerate}

After estimating parameters, the calibrated model can be used for forecasting. The forecasting uncertainty can also be obtained using bootstrap method. 
For each of the bootstrap estimates $\{\hat{\theta}^b: b = 1, \ldots, B\}$ obtained in step 2, generate the $h$-units ahead forecast $\hat{y}^b\left(t_{n\text{+}h}\right)$ by solving the ODE with given $\hat{\theta}^b$, and then use the 2.5th and 97.5th percentiles of $\{\hat{y}^b\left(t_{n\text{+}h}\right): b=1, \ldots,B\}$ to construct the 95\% PI for $y\left(t_{n \text{+} h}\right)$
, $h\ge 1$.\\

\noindent \textbf{Computer code}. For fitting and forecasting with the Frequentist method, we employed the MATLAB toolbox \textit{QuantDiffForecast}~\cite{chowell2024parameter}, which was developed for parameter estimation and short-term forecasting with quantified uncertainty using ODE models. It provides a comprehensive, easy-to-use framework for estimating model parameters and generating forecasts through a parametric bootstrapping approach. The toolbox is suitable for a diverse audience, including students and researchers in dynamic systems, and it supports various ODE models with different estimation methods and error structures. This toolbox offers several optimization methods; however, as previously noted, our study employs the NLS optimization method.

\section{Case Studies}\label{sec:data}
In this section, we systematically evaluate the forecasting performance of Bayesian and Frequentist estimation methods by applying them to both simulated and real epidemic data. This approach allows us to assess the robustness and applicability of each method across different contexts and datasets, thereby providing a more comprehensive understanding of their strengths and limitations.

\subsection{Simulated Data 1}
Simulated data offers a controlled environment with known true parameter values, enabling direct comparison of estimated parameters and forecasting performance. In this study, we generated simulated time-series data by numerically solving the SEIR model using the MATLAB function \textit{ode45}, with the following model parameters: population size $N = 100,000$, transmission rate $\beta = 0.5$, incubation rate $\kappa = 1$, recovery rate $\gamma = 0.25$, and reporting proportion $\rho = 0.5$. Thus, the basic reproduction number, $\mathcal{R}_0=\frac{\beta}{\gamma}=2$. The observations correspond to the curve of newly reported cases given by $\frac{dC}{dt}$. We assume that the initial conditions of the model are known ($S=N-1$, $E=0$, $I=1$, $R=0$, $C=1$), making the model parameters structurally identifiable from the observations~\cite{chowell2023structural}. Moreover, we added normally distributed noise with a standard deviation of 5 to the simulated curve $\frac{dC}{dt}$. The simulated time series covering 120 days is shown in Figure~\ref{fig:datasets}(A).

For Bayesian estimation, for parameters $\beta$, $\gamma$, and $\kappa$, we use normal distributions with true values as the means and large variances (100) for weakly informative priors. We also consider uniform distributions as priors, where the ranges are the same as those used in the Frequentist method for comparison.
The proportion parameter $\rho$ is only estimated using a uniform prior. These priors are given in Table~\ref{tab:simpriors}. For the Frequentist method, a range of 0--25 was used to infer the model parameters ($\beta$, $\gamma$, and $\kappa$) and 0--1 for the reporting proportion parameter ($\rho$). We fit the number of new daily cases in the model ($\frac{dC}{dt}$) to the simulated data using an increasing length of calibration periods: 50, 60, 70, 80, and 90 days. For each calibration period, we evaluated two forecasting horizons: 10 days and 30 days.

In Section~\ref{sec:discussion}, we discuss parameter identifiability at the 90-day calibration mark, crucial for precise forecasting and understanding epidemic dynamics. These periods and horizons enable us to assess differences in forecasting performance between Frequentist and Bayesian methods using simulated data generated from the same model.

\begin{table}[H]
\small\sf\centering
\caption{\footnotesize Prior distributions utilized in the Bayesian methods while analyzing the simulated data.}
\label{tab:simpriors}
\begin{tabular}{|c|c|c|}
\toprule
         & Bayesian (Prior1) & Bayesian (Prior2) \\
\midrule
$\beta$  & Uniform(0,25)     & Normal(0.5,100)    \\
$\gamma$ & Uniform(0,25)     & Normal(0.25,100)   \\
$\kappa$ & Uniform(0,25)     & Normal(1,100)      \\
$\rho$   & Uniform(0,1)      & Uniform(0,1)       \\
\bottomrule
\end{tabular}
\end{table}
\subsection{Simulated Data 2}
To assess the robustness of the inference methods under different epidemic dynamics, we generated a second simulated dataset using the SEIR model with a lower basic reproduction number, $\mathcal{R}_0 = \frac{\beta}{\gamma} = \frac{0.375}{0.25} = 1.5$. All model parameters remain the same as in the previous scenario, except for the transmission rate, which is set to $\beta = 0.375$. The initial conditions, noise structure, and observation process ($\frac{dC}{dt}$) are unchanged. The simulated time series spans 170 days and is shown in Figure~\ref{fig:datasets}(B).

For inference, we use the same prior distributions and parameter bounds as in the previous scenario, except for the normal prior on $\beta$, which is updated to reflect the new true value. The prior distributions used for this dataset are summarized in Table~\ref{tab:simpriors2}. Model fitting is performed over increasing calibration periods of 90, 100, 110, 120, and 130 days, with forecasting horizons of 10 and 30 days.

\begin{table}[H]
\small\sf\centering
\caption{\footnotesize Prior distributions utilized in the Bayesian methods for the second simulated dataset.}
\label{tab:simpriors2}
\begin{tabular}{|c|c|c|}
\hline
         & Bayesian (Prior1) & Bayesian (Prior2) \\ \hline
$\beta$  & Uniform(0,25)     & Normal(0.375,100) \\ \hline
$\gamma$ & Uniform(0,25)     & Normal(0.25,100)  \\ \hline
$\kappa$ & Uniform(0,25)     & Normal(1,100)     \\ \hline
$\rho$   & Uniform(0,1)      & Uniform(0,1)      \\ \hline
\bottomrule
\end{tabular}
\end{table}

\subsection{San Francisco 1918 Flu}
The 1918 influenza pandemic, or Spanish Flu, was one of the deadliest pandemics, causing millions of deaths worldwide. We analyzed the trajectory of the daily reported cases of the 1918 influenza pandemic in San Francisco using the previously defined SEIR model incorporating a reporting proportion $\rho$. The observations correspond to the number of new reported cases given by $\frac{dC}{dt}$ in the SEIR model.

Previous studies have provided critical estimates of key epidemiological parameters, such as the basic reproduction number ($\mathcal{R}_0$) and incubation rates, which are directly relevant to our analysis. Specifically, the estimate of $\mathcal{R}_0$ in the range of 2-3 for the 1918 influenza pandemic in San Francisco~\cite{cox2000global,ganyani2018assessing,chowell2007comparative,biggerstaff2014estimates} and the average incubation rate of 2 days~\cite{czumbel2018management,richardson2001evidence,book2003report,center1987health} influenced our choice of priors. These values guided our selection of priors in our Bayesian analysis, as detailed in Table~\ref{tab:sanpriors}. By grounding our prior distributions in these established findings, we ensure that the model parameters are both biologically plausible and aligned with historical data. For the Frequentist method, we considered a sufficiently wide range of 0--2 for parameters $\beta$ and $\kappa$ and 0--1 for parameter $\gamma$. The population size of San Francisco at the time was 550,000~\cite{outbreak_datasets}.

We fit the number of newly infected individuals $\frac{dC}{dt}$ in the SEIR model to the time series data of new cases. The dataset spans 62 days, capturing the progression of the fall wave in the city. The outbreak peaked on day 32, with a maximum of 2319 newly reported cases. We employ several calibration periods to conduct a comprehensive analysis: 10, 15, 20, 25, and 30 days (Figure~\ref{fig:datasets}(C)). Additionally, we consider two forecasting horizons: 10 days and 30 days. These horizons enable us to evaluate the model's predictive capabilities over both short-term and medium-term scenarios. 
\begin{table}[H]
\small\sf\centering
\caption{\footnotesize Prior distributions utilized in the Bayesian methods while analyzing the San Francisco and Cumberland flu 1918 data.}
\label{tab:sanpriors}
\begin{tabular}{|c|c|c|}
\toprule
         & Bayesian (Prior1)   &  Bayesian (Prior2)             \\
\midrule
$\beta$  & Uniform(0, 2)      & Normal(1, 0.25)     \\
$\gamma$ & Uniform(0, 1)      & Normal(0.5, 0.0625)  \\
$\kappa$ & Uniform(0, 2)      & Normal(1, 0.25)     \\
$\rho$   & Uniform(0, 1)      & Uniform(0, 1)      \\
\bottomrule
\end{tabular}
\end{table}

\subsection{Cumberland 1918 Flu}
We also analyzed the curve of daily deaths of the 1918 influenza pandemic in Cumberland, Maryland, which was also analyzed previously in~\cite{frost1919epidemiology,glezen1996emerging,andrade2021bayesian}, using the SEIR model~(\ref{eq:SEIR}). The observations correspond to the daily number of new cases given by $\frac{dC}{dt}$ in the SEIR model~(\ref{eq:SEIR}).

The basic reproduction number of the 1918 epidemic in Cumberland has been previously estimated in~\cite{okland2019race,vynnycky2007estimates} to be between 2 and 3. Therefore, with an incubation period averaging two days for influenza~\cite{czumbel2018management,richardson2001evidence,book2003report,center1987health}, we employed the same prior distributions as for the case study of the 1918 influenza pandemic in San Francisco for the Bayesian estimation approach. Moreover, Cumberland's population during the 1918 pandemic was about 10,000~\cite{andrade2021bayesian}, providing a contrast to larger urban areas and illustrating the epidemic's impact on smaller communities. The data set spans 90 days and the peak of the outbreak occurred on day 30, with a maximum of 138 newly reported cases. However, the epidemic curve shown in Figure~\ref{fig:datasets}(D) indicates the presence of two peaks which makes the comparison more interesting. Capturing the peak can be challenging when calibration lines fall between the two peaks. Those calibration lines are at 35 and 40. Indeed, significant demographic noise in the data helps compare Bayesian and Frequentist methods under noisy conditions in a small community.

We fit the number of new reported cases given by $\frac{dC}{dt}$ in the SEIR model to the time-series data of new cases. For this dataset, we use several calibration periods: 10, 15, 20, 25, 30, 35, 40, 45, 50, 55, and 60 days, which allow us to assess the model performance and accuracy over varying lengths of historical data. Additionally, we consider two forecasting horizons: 10 days and 30 days.

\subsection{Bombay Plague 1896-97}
We analyzed the weekly curve of deaths from Bombay's 1896-97 plague outbreak, which has been examined in previous publications~\cite{white2020modeling,bacaer2012model,monecke2009modelling,pell2018simple,mangiarotti2015low}. The 1896-97 Bombay plague is particularly relevant for modeling studies due to its significant impact on public health and the development of epidemiological models. This outbreak was one of the earliest instances where mathematical models, like the SIR model, were applied to understand the dynamics of infectious diseases. The work of Kermack and McKendrick on the SIR model~\cite{kermack1927contribution} was partly inspired by the need to explain the epidemic's progression. To that end, we use the SEIRD model that incorporates a class for disease related deaths. In the work of Bacaer~\cite{bacaer2012model}, it is mentioned that the average value for the incubation rate is between 3-5.5 days. Also, the basic reproduction number has been previously estimated at 1.09. Accordingly, the prior distributions are shown in Table~\ref{tab:plaguepriors}. Moreover, ranges of the parameters $\beta$, $\gamma$, and $\kappa$ are set between 0 and 10 in the Frequentist method. As already explained, the lower and upper bounds for the parameter $\rho$ are set between 0 and 1.

The population during the outbreak was around 100,000~\cite{bacaer2012model} whereas the epidemic curve comprises 34 biweekly data points, with the outbreak peaking in biweek 19 at 925 new cases. This sharp peak and temporal resolution of the data create a unique case study for comparing estimation approaches. 

We fit the number of new deaths, $\frac{dD}{dt}$, in the SEIRD model to the mortality curve. For our analysis, we investigate several calibration biweekly periods: 12, 14, 16, 18, 20, 22, 24, 26, and 28 weeks (See Figure~\ref{fig:datasets}(E)). Additionally, we consider three forecasting horizons: 2, 4, and 6 biweeks.

The 1896-97 plague in Bombay, part of the third plague pandemic, had significant public health impacts. The sharp post-peak decrease in deaths presents a forecasting challenge (See Figure~\ref{fig:datasets}(E)). Calibration lines immediately after the peak were used to thoroughly evaluate forecasting performance.

\begin{table}[H]
\small\sf\centering
\caption{\footnotesize Prior distributions utilized in the Bayesian methods while analyzing Bombay plague 1896-97.}
\label{tab:plaguepriors}
\begin{tabular}{|c|c|c|}
\toprule
         & Bayesian (Prior1)   & Bayesian (Prior2)           \\
\midrule
$\beta$  & Uniform(0, 10)     & Normal(5, 6.76)     \\
$\gamma$ & Uniform(0, 10)     & Normal(5, 6.76)     \\
$\kappa$ & Uniform(0, 10)     & Normal(5, 6.76)     \\
$\rho$   & Uniform(0, 1)      & Uniform(0, 1)      \\
\bottomrule
\end{tabular}
\end{table}

\subsection{Switzerland COVID-19}
We analyzed daily reported cases of the COVID-19 pandemic in Switzerland, which has been investigated in previous studies~\cite{lopez2024challenges,grinsztajn2021bayesian,stringhini2020seroprevalence}. For instance, the basic reproduction number ($\mathcal{R}_0$) was estimated to be 2.7 using an SEIR-type model~\cite{grinsztajn2021bayesian}, while the literature reports an average incubation period of 5-7 days for SARS-CoV-2~\cite{grinsztajn2021bayesian,lauer2020incubation}. Accordingly, we use prior distributions given in Table~\ref{tab:covpriors}. Accordingly, we assume the lower and upper bounds for the parameter $\beta$ to be 0 and 4 in the Frequentist method while they are 0 and 1 for the parameter $\rho$ as expected.

\begin{figure}[H]
    \centering
    \includegraphics[width=0.4\textwidth]{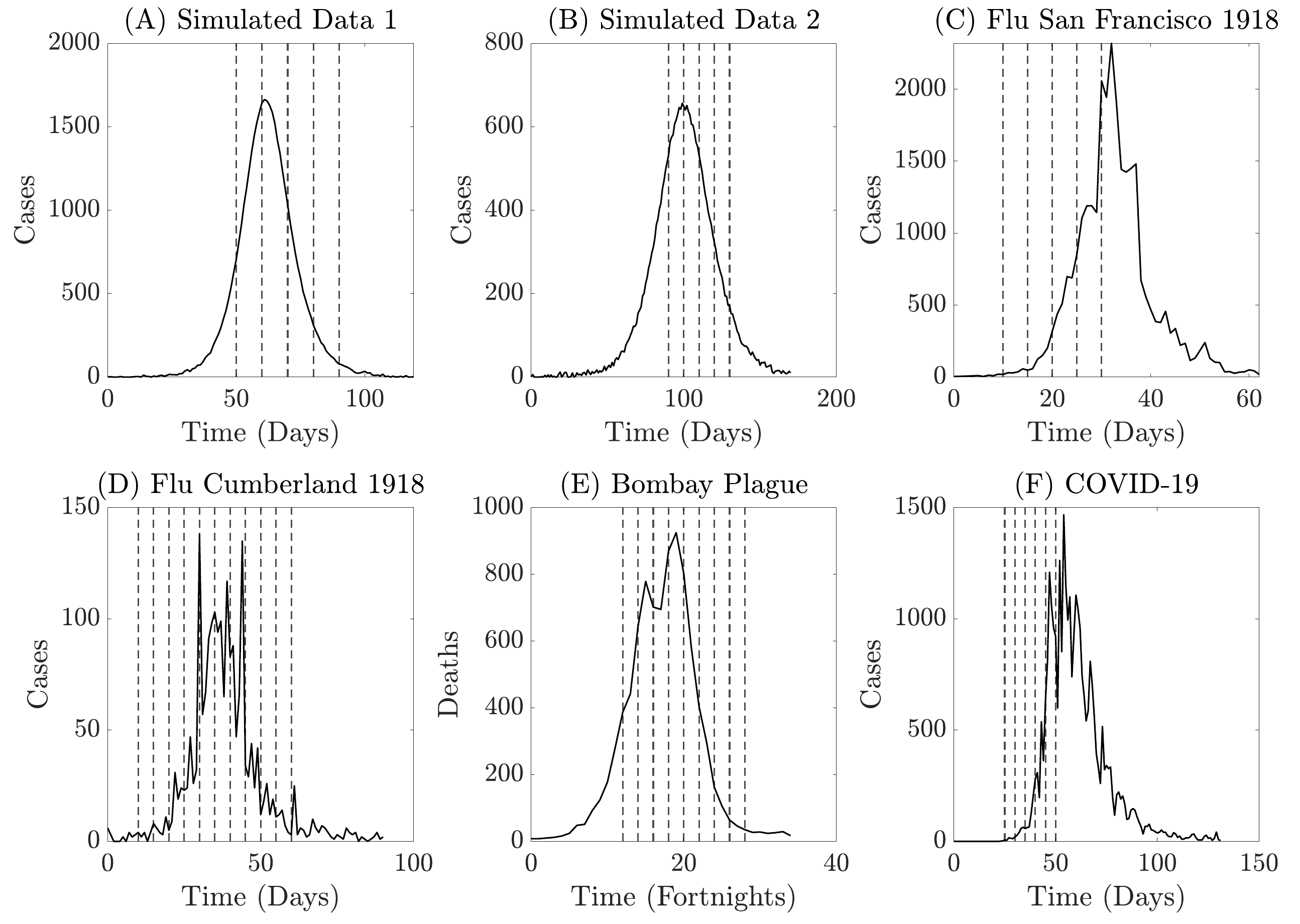}
    \caption{\footnotesize Epidemic trajectories for simulated and real epidemics analyzed in our study: A) the first simulated curve of daily new cases from the SEIR model~(\ref{eq:SEIR}) with $\mathcal{R}_0=2$, B) the second simulated curve of daily new cases from the SEIR model~(\ref{eq:SEIR}) with $\mathcal{R}_0=1.5$,  C) new daily cases of the 1918 influenza pandemic in San Francisco, D) new daily cases of the 1918 pandemic in Cumberland, Maryland, E) biweekly curve of plague deaths in Bombay,  F) new daily cases of COVID-19 in Switzerland. The timing of the calibration periods investigated for each dataset are indicated with dashed vertical lines.}
    \label{fig:datasets}
\end{figure}

The population size was set at 47,332,614, and the epidemic curve comprised over 131 days as shown in Figure~\ref{fig:datasets}(F). The outbreak peaked on day 54 with 1468 new cases. The first 22 data points are zero, indicating no reported cases initially, likely before widespread virus detection.

We fit the number of new reported cases, $\frac{dC}{dt}$, in the SEIUR model to the epidemic curve. Calibration periods of 30, 35, 40, 45, and 50 days were used to assess calibration and forecasting performance with two forecasting horizons: 10 and 30 days as shown in Figure~\ref{fig:datasets}(F).

The presence of an initial period with zero cases can significantly enhance the analysis of the comparison between Bayesian and Frequentist methods. This initial phase allows for the assessment of how each method handles the introduction and early spread of the infection, providing insights into their sensitivity and robustness in the face of an outbreak's onset. Additionally, a sharp increase immediately following the zero cases, as shown in Figure~\ref{fig:datasets}(E), presents another challenge for methods aiming to produce accurate forecasts in the early stages of the epidemic.

\begin{table}[H]
\small\sf\centering
\caption{\footnotesize Prior distributions utilized in the Bayesian methods while analyzing COVID-19 in Switzerland.}
\label{tab:covpriors}
\begin{tabular}{|c|c|c|}
\toprule
          & Bayesian (Prior1)  & Bayesian (Prior2)      \\
\midrule
$\beta$   & Uniform(0, 4)    & Normal(2, 1)     \\
$\rho$    & Uniform(0, 1)    & Uniform(0, 1)    \\
\bottomrule
\end{tabular}
\end{table}

\section{Performance Metric}\label{sec:metrics}
To comprehensively evaluate the forecasting performance, we use four performance metrics: MAE, RMSE, WIS, and the coverage of the 95\%PI~\cite{gneiting2007strictly}. While it is possible to generate $h$-time units ahead forecasts of an evolving process, where $h$ is a positive integer, those forecasts looking into the future cannot be evaluated until sufficient data for the $h$-time units ahead has been collected.

MAE is given by
\begin{equation}
\label{eq:mae}
    {\text{MAE}} = \frac{1}{N}\sum _{i=1}^{N}\left\vert f\left(t_i,\hat{\Theta}\right)-y_{t_i}\right\vert,
\end{equation}
where $t_i$, $i=1,\cdots,N$ are the time points of the time series data~\cite{kuhn2013applied}, $N$ is the number of observed data in the calibration period or forecasting period. Similarly, RMSE is defined as:
\begin{equation}
    \label{eq:rmse}
    \text{RMSE} = \sqrt{\frac{1}{N} \sum_{i=1}^{N} \left( f\left(t_i, \hat{\Theta}\right) - y_{t_i} \right)^2 }.
\end{equation}

The coverage of the 95\% PI 
corresponds to the fraction of data points that fall within the 95\% PI, calculated as:
\begin{equation}
    \label{eq:cov}
    \text{95\% PI coverage} = \frac{1}{N} \sum_{i=1}^N \mathbf{1}\left(y_{t_i}>L_{t_i}\cap y_{t_i}<U_{t_i}\right),
\end{equation}
where $L_{t_i}$ and $U_{t_i}$ are the lower and upper bounds of the 95\% PIs, respectively, $Y_{t_i}$ are the data, and $\mathbf{1}$ is an indicator variable that equals 1 if $Y_{t_i}$ is in the specified interval and 0 otherwise.

WIS~\cite{gneiting2007strictly,universitynicosia2018}, is a proper score that provides quantiles of predictive forecast distribution by
combining a set of Interval Scores (IS) for probabilistic forecasts. An IS is a simple proper score that requires only a central $(1-\alpha)\times 100\%$ PI~\cite{gneiting2007strictly} and is described as:
\begin{align}
    \label{eq:IS}
    \text{IS}_\alpha(F,y) &= \left( u - l \right) + \frac{2}{\alpha} \left( l - y \right) \times \mathbf{1}\left( y < l \right) \nonumber \\
    &\quad + \frac{2}{\alpha} \left( y - u \right) \times \mathbf{1}\left( y > u \right),
\end{align} 
where $l$ and $u$ represent the $\frac{\alpha}{2}$ and $\left(1-\frac{\alpha}{2}\right)$ quantiles of the forecast $F$, respectively. The IS consists of three distinct quantities:
\begin{itemize}
    \item The sharpness of $F$, given by the width $u - l$ of the central $(1 - \alpha ) \times  100\%$ PI.
    \item  A penalty term $\frac{2}{\alpha}\left(y-u\right)\times \mathbf{1}\left(y<l\right)$ for the observations that fall below the lower end point $l$ of the $\left(1 - \alpha \right) \times  100\%$ PI. This penalty term is directly proportional to the distance between $y$ and the lower end $l$ of PI. The strength of the penalty depends on the level $\alpha$.
    \item An analogous penalty term $\frac{2}{\alpha}\left(y-u\right)\times \mathbf{1}\left(y>u\right)$ for the observations falling above the upper limit $u$ of PI.
\end{itemize}
To provide more detailed and accurate information on the entire predictive distribution, we report several central PIs at different levels $\left(1-\alpha _1\right)<\left(1-\alpha _2\right)<\cdots<\left(1-\alpha _k\right)$ along with the predictive median, $\tilde{y}$, which can be
seen as a central prediction interval at level $\left(1-\alpha_0\right)\rightarrow 0$. This is referred to as the WIS, and it can be evaluated
as follows:
\begin{align}
    \label{eq:wis}
    \text{WIS}_{\alpha_{0:K}}(F,y) &= \frac{1}{K + \frac{1}{2}}\left( w_0 |y - \tilde{y}| \right. \nonumber \\
    &\quad + \left. \sum_{k=1}^K w_k IS_{\alpha_k}(F,y) \right),
\end{align}

where $w_k=\frac{\alpha_k}{2}$ for $k=1,\cdots,K$ and $w_0=\frac{1}{2}$. Hence, WIS can be interpreted as a measure of how close the entire
distribution is to the observation in units on the scale of the observed data~\cite{bracher2021evaluating,cramer2022evaluation}.

\section{Results}\label{sec:result}
This section presents the results of comparing the Bayesian and Frequentist methods. We provide tables and figures to show the fitting and forecasting, the estimation of the parameters, the performance metrics, and the average error bar charts. Moreover, a panel showing the comparison of true value parameters and the estimation of them is given for the simulated data. As we explained, we consider the SEIR model with the simulated data, San Francisco 1918 flu, and Cumberland 1918 flu; the SEIRD model with the Bombay plague 1896-97; and the SEIUR model with the Switzerland COVID-19. We first start the results for the simulated data. Then, we are going to have a comparison for real-world data sets.

\subsection{Simulated Data}
In this section, we analyze the comparative performance of Frequentist and Bayesian methods using two simulated datasets with different basic reproduction numbers ($\mathcal{R}_0 = 2$ and $\mathcal{R}_0 = 1.5$) to examine how epidemic intensity affects the relative performance of these estimation approaches.


\subsubsection{Simulated Data 1}

Here, we present an analysis of the first simulated data, which is generated by the forward solution of the SEIR model~(\ref{eq:SEIR}), as detailed earlier, using the parameters $\beta = 0.5$, $\kappa = 1$, $\gamma = 0.25$, $\rho = 0.5$, and $N = 100,000$. The primary goal of this analysis is to compare the forecasting performance of three methods: Bayesian (Prior1), Bayesian (Prior2), and the Frequentist method. We evaluate these methods using several calibration periods (50, 60, 70, 80, and 90 days) and two forecasting horizons (10 and 30 days).\\

Figures~\ref{fig:S-freq-bayes-sim-10} and \ref{fig:S-freq-bayes-sim-30} show that all methods fit the data well during the calibration periods, and the forecasts are mostly good. However, when using a calibration period of 50, all methods struggle for 30 days ahead forecasts because there is not enough data from the early stage of the epidemic to estimate the parameters accurately(Table~\ref{tab:S-paramessimnew30}). Bayesian methods tend to reduce errors more effectively, particularly for MAE, RMSE, and WIS, especially when forecasting over longer horizons. In contrast, the Frequentist method shows poor performance at the 70-day calibration period, occurring just after the peak of the epidemic. At the 90-day calibration period, near the epidemic's end, the Frequentist method again underperforms compared to the Bayesian methods, with higher uncertainty in the fitted curves. It is interesting to note that at calibration period 60, just before the peak, all methods correctly predict the peak and provide reliable forecasts.\\

Tables~\ref{tab:S-freq-bayes-sim-10} and~\ref{tab:S-freq-bayes-sim-30} further confirm these trends by summarizing the performance metrics across all calibration periods. Bayesian methods achieve $100\%$ coverage of the 95\% prediction interval (PI) at calibration periods 50 and 60, while the Frequentist methods minimize errors effectively, which is particularly notable before the peak. For the 30-day forecasting horizon, the Frequentist method's forecasts with longer calibration periods (e.g., 80 and 90 days) and the Bayesian method at the 90-day calibration period, display coverage rates close to 95\%. In contrast, the Frequentist method's weak performance at calibration period 70 for both forecasting horizons is evident. Achieving coverage rates of 0\% and 30\% at this stage is a clear limitation, whereas Bayesian methods consistently cover all actual cases.\\

Table~\ref{tab:S-paramessimnew30} provides further key insights. While all methods perform reasonably well in terms of forecasting, the Frequentist estimates exhibit larger deviations from the true values. For calibration periods 70 and 90, there is noticeable bias in the Frequentist estimates, resulting in significant forecast errors, particularly in terms of MAE and MSE. However, at the 50-day calibration period, despite the large deviations in Frequentist estimates from the true values, short-term forecast errors (MAE and RMSE) remain relatively low.\\

Conversely, the Bayesian methods, although displaying some uncertainty in parameter estimates with limited data (e.g., calibration period 50), improve in accuracy and converge toward the true values as the calibration period increases. The Frequentist method shows more variability in its parameter estimates, without providing a clear indication of uncertainty (as shown in Figure ~\ref{fig:S-params-sim-30}). Nevertheless, these fluctuations contribute to larger forecast errors as the calibration period extends.

\subsubsection{Simulated Data 2}

Following the same analytical framework, we present results for the second simulated dataset, which uses identical model parameters except for $\beta = 0.375$ (corresponding to $\mathcal{R}_0 = 1.5$), and evaluates the three methods across calibration periods of 90, 100, 110, 120, and 130 days with the same forecasting horizons of 10 and 30 days.

Figures~\ref{fig:S-freq-bayes-sim2-10} and \ref{fig:S-freq-bayes-sim2-30} demonstrate that all methods achieve good fits during the calibration periods for the second simulated dataset with $\mathcal{R}_0 = 1.5$. The forecasts generally perform well across the extended calibration periods, though some notable differences emerge between the methods. At the 90-day calibration period, corresponding to the early epidemic phase, all methods show reasonable forecasting ability, though the Bayesian approaches tend to provide slightly more conservative uncertainty bounds.  Interestingly, at the 110-day calibration period, right after the epidemic peak, all methods successfully capture the disease dynamics and provide reliable short-term forecasts. However, for longer forecasting horizons, the Bayesian methods demonstrate better performance in terms of MAE, RMSE, and WIS metrics, particularly evident in the 120 and 130-day calibration periods during the post-peak phase. The Frequentist method shows increased uncertainty in its predictions during these later calibration periods, while the Bayesian approaches maintain more stable forecasting performance. However, the performance of the Frequentist method remains consistently strong.

Tables~\ref{tab:S-freq-bayes-sim2-10} and \ref{tab:S-freq-bayes-sim2-30} reveal that Bayesian methods achieve adequate coverage at early calibration periods with a shorter forecasting horizon, but experience a dramatic drop to only 40\% coverage at the peak, before recovering to 100\% coverage immediately after the peak and maintaining 90\% coverage in subsequent periods. The Frequentist approach shows more consistent coverage performance for short horizons, ranging from 80--100\% across all calibration periods. For long horizons, Bayesian methods demonstrate more stable coverage performance, maintaining perfect or near-perfect coverage (86.67--100\%) throughout most periods (except calibration 100), while the Frequentist method shows variable performance, particularly declining to 83.33\% coverage immediately after the peak, but it still looks great. Put simply, the Frequentist approach makes more accurate forecasts, but its confidence intervals are not as trustworthy. In contrast, Bayesian methods are better at capturing the data points within uncertainty, though sometimes at the cost of slightly less precise predictions.

Table~\ref{tab:S-paramessim2new30} and Figure~\ref{fig:S-params-sim2-10} show that the Frequentist method generally provides more accurate estimates closer to the true values, particularly for $\beta$ and $\gamma$ parameters across most calibration periods. However, at calibration periods 90, the Frequentist estimates show notable deviations for $\kappa$, where estimates range significantly from the true value of 1. It is clearly seen that all methods are unable to provide consistent accurate estimation of parameters, where Bayesian methods show a larger deviation from the true values than the Frequentist method. 

Overall, our results based on simulated datasets indicate that for the higher $\mathcal{R}_0$ scenario (e.g., $2$), Frequentist methods may yield more accurate short-term forecasts during early epidemic stages, but their performance can become unstable as the calibration window grows. In contrast, Bayesian methods demonstrate more consistent and reliable performance across stages, except exactly at the peak of the pandemic. For the lower $\mathcal{R}_0$ scenario (e.g., $1.5$), Bayesian methods consistently outperform Frequentist approaches across all metrics and calibration periods, again except exactly at the peak of the pandemic, highlighting their robustness in slower epidemic scenarios where uncertainty plays a larger role.

\subsection{Real-World Data}
Although simulated data comparisons offer important insights, real-world datasets often present additional complexities that can challenge or even reverse conclusions drawn from simulations. For this reason, we conduct a diverse set of analyses using well-established real-world datasets. These include the San Francisco 1918 influenza pandemic, the Cumberland 1918 flu, the 1896-97 Bombay plague, and the Switzerland COVID-19 pandemic datasets. By leveraging these historical and contemporary datasets, we aim to evaluate the robustness of each method in practical settings, where data variability and noise are more pronounced.

\subsubsection{San Francisco 1918 Flu}

In this section, we present a comparison of the forecasting performance of the methods using the SEIR model ~(\ref{eq:SEIR}) applied to the San Francisco 1918 flu dataset. We evaluate the methods across several calibration periods: 10, 15, 20, 25, and 30 days, and two forecasting horizons: 10 and 30 days.\\

For calibration periods of 10 and 15 days, the overall trends are similar across all methods (Figures~\ref{fig:S-freq-bayes-san-10} and \ref{fig:S-freq-bayes-san-30}). The Bayesian methods outperform the Frequentist method for the 10-day forecasting horizon, while the Frequentist method achieves better performance for the 30-day horizon (as shown in Tables ~\ref{tab:S-freq-bayes-san-10} and \ref{tab:S-freq-bayes-san-30}). As the calibration period increases to 20, 25, and 30 days, the Bayesian method with Prior 2 demonstrates superior forecasting accuracy, consistently achieving the lowest MAE, RMSE, and WIS across all cases. Notably, at calibration periods 25 and 30, the Bayesian method with Prior 2 predicts the post-peak decline, which significantly reduces forecast errors. In contrast, the Frequentist method and Bayesian Prior 1 only capture this trend at calibration periods 25 and 30, respectively.\\

In terms of coverage rates, all methods perform similarly across most cases. Except for the calibration period of 10 and the 10 days ahead forecasts at calibration period of 15, all methods have low coverage rates of the 95\% PIs, where the peak is in the forecasting period.  These methods either predict an earlier occurrence of a lower peak (e.g., Bayesian with Prior 2), or predict a delayed occurrence of a much higher peak (e.g., Frequentist at calibration 30), and hence lead to PIs with lower coverage rates. \\

Table~\ref{tab:S-paramessannew30} reveals an intriguing result for the 30-day calibration period: the Bayesian methods, which outperform the Frequentist method, estimate the reporting proportion parameter $\rho$ as 0.03, suggesting that only 3\% of infected individuals are reported. In contrast, the Frequentist method estimates $\rho$ to be 1. Another notable observation is that the estimated parameter values are fairly consistent in the Frequentist method at calibration periods 10, 15, and 30 days. In contrast, the estimates for calibration periods 20 and 25 show noticeable differences from those at 10, 15, and 30. For instance, the transmission rate ($\beta$) is estimated to be around 1.3 during the 10, 15, and 30-day calibration periods, whereas at calibration periods 20 and 25, the estimates change to 0.71 and 1.77, respectively. It is also important to note that the estimate for $\rho$ is exceptionally small for Prior 2 and the Frequentist method at the 25-day calibration period, as well as for Prior 1 and 2 at the 30-day calibration period.

\subsubsection{Cumberland 1918 Flu}

In this section, we continue comparing the forecasting performance of the methods using the SEIR model~(\ref{eq:SEIR}) applied to the Cumberland 1918 flu dataset. We evaluate the methods using several calibration periods: 10, 15, 20, 25, 30, 35, 40, 45, 50, 55, and 60 days, and two forecasting horizons: 10 and 30 days.\\

Figures~\ref{fig:S-freq-bayes-cum-10} and \ref{fig:S-freq-bayes-cum-30} demonstrate that the trends across all methods at calibration periods of 20, 40, 50, and 60 are generally similar. However, at the 30-day calibration period (the second column), the Bayesian methods maintain an increasing trend immediately after the calibration line before decreasing, whereas the Frequentist method shows an immediate decrease.
At calibration period 30, the prediction intervals of the Bayesian methods are noticeably wider, encompassing more observations. As shown in Tables~\ref{tab:S-freq-bayes-cum-10} and~\ref{tab:S-freq-bayes-cum-30}, the Bayesian methods outperform the Frequentist method in terms of performance metrics at both forecasting horizons for the calibration periods of 25, 30, and 45 days. At other calibration periods, such as 15, 20, 35, and 40 days, the Frequentist method exhibits better forecasting performance, particularly at calibration periods 35 and 40. For longer calibration periods, such as 50, 55, and 60 days, the results from all methods converge, with no single method consistently dominating the others in terms of performance metrics.\\

Table~\ref{tab:S-paramescumnew30} provides the parameter estimates for each calibration period. The lowest estimate for the reporting proportion parameter ($\rho$), 0.05, was obtained at the 30-day calibration period by the Frequentist method. This period, just before the epidemic peak, is where the Frequentist method underperforms compared to the Bayesian methods, indicating its inability to accurately capture the peak in this dataset, despite its better average performance over most of other calibration periods.

\subsubsection{Bombay Plague 1896-97}

Here, we use the SEIRD model~(\ref{eq:SEIRD}) to analyze the Bombay plague epidemic. The forecast performance is evaluated over several calibration bi weekly periods: 12, 14, 16, 18, 20, 22, 24, 26, and 28, with forecasting horizons of 2, 4, and 6 biweekly. \\

Figures~\ref{fig:S-freq-bayes-plague-2}--\ref{fig:S-freq-bayes-plague-6} shows that the forecast trends are generally similar across all methods at calibration periods 12, 16, 24, and 28, with one key exception at the 20-period calibration (the third column). At this point, the Bayesian method correctly forecasts a mild decrease, while the Frequentist method incorrectly predicts continued growth. This observation aligns with earlier findings, suggesting that the Frequentist method struggles to accurately identify the pandemic's peak.   Tables~\ref{tab:S-freq-bayes-plague-2}--\ref{tab:S-freq-bayes-plague-6} further illustrate that the Frequentist method performs better before the peak or once the peak has passed, such as calibration periods 16, 22, 24, 26, and 28. \\

Table~\ref{tab:S-paramesplaguenew6} presents the parameter estimates. A notable finding is that after the pandemic's peak (calibration period 20), both the Bayesian methods and the Frequentist method converge on similar estimates for the reporting proportion parameter. However, based on our earlier discussion, we might conclude that the Frequentist method provides more accurate estimates for other parameters after the peak, whereas the Bayesian method yields more accurate estimates around the peak of the pandemic.

\subsubsection{Switzerland COVID-19}
In this section, we apply the SEIUR model with $\kappa = 0.2$ and $\gamma = 0.25$ to analyze Switzerland's COVID-19 data. Forecast performance is evaluated over several calibration periods: 30, 35, 40, 45, and 50, with forecasting horizons of 10 and 30 days. \\

Figure~\ref{fig:S-fit-covid-10} highlights the weak performance of the Bayesian methods in the early stages (calibration period 30), where the Frequentist method provides an excellent forecast. However, all methods fail to accurately forecast the trend at calibration period 50. On the other hand, Figure ~\ref{fig:S-fit-covid-30} illustrate the weak performance of all methods for the 30-day forecasting horizon, likely due to the large forecast horizon relative to the calibration period. Additionally, this poor performance may be influenced by the choice of model (ODE system) or the limited number of parameters being estimated ($\beta$ and $\rho$), which restricts the flexibility of the methods. Overestimation of case incidences is observed in most cases. Nevertheless, Tables~\ref{tab:S-fit-covid-10} and \ref{tab:S-fit-covid-30} show that the Bayesian methods outperform the Frequentist method in terms of performance metrics.\\

Table~\ref{tab:S-paramescovidnew30} provides the parameter estimates. The better performance of the Frequentist method at calibration period 30 might be due to its handling of the reporting proportion parameter in the model, even though this parameter is very small. In contrast, the Bayesian methods estimate this parameter to be approximately zero, resulting in an almost zero forecast for case incidences.

\subsubsection{Summary Across Datasets and Phases}
No single forecasting method works best in every situation. The Frequentist method does well around the peak of simulated data and after the peak using real datasets. However, it does not do as well right before the peak, where Bayesian methods, especially those with uniform priors, tend to give better predictions. This shows that the best method depends on the stage of the outbreak and the type of data available. Moreover, Bayesian methods are better at capturing the data points within the prediction interval in most cases, which is especially useful when the data are limited or noisy. On the other hand, Frequentist methods usually give more accurate single-number predictions, but their confidence in those predictions can be off during certain parts of an outbreak. Overall, how well each method works depends a lot on how the disease spreads—each has its strengths at different times in the epidemic.

\section{Discussion}\label{sec:discussion}

In this study, we evaluated the performance of Bayesian and Frequentist methods for epidemic forecasting across both simulated and real-world datasets. Our findings highlight that the relative effectiveness of these approaches is highly context-dependent. Bayesian methods, in most cases, offer superior uncertainty quantification throughout the epidemic timeline, providing key advantages in specific forecasting scenarios, particularly when data are sparse or noisy. Meanwhile, Frequentist methods generally achieve greater point prediction accuracy, reflected in lower MAE, RMSE, and WIS metrics across most settings, though their confidence intervals may be less reliable during certain epidemic phases. Notably, performance varies considerably with epidemic characteristics such as the basic reproduction number, with each method demonstrating strengths at different stages of the epidemic based on underlying transmission dynamics. In fact, in terms of datasets and phases, Bayesian methods, especially those with uniform priors, tend to produce more accurate predictions in the early phase before the peak. In contrast, the Frequentist approach shows better performance near the peak in simulated outbreaks and after the peak in real-world scenarios, but is less reliable before the peak

These conclusions are substantiated by key metrics such as MAE and RMSE for point prediction accuracy, alongside 95\% prediction interval coverage and WIS for evaluating uncertainty quantification. For ease of reference, we have also provided two summary tables (\ref{tab:S-epidemic-phases} and \ref{tab:S-summary}).

Parameter identifiability is crucial for both Bayesian and Frequentist approaches, impacting the reliability and interpretability of models. In Bayesian methods, poor identifiability can lead to diffuse or multimodal posterior distributions, creating high uncertainty and computational challenges in MCMC simulations, though priors can sometimes mitigate these issues. In Frequentist methods, non-identifiability results in high variance or biased estimators, compromising the reliability of point estimates and confidence intervals and undermining desirable asymptotic properties like consistency and efficiency. Therefore, ensuring parameter identifiability is essential for the effectiveness and credibility of both statistical frameworks. This is particularly relevant to our study's findings, as the identifiability of key parameters can directly influence the performance of both approaches during different epidemic phases.

The bias observed in Figure~\ref{fig:S-freq-bayes-sim-10} at the 90-day calibration can be explained by the Frequentist method's systematic overestimation of key parameters, particularly the transmission rate $\beta$ (10.09 vs. 0.5 the true value), which compounds over longer calibration periods. While Bayesian methods benefit from prior regularization that constrains parameters to more realistic ranges, the Frequentist approach lacks this constraint and converges to parameter values that overfit the calibration data, leading to increasingly biased projections as the forecast horizon extends.

When using Bayesian methods, we recommend relying on parameter estimates where Rhat, the convergence diagnostic value, is below 1.1~\cite{margossian2021nested,burkner2017brms}. In this study, we increased the number of iterations (50{,}000 for most cases, except where indicated in Table~\ref{tab:S-niter}) to ensure convergence in all cases. It is important to note that while different forecasting horizons can lead to slightly varying results, this variation becomes negligible once the chains have converged. This methodological enhancement, including using larger forecasting horizons and applying a zoomed version of those results for shorter horizons, ensured more robust and computationally efficient results. For example, in the simulated data with a 10-day forecasting horizon, we extracted the results from a 30-day forecast by focusing on the first 10 steps beyond the calibration line. This approach led to convergent chains with fewer Bayesian method iterations, further improving computational efficiency. When using the Frequentist method, to mitigate the impact of initial guesses, we increased the number of initial guesses for optimization to 60.

Accurate epidemic forecasting is crucial for informed public health decision-making, particularly when allocating resources and implementing interventions to control the spread of infectious diseases. The public health relevance of our study lies in its potential to guide policymakers in selecting the most appropriate forecasting method based on their specific priorities and epidemic context. Bayesian methods offer advantages when robust uncertainty quantification is paramount, particularly valuable when decisions must be made with careful consideration of prediction intervals and prior knowledge. In contrast, Frequentist methods provide consistently more accurate point predictions and reliable parameter estimates, making them valuable when precise forecasts of key epidemiological parameters are the primary concern. This distinction is crucial for optimizing the timing and effectiveness of public health interventions based on whether decision-makers prioritize prediction accuracy or uncertainty assessment.

A key strength of this study is the rigorous comparative analysis of two widely used estimation methods across multiple datasets with varying epidemic parameters. By using both simulated and real-world epidemic data, we evaluated the performance of Bayesian and Frequentist methods under various conditions, enhancing the robustness and credibility of our findings. Additionally, we employed multiple performance metrics, such as MAE, RMSE, 95\% PI coverage, and WIS, to comprehensively assess the strengths and weaknesses of each method. This multi-metric evaluation provides deeper insights into the trade-offs between point prediction accuracy and uncertainty quantification.

However, our study is not exempt from limitations. First, we acknowledge that our analysis is based on deterministic compartmental models, which represent a simplification of the complex and inherently stochastic nature of real epidemics. Second, our study relies on specific compartmental models (SEIR, SEIRD, SEIUR) for each dataset, which may not capture the full complexity of disease dynamics. Exploring alternative models and validating these methods across a broader range of datasets would enhance the generalizability of our findings. Third, the performance of each method may vary based on factors such as population size, data quality, transmission dynamics, and epidemic parameters like the basic reproduction number, which should be considered in future studies. Fourth, Frequentist estimates sometimes exhibited persistent bias even with longer calibration periods, likely due to convergence issues and the presence of non-convex error landscapes. This underscores the importance of robust initialization, parameter constraints, and diagnostic checks when using optimization-based methods for epidemic forecasting. Finally, future research should extend this comparative framework to stochastic models and explore more complex error structures that better reflect the realities of epidemic data. Additionally, further investigation into the sensitivity of Bayesian forecasts to the choice of priors and the conditions under which the likelihood sufficiently dominates the prior will be essential to refining model-based forecasting strategies.

It is important to emphasize that our comparison is not intended to be an exhaustive evaluation of all possible Bayesian or Frequentist approaches. Instead, we focus on a representative and commonly used setup involving a normal error structure applied to compartmental ODE models, estimated using Stan (Bayesian) and nonlinear least squares (Frequentist). By holding the error structure constant across paradigms, we aimed to isolate and better understand differences in performance attributable to the estimation frameworks themselves, rather than variations in model formulation or error assumptions. This constrained comparison highlights practical trade-offs that may arise when choosing an estimation method in real-world forecasting scenarios, but future work should extend this analysis to incorporate alternative likelihoods, estimation techniques, and model structures. While our results reveal distinct strengths of each method in different epidemic phases, these findings reflect our specific model and estimation choices, which were guided by epidemiological relevance and common usage in the literature.

In summary, our findings highlight that neither estimation framework is universally superior; each has distinct strengths and limitations shaped by data availability, epidemic phase, model parameters, and underlying assumptions. Frequentist methods generally yield more accurate point forecasts, while Bayesian methods often provide more reliable uncertainty quantification. The optimal approach depends on whether the primary goal is forecast accuracy or capturing uncertainty comprehensively. As discussed, ensuring parameter identifiability is a critical prerequisite for the success of both frameworks. These insights can help researchers and public health practitioners align estimation strategies with specific forecasting goals to support more effective epidemic response efforts.


\begin{verbatim}
\begin{dci}
Not applicable.
\end{dci}
\end{verbatim}

\begin{verbatim}
\begin{funding}
HK is supported by a 2CI Fellowship
from Georgia State University. G.C.
is partially supported by NSF grants
2125246 and 2026797.
\end{funding}
\end{verbatim}

\begin{verbatim}
\begin{sm}
To typeset a
  "Supplemental material" section.
\end{sm}
\end{verbatim}

\begin{verbatim}
HK, RL, and GC conceived the study.
HK conducted the modeling analysis
and analyzed the data. HK, RL, and
GC wrote the first draft of the
manuscript. HK, RL, PS, and GC
contributed to writing and reviewing
subsequent drafts of the paper.
\end{verbatim}

\clearpage  
\onecolumn  
\appendix
\renewcommand{\thefigure}{S\arabic{figure}}
\renewcommand{\thetable}{S\arabic{table}}
\renewcommand{\theequation}{S\arabic{equation}}
\setcounter{figure}{0}
\setcounter{table}{0}
\setcounter{equation}{0}

\section{Supplementary Material}

\section*{Simulated Data 1}

\begin{figure}[H]
    \centering
    \includegraphics[width=1\textwidth]{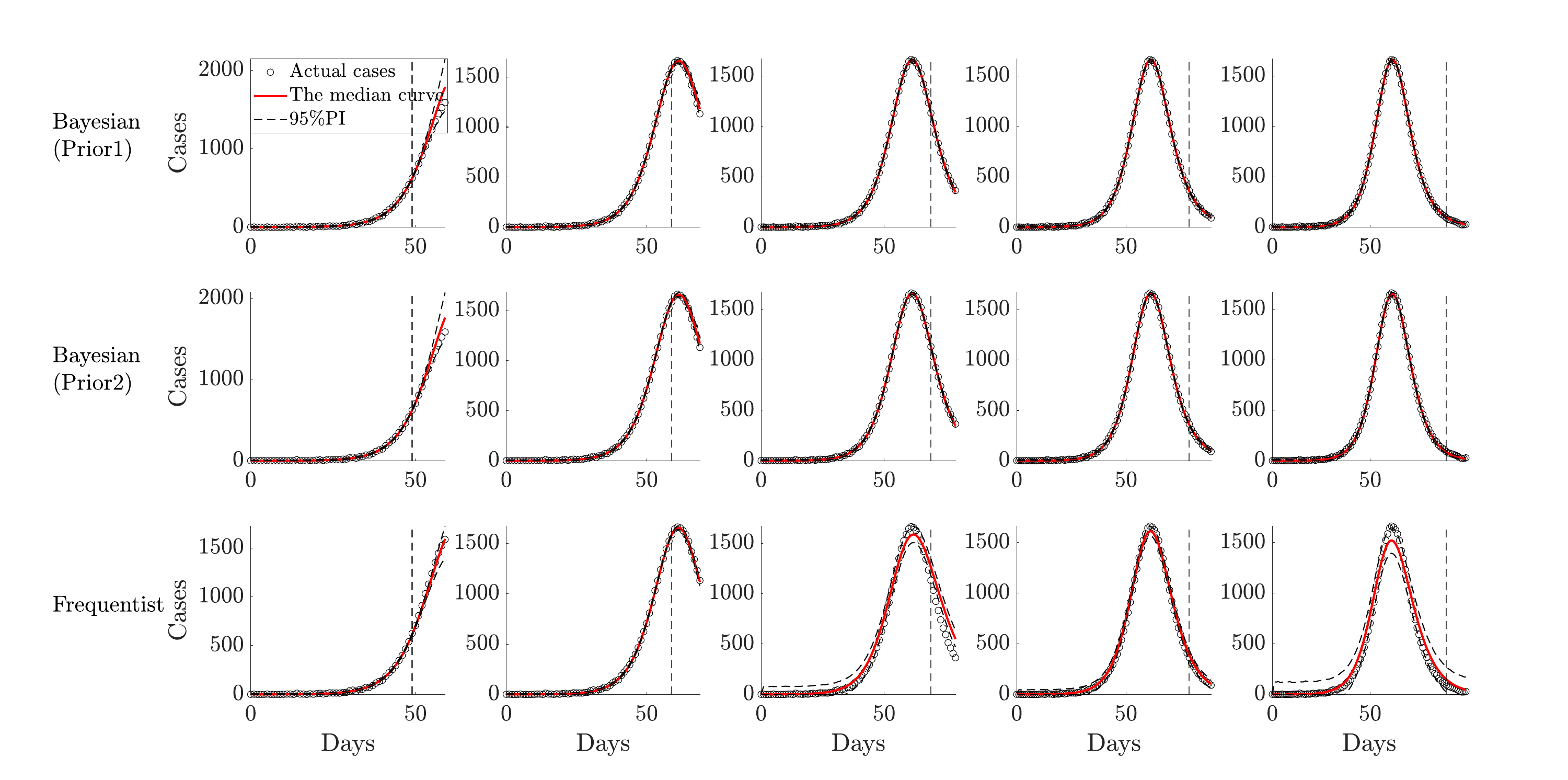}
    \caption{\footnotesize Panel showcasing the fitting of three different methods for the first simulated data, using several calibration periods: 50, 60, 70, 80, and 90 days, with a forecasting horizon of 10 days. The SEIR model is utilized, fitting the data to the newly infected people $\frac{dC}{dt}$. The population size is 100,000, assuming a normal error structure and the initial condition $(99999,0,1,0,1)$.}
    \label{fig:S-freq-bayes-sim-10}
\end{figure}

\begin{table}[H]
\centering
\begin{tabular}{|c|c|c|c|c|}
\hline
\textbf{Calibration} & \textbf{Metrics} & \textbf{Bayesian (Prior1)} & \textbf{Bayesian (Prior2)} & \textbf{Frequentist} \\ \hline
\multirow{4}{*}{50}  & MAE              & 68.63                      & 58.92                      & \textbf{26.65}       \\ \cline{2-5} 
                     & RMSE             & 92.89                      & 81.15                      & \textbf{29.72}       \\ \cline{2-5} 
                     & WIS              & 37.45                      & 32.23                      & \textbf{19.41}       \\ \cline{2-5} 
                     & 95\%PI           & \textbf{100}               & \textbf{100}               & 60.00                \\ \hline
\multirow{4}{*}{60}  & MAE              & 19.22                      & 16.05                      & \textbf{8.83}        \\ \cline{2-5} 
                     & RMSE             & 24.72                      & 20.71                      & \textbf{10.54}       \\ \cline{2-5} 
                     & WIS              & 10.4                       & 8.88                       & \textbf{5.65}        \\ \cline{2-5} 
                     & 95\%PI           & 100                        & 100                        & 100.00               \\ \hline
\multirow{4}{*}{70}  & MAE              & 7.45                       & \textbf{6.34}              & 196.71               \\ \cline{2-5} 
                     & RMSE             & 9.59                       & \textbf{8.46}              & 197.60               \\ \cline{2-5} 
                     & WIS              & 4.29                       & \textbf{3.88}              & 166.38               \\ \cline{2-5} 
                     & 95\%PI           & \textbf{100}               & \textbf{100}               & 0.00                 \\ \hline
\multirow{4}{*}{80}  & MAE              & \textbf{6.78}              & 6.89                       & 19.10                \\ \cline{2-5} 
                     & RMSE             & \textbf{7.82}              & 7.92                       & 20.48                \\ \cline{2-5} 
                     & WIS              & \textbf{4.28}              & 4.34                       & 10.83                \\ \cline{2-5} 
                     & 95\%PI           & 80                         & 70                         & \textbf{100.00}      \\ \hline
\multirow{4}{*}{90}  & MAE              & \textbf{4.85}              & 6.87                       & 34.50                \\ \cline{2-5} 
                     & RMSE             & \textbf{6.57}              & 7.86                       & 35.75                \\ \cline{2-5} 
                     & WIS              & \textbf{3.37}              & 4.14                       & 19.93                \\ \cline{2-5} 
                     & 95\%PI           & 80                         & \textbf{100}               & \textbf{100.00}      \\ \hline
\end{tabular}
\caption{\footnotesize The performance metrics of three different methods for the first simulated data, using several calibration periods: 50, 60, 70, 80, and 90 days, with a forecasting horizon of 10 days. The SEIR model is utilized, fitting the data to the newly infected people $\frac{dC}{dt}$. The population size is 100,000, assuming a normal error structure and the initial condition $(99999,0,1,0,1)$.}
\label{tab:S-freq-bayes-sim-10}
\end{table}

\begin{figure}[H]
    \centering
    \includegraphics[width=1\textwidth]{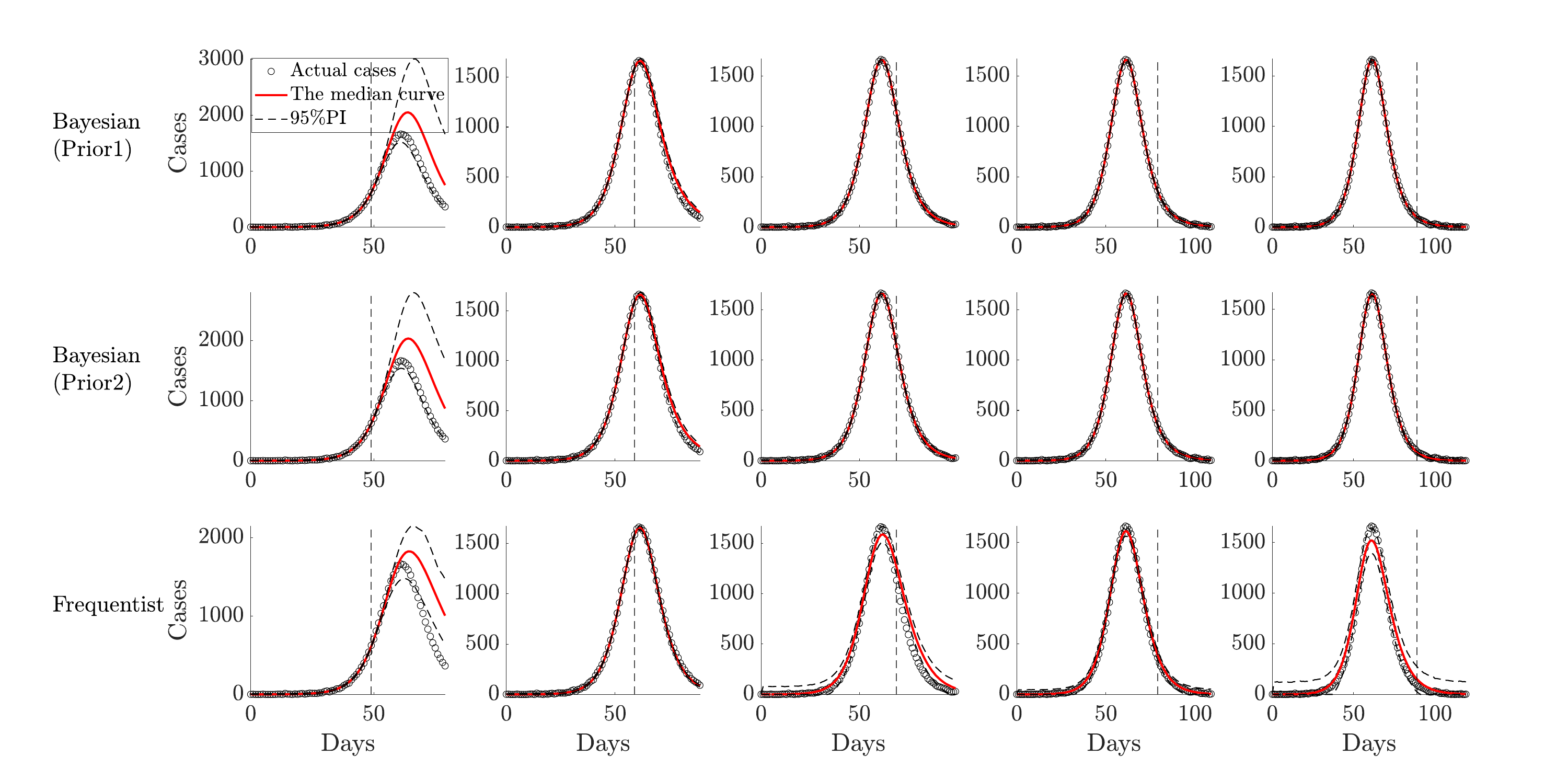}
    \caption{\footnotesize Panel showcasing the fitting of three different methods for the first simulated data, using several calibration periods: 50, 60, 70, 80, and 90 days, with a forecasting horizon of 30 days. The SEIR model is utilized, fitting the data to the newly infected people $\frac{dC}{dt}$. The population size is 100,000, assuming a normal error structure and the initial condition $(99999,0,1,0,1)$.}
    \label{fig:S-freq-bayes-sim-30}
\end{figure}

\begin{table}[H]
\centering
\begin{tabular}{|c|c|c|c|c|}
\hline
\textbf{Calibration} & \textbf{Metrics} & \textbf{Bayesian (Prior1)} & \textbf{Bayesian (Prior2)} & \textbf{Frequentist} \\ \hline
\multirow{4}{*}{50}  & MAE              & 357.94                     & 387.56                     & \textbf{325.35}      \\ \cline{2-5} 
                     & RMSE             & \textbf{424.47}            & 468.50                     & 430.90               \\ \cline{2-5} 
                     & WIS              & \textbf{210.91}            & 228.93                     & 246.22               \\ \cline{2-5} 
                     & 95\%PI           & \textbf{100}               & \textbf{100}               & 43.33                \\ \hline
\multirow{4}{*}{60}  & MAE              & 49.65                      & 41.98                      & \textbf{14.50}       \\ \cline{2-5} 
                     & RMSE             & 55.13                      & 46.67                      & \textbf{16.19}       \\ \cline{2-5} 
                     & WIS              & 25.51                      & 20.84                      & \textbf{9.85}        \\ \cline{2-5} 
                     & 95\%PI           & \textbf{100}               & \textbf{100}               & 60.00                \\ \hline
\multirow{4}{*}{70}  & MAE              & 5.85                       & \textbf{5.05}              & 131.62               \\ \cline{2-5} 
                     & RMSE             & 7.43                       & \textbf{6.56}              & 144.89               \\ \cline{2-5} 
                     & WIS              & 3.67                       & \textbf{3.46}              & 103.42               \\ \cline{2-5} 
                     & 95\%PI           & \textbf{100}               & \textbf{100}               & 30.00                \\ \hline
\multirow{4}{*}{80}  & MAE              & \textbf{6.7}               & 6.77                       & 9.84                 \\ \cline{2-5} 
                     & RMSE             & \textbf{7.82}              & 7.88                       & 12.94                \\ \cline{2-5} 
                     & WIS              & \textbf{4.28}              & 4.32                       & 7.09                 \\ \cline{2-5} 
                     & 95\%PI           & 80                         & 76.67                      & \textbf{100.00}      \\ \hline
\multirow{4}{*}{90}  & MAE              & \textbf{4.39}              & 4.97                       & 16.86                \\ \cline{2-5} 
                     & RMSE             & \textbf{5.46}              & 6.69                       & 21.97                \\ \cline{2-5} 
                     & WIS              & \textbf{2.79}              & 3.35                       & 13.11                \\ \cline{2-5} 
                     & 95\%PI           & 90                         & 96.67                      & \textbf{100.00}      \\ \hline
\end{tabular}
\caption{\footnotesize The performance metrics of three different methods for the first simulated data, using several calibration periods: 50, 60, 70, 80, and 90 days, with a forecasting horizon of 30 days. The SEIR model is utilized, fitting the data to the newly infected people $\frac{dC}{dt}$. The population size is 100,000, assuming a normal error structure and the initial condition $(99999,0,1,0,1)$.}
\label{tab:S-freq-bayes-sim-30}
\end{table}

\begin{table}[H]
\centering
\begin{tabular}{|c|c|c|c|c|c|}
\hline
\multirow{21}{*}{\textbf{\rotatebox[origin=c]{90}{Estimate (CI)}}} & \textbf{calibration} & \textbf{parameter} & \textbf{Bayesian(Prior1)} & \textbf{Bayesian (Prior2)} & \textbf{Frequentist} \\ \cline{2-6} 
                                         & \multirow{4}{*}{50}  & $\beta$            & 0.29 (0.23,2.92)          & 1.1 (0.25,3.36)            & 20.51 (17.00,22.50)  \\ \cline{3-6} 
                                         &                      & $\gamma$           & 0.11 (0.06,0.78)          & 0.38 (0.07,0.93)           & 8.32 (4.38,9.61)     \\ \cline{3-6} 
                                         &                      & $\kappa$           & 6.06 (0.06,23.94)         & 0.23 (0.05,18.55)          & 0.12 (0.10,0.16)     \\ \cline{3-6} 
                                         &                      & $\rho$             & 0.61 (0.43,0.91)          & 0.64 (0.44,0.92)           & 0.67 (0.53,0.86)     \\ \cline{2-6} 
                                         & \multirow{4}{*}{60}  & $\beta$            & 0.29 (0.27,0.63)          & 0.31 (0.27,0.73)           & 0.55 (0.53,0.56)     \\ \cline{3-6} 
                                         &                      & $\gamma$           & 0.11 (0.09,0.28)          & 0.12 (0.09,0.33)           & 0.28 (0.26,0.29)     \\ \cline{3-6} 
                                         &                      & $\kappa$           & 5.46 (0.46,23.76)         & 3.06 (0.39,19.84)          & 0.83 (0.78,0.89)     \\ \cline{3-6} 
                                         &                      & $\rho$             & 0.46 (0.44,0.48)          & 0.46 (0.45,0.48)           & 0.48 (0.48,0.51)     \\ \cline{2-6} 
                                         & \multirow{4}{*}{70}  & $\beta$            & 0.49 (0.38,0.64)          & 0.48 (0.37,0.63)           & 16.16 (16.15,16.17)  \\ \cline{3-6} 
                                         &                      & $\gamma$           & 0.22 (0.16,0.28)          & 0.21 (0.15,0.28)           & 9.54 (8.61,10.54)    \\ \cline{3-6} 
                                         &                      & $\kappa$           & 0.71 (0.44,1.25)          & 0.74 (0.45,1.34)           & 0.25 (0.21,0.32)     \\ \cline{3-6} 
                                         &                      & $\rho$             & 0.47 (0.46,0.47)          & 0.47 (0.46,0.47)           & 0.65 (0.6,0.74)      \\ \cline{2-6} 
                                         & \multirow{4}{*}{80}  & $\beta$            & 0.43 (0.4,0.45)           & 0.42 (0.4,0.45)            & 1.74 (1.61, 1.97)    \\ \cline{3-6} 
                                         &                      & $\gamma$           & 0.18 (0.17,0.2)           & 0.18 (0.17,0.2)            & 0.92 (0.82, 1.06)    \\ \cline{3-6} 
                                         &                      & $\kappa$           & 0.94 (0.83,1.05)          & 0.94 (0.84,1.06)           & 0.29 (0.26, 0.34)    \\ \cline{3-6} 
                                         &                      & $\rho$             & 0.46 (0.46,0.47)          & 0.46 (0.46,0.47)           & 0.53 (0.51, 0.56)    \\ \cline{2-6} 
                                         & \multirow{4}{*}{90}  & $\beta$            & 0.46 (0.44,0.47)          & 0.79 (0.77,0.81)           & 10.09 (9.94, 11.7)   \\ \cline{3-6} 
                                         &                      & $\gamma$           & 0.2 (0.19,0.21)           & 0.3 (0.27,0.32)            & 5.91 (5.69, 7.31)    \\ \cline{3-6} 
                                         &                      & $\kappa$           & 0.83 (0.78,0.89)          & 0.27 (0.25,0.3)            & 0.26 (0.23, 0.29)    \\ \cline{3-6} 
                                         &                      & $\rho$             & 0.47 (0.46,0.47)          & 0.43 (0.43,0.45)           & 0.6 (0.58, 0.63)     \\ \hline
\end{tabular}
\caption{\footnotesize The parameter estimation of three different methods for the first simulated data, using several calibration periods: 50, 60, 70, 80, and 90 days. The SEIR model is utilized, fitting the data to the newly infected people $\frac{dC}{dt}$. The population size is 100,000, assuming a normal error structure and the initial condition $(99999,0,1,0,1)$.}
\label{tab:S-paramessimnew30}
\end{table}

\begin{figure}[H]
    \centering
    \includegraphics[width=0.9\textwidth]{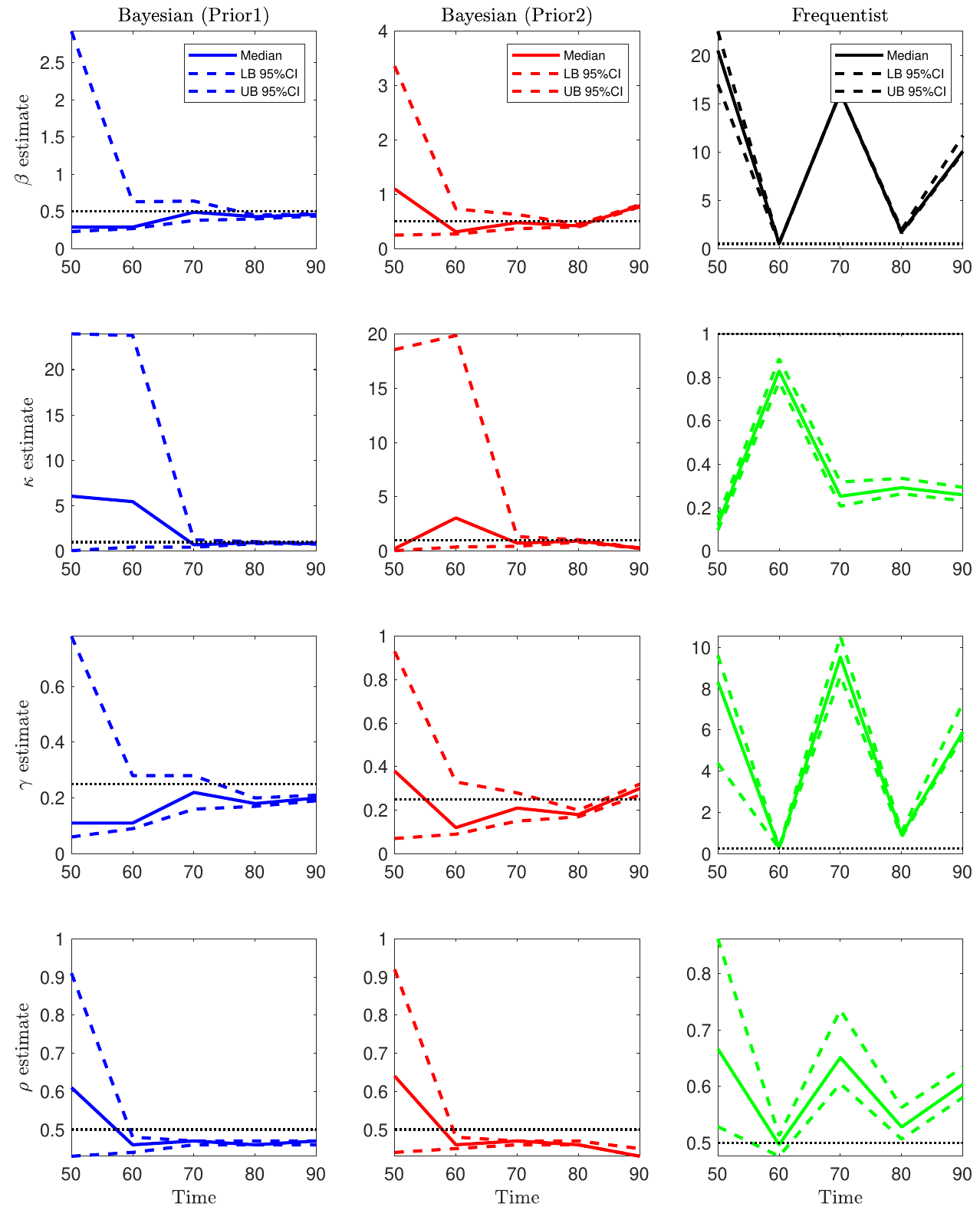}
    \caption{\footnotesize Parameter estimation panel of three different methods for the first simulated data, using several calibration periods: 50, 60, 70, 80, and 90 days. The SEIR model is utilized, fitting the data to the newly infected people $\frac{dC}{dt}$. The population size is 100,000, assuming a normal error structure and the initial condition $(99999,0,1,0,1)$.}
    \label{fig:S-params-sim-30}
\end{figure}

\pagebreak

\section*{Simulated Data 2}
\begin{figure}[H]
    \centering
    \includegraphics[width=1\textwidth]{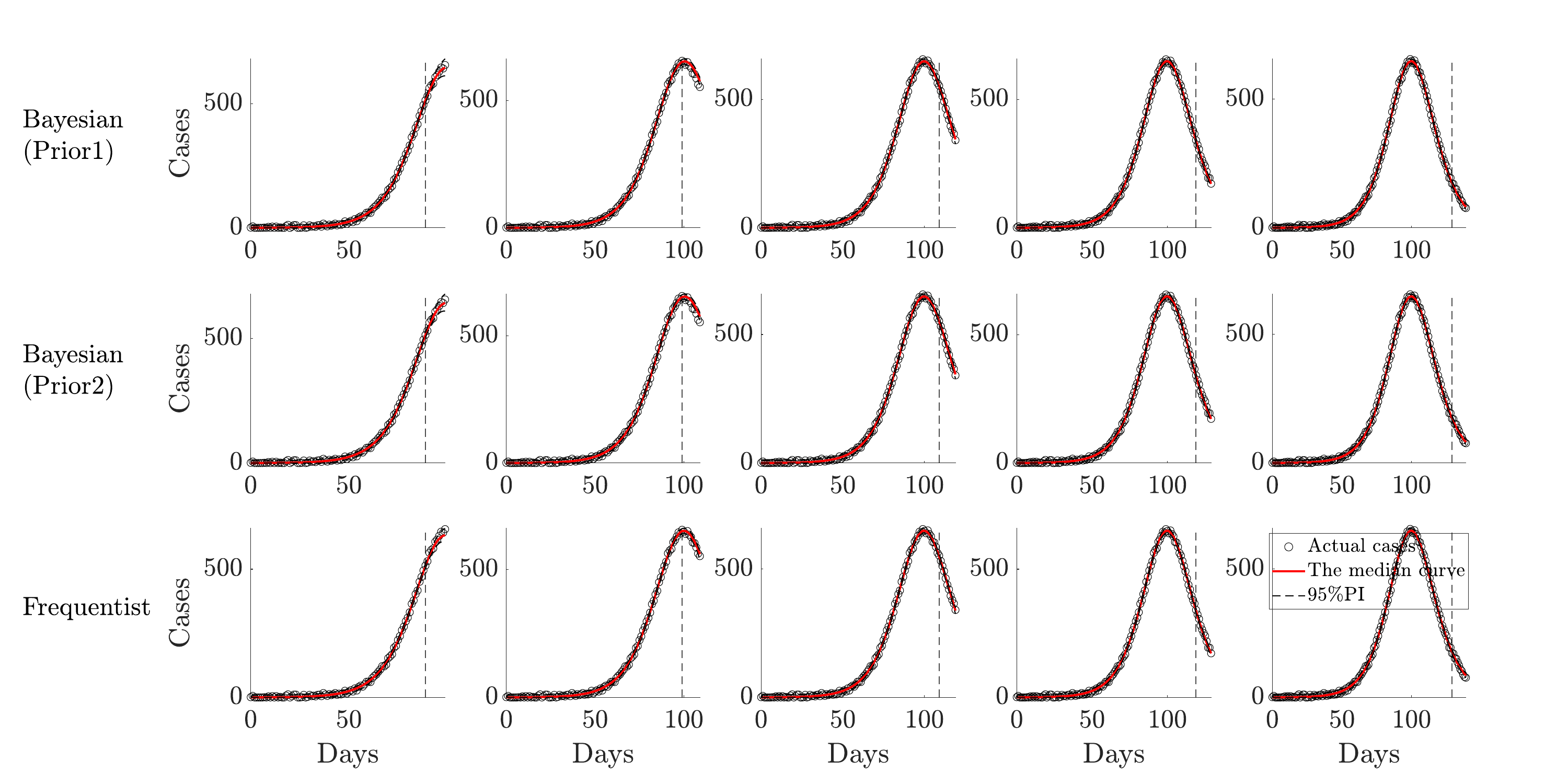}
    \caption{\footnotesize Panel showcasing the fitting of three different methods for the second simulated data, using several calibration periods: 90, 100, 110, 120, and 130 days, with a forecasting horizon of 10 days. The SEIR model is utilized, fitting the data to the newly infected people $\frac{dC}{dt}$. The population size is 100,000, assuming a normal error structure and the initial condition $(99999,0,1,0,1)$.}
    \label{fig:S-freq-bayes-sim2-10}
\end{figure}

\begin{table}[H]
\centering
\begin{tabular}{|c|c|c|c|c|}
\hline
\textbf{Calibration} & \textbf{Metrics} & \textbf{Bayesian (Prior1)} & \textbf{Bayesian (Prior2)} & \textbf{Frequentist} \\ \hline
\multirow{4}{*}{90}  & MAE              & \textbf{5.31}              & 6.56                       & 10.45                \\ \cline{2-5} 
                     & RMSE             & \textbf{6.56}              & 7.75                       & 12.34                \\ \cline{2-5} 
                     & WIS              & \textbf{3.67}              & 4.02                       & 6.73                 \\ \cline{2-5} 
                     & 95\%PI           & \textbf{100}               & \textbf{100}               & 90                   \\ \hline
\multirow{4}{*}{100} & MAE              & 14.08                      & 14.68                      & \textbf{7.95}        \\ \cline{2-5} 
                     & RMSE             & 15.98                      & 16.96                      & \textbf{9.15}        \\ \cline{2-5} 
                     & WIS              & 10.18                      & 10.82                      & \textbf{4.84}        \\ \cline{2-5} 
                     & 95\%PI           & 40                         & 40                         & \textbf{80}          \\ \hline
\multirow{4}{*}{110} & MAE              & 4.65                       & 5.00                       & \textbf{3.77}        \\ \cline{2-5} 
                     & RMSE             & 6.04                       & 6.38                       & \textbf{4.30}        \\ \cline{2-5} 
                     & WIS              & 3.10                       & 3.25                       & \textbf{2.20}        \\ \cline{2-5} 
                     & 95\%PI           & \textbf{100}               & \textbf{100}               & \textbf{100}         \\ \hline
\multirow{4}{*}{120} & MAE              & 5.18                       & \textbf{5.06}              & 5.23                 \\ \cline{2-5} 
                     & RMSE             & 6.17                       & \textbf{6.04}              & 6.25                 \\ \cline{2-5} 
                     & WIS              & 3.20                       & \textbf{3.12}              & 3.41                 \\ \cline{2-5} 
                     & 95\%PI           & \textbf{90}                & \textbf{90}                & \textbf{90}          \\ \hline
\multirow{4}{*}{130} & MAE              & 5.29                       & \textbf{5.26}              & 5.91                 \\ \cline{2-5} 
                     & RMSE             & 6.31                       & \textbf{6.29}              & 6.77                 \\ \cline{2-5} 
                     & WIS              & 3.35                       & \textbf{3.34}              & 3.69                 \\ \cline{2-5} 
                     & 95\%PI           & \textbf{90}                & \textbf{90}                & \textbf{90}          \\ \hline
\end{tabular}
\caption{\footnotesize The performance metrics of three different methods for the second simulated data, using several calibration periods: 90, 100, 110, 120, and 130 days, with a forecasting horizon of 10 days. The SEIR model is utilized, fitting the data to the newly infected people $\frac{dC}{dt}$. The population size is 100{,}000, assuming a normal error structure and the initial condition $(99999,0,1,0,1)$.}
\label{tab:S-freq-bayes-sim2-10}
\end{table}

\begin{figure}[H]
    \centering
    \includegraphics[width=1\textwidth]{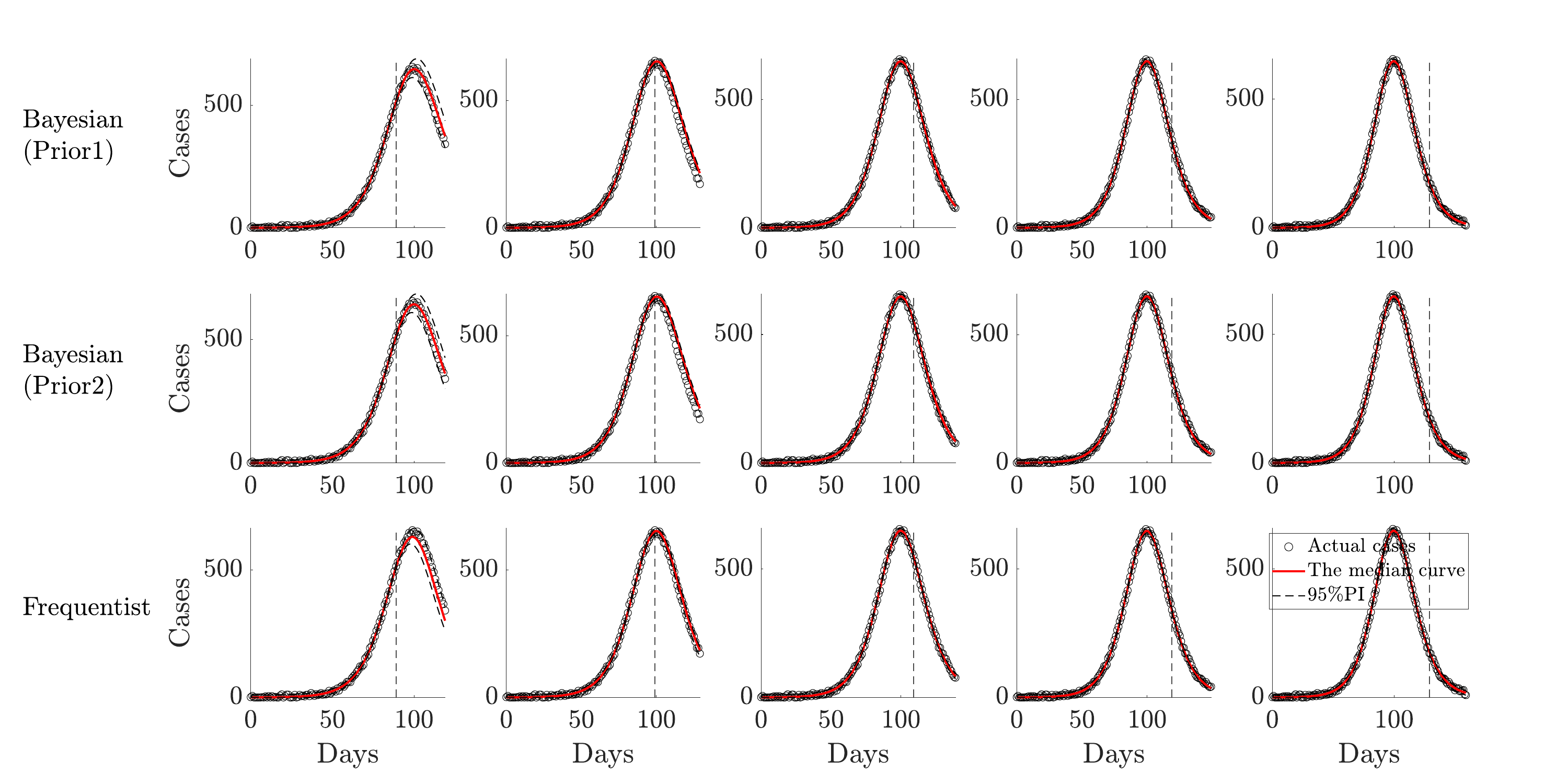}
    \caption{\footnotesize Panel showcasing the fitting of three different methods for the second simulated data, using several calibration periods: 90, 100, 110, 120, and 130 days, with a forecasting horizon of 30 days. The SEIR model is utilized, fitting the data to the newly infected people $\frac{dC}{dt}$. The population size is 100,000, assuming a normal error structure and the initial condition $(99999,0,1,0,1)$.}
    \label{fig:S-freq-bayes-sim2-30}
\end{figure}

\begin{table}[H]
\centering
\begin{tabular}{|c|c|c|c|c|}
\hline
\textbf{Calibration} & \textbf{Metrics} & \textbf{Bayesian (Prior1)} & \textbf{Bayesian (Prior2)} & \textbf{Frequentist} \\ \hline
\multirow{4}{*}{90}  & MAE              & 13.41                      & \textbf{9.72}              & 12.14                \\ \cline{2-5} 
                     & RMSE             & 17.57                      & \textbf{12.08}             & 13.36                \\ \cline{2-5} 
                     & WIS              & 8.53                       & \textbf{6.77}              & 6.95                 \\ \cline{2-5} 
                     & 95\%PI           & \textbf{100}               & \textbf{100}               & 93.33                \\ \hline
\multirow{4}{*}{100} & MAE              & 31.44                      & 28.47                      & \textbf{4.63}        \\ \cline{2-5} 
                     & RMSE             & 34.31                      & 32.42                      & \textbf{5.46}        \\ \cline{2-5} 
                     & WIS              & 26.26                      & 23.43                      & \textbf{2.84}        \\ \cline{2-5} 
                     & 95\%PI           & 13.33                      & 13.33                      & \textbf{93.33}       \\ \hline
\multirow{4}{*}{110} & MAE              & \textbf{4.67}              & 4.82                       & 5.08                 \\ \cline{2-5} 
                     & RMSE             & \textbf{5.78}              & 5.92                       & 6.26                 \\ \cline{2-5} 
                     & WIS              & 3.20                       & \textbf{3.19}              & 3.25                 \\ \cline{2-5} 
                     & 95\%PI           & \textbf{100}               & \textbf{100}               & 83.33                \\ \hline
\multirow{4}{*}{120} & MAE              & 5.42                       & \textbf{5.27}              & 4.73                 \\ \cline{2-5} 
                     & RMSE             & 6.57                       & 6.42                       & \textbf{5.59}        \\ \cline{2-5} 
                     & WIS              & 3.39                       & 3.30                       & \textbf{2.92}        \\ \cline{2-5} 
                     & 95\%PI           & 86.67                      & 86.67                      & \textbf{90.00}       \\ \hline
\multirow{4}{*}{130} & MAE              & 4.84                       & \textbf{4.82}              & 25.61                \\ \cline{2-5} 
                     & RMSE             & 5.83                       & \textbf{5.81}              & 28.50                \\ \cline{2-5} 
                     & WIS              & 3.05                       & \textbf{3.04}              & 16.21                \\ \cline{2-5} 
                     & 95\%PI           & 86.67                      & 86.67                      & \textbf{96.67}       \\ \hline
\end{tabular}
\caption{\footnotesize The performance metrics of three different methods for the second simulated data, using several calibration periods: 90, 100, 110, 120, and 130 days, with a forecasting horizon of 30 days. The SEIR model is utilized, fitting the data to the newly infected people $\frac{dC}{dt}$. The population size is 100{,}000, assuming a normal error structure and the initial condition $(99999,0,1,0,1)$.}
\label{tab:S-freq-bayes-sim2-30}
\end{table}

\begin{table}[H]
\centering
\begin{tabular}{|c|c|c|c|c|c|}
\hline
\multirow{21}{*}{\rotatebox[origin=c]{90}{\textbf{Estimate (CI)}}} & \textbf{Calibration} & \textbf{Parameter} & \textbf{Bayesian (Prior1)} & \textbf{Bayesian (Prior2)} & \textbf{Frequentist} \\ \cline{2-6}
& \multirow{4}{*}{90}  & $\beta$  & 19.22 (6.86, 24.73) & 8.86 (1.76, 22.22) & 0.4 (0.38, 6.11) \\ \cline{3-6}
&                        & $\gamma$ & 13.67 (4.89, 17.68) & 6.37 (1.29, 15.81) & 0.3 (0.28, 4.45) \\ \cline{3-6}
&                        & $\kappa$ & 0.25 (0.21, 0.27)   & 0.26 (0.23, 0.35)  & 5.76 (0.3, 7.43) \\ \cline{3-6}
&                        & $\rho$   & 0.59 (0.51, 0.69)   & 0.59 (0.52, 0.68)  & 0.61 (0.55, 0.68) \\ \cline{2-6}
& \multirow{4}{*}{100} & $\beta$  & 18.33 (5.26, 24.64) & 4.33 (3.74, 5.07)  & 4.79 (0.9, 5.29) \\ \cline{3-6}
&                        & $\gamma$ & 13.13 (3.84, 17.69) & 14.5 (3.66, 29.17) & 3.57 (0.69, 3.91) \\ \cline{3-6}
&                        & $\kappa$ & 0.25 (0.23, 0.28)   & 2.65 (0.88, 14.88) & 0.31 (0.29, 0.63) \\ \cline{3-6}
&                        & $\rho$   & 0.61 (0.57, 0.65)   & 0.12 (0.01, 0.3)   & 0.64 (0.61, 0.67) \\ \cline{2-6}
& \multirow{4}{*}{110} & $\beta$  & 1.67 (0.35, 7.07)   & 0.7 (0.58, 1)      & 0.45 (0.44, 2.34) \\ \cline{3-6}
&                        & $\gamma$ & 1.24 (0.26, 4.99)   & 4.82 (4.03, 5.84)  & 0.34 (0.33, 1.75) \\ \cline{3-6}
&                        & $\kappa$ & 0.37 (0.25, 14.76)  & 2.06 (0.78, 6.25)  & 2.55 (0.35, 2.7) \\ \cline{3-6}
&                        & $\rho$   & 0.62 (0.56, 0.66)   & 1.52 (0.59, 4.44)  & 0.66 (0.63, 0.68) \\ \cline{2-6}
& \multirow{4}{*}{120} & $\beta$  & 1.33 (0.42, 2.03)   & 0.34 (0.25, 0.66)  & 1.88 (0.88, 2) \\ \cline{3-6}
&                        & $\gamma$ & 1.00 (0.31, 1.49)   & 0.61 (0.57, 0.65)  & 1.43 (0.69, 1.53) \\ \cline{3-6}
&                        & $\kappa$ & 0.43 (0.33, 2.31)   & 4.65 (4.2, 5.27)   & 0.4 (0.39, 0.72) \\ \cline{3-6}
&                        & $\rho$   & 0.63 (0.59, 0.66)   & 1.34 (0.42, 2.12)  & 0.66 (0.64, 0.68) \\ \cline{2-6}
& \multirow{4}{*}{130} & $\beta$  & 1.85 (1.46, 2.29)   & 1 (0.31, 1.54)     & 0.41 (0.41, 0.42) \\ \cline{3-6}
&                        & $\gamma$ & 1.36 (1.10, 1.66)   & 0.42 (0.32, 2.39)  & 0.31 (0.3, 0.31) \\ \cline{3-6}
&                        & $\kappa$ & 0.35 (0.31, 0.41)   & 0.63 (0.59, 0.66)  & 4.13 (3.91, 4.47) \\ \cline{3-6}
&                        & $\rho$   & 0.61 (0.58, 0.64)   & 4.58 (4.05, 5.19)  & 0.64 (0.63, 0.64) \\ \hline
\end{tabular}
\caption{\footnotesize The parameter estimation of three different methods for the second simulated data, using several calibration periods: 90, 100, 110, 120, and 130 days. The SEIR model is utilized, fitting the data to the newly infected people $\frac{dC}{dt}$. The population size is 100{,}000, assuming a normal error structure and the initial condition $(99999,0,1,0,1)$.}
\label{tab:S-paramessim2new30}
\end{table}

\begin{figure}[H]
    \centering
    \includegraphics[width=0.9\textwidth]{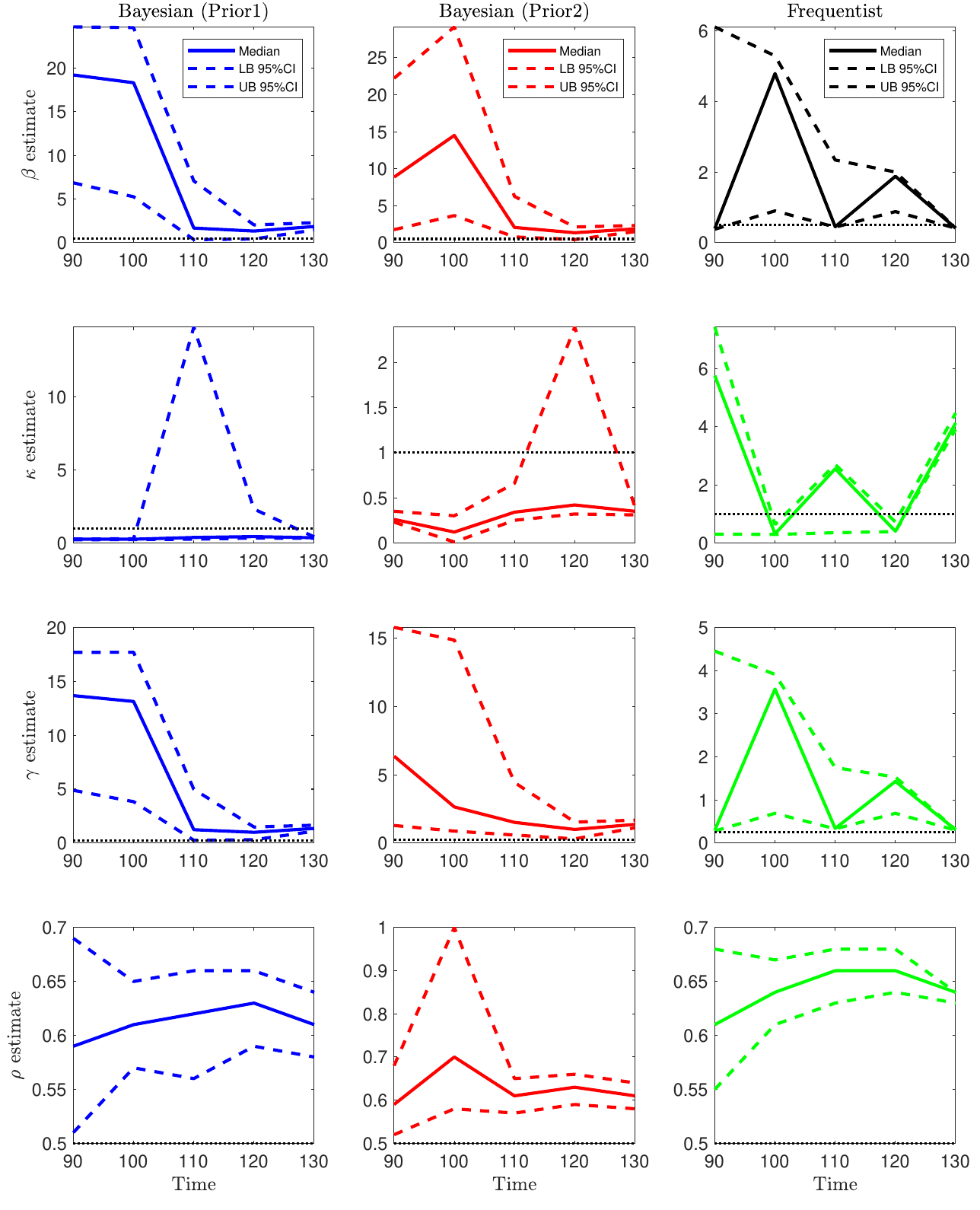}
    \caption{\footnotesize Parameter estimation panel of three different methods for the first simulated data, using several calibration periods: 50, 60, 70, 80, and 90 days. The SEIR model is utilized, fitting the data to the newly infected people $\frac{dC}{dt}$. The population size is 100,000, assuming a normal error structure and the initial condition $(99999,0,1,0,1)$.}
    \label{fig:S-params-sim2-10}
\end{figure}

\pagebreak


\section*{Flu San Francisco 1918}
\begin{figure}[H]
    \centering
    \includegraphics[width=1\textwidth]{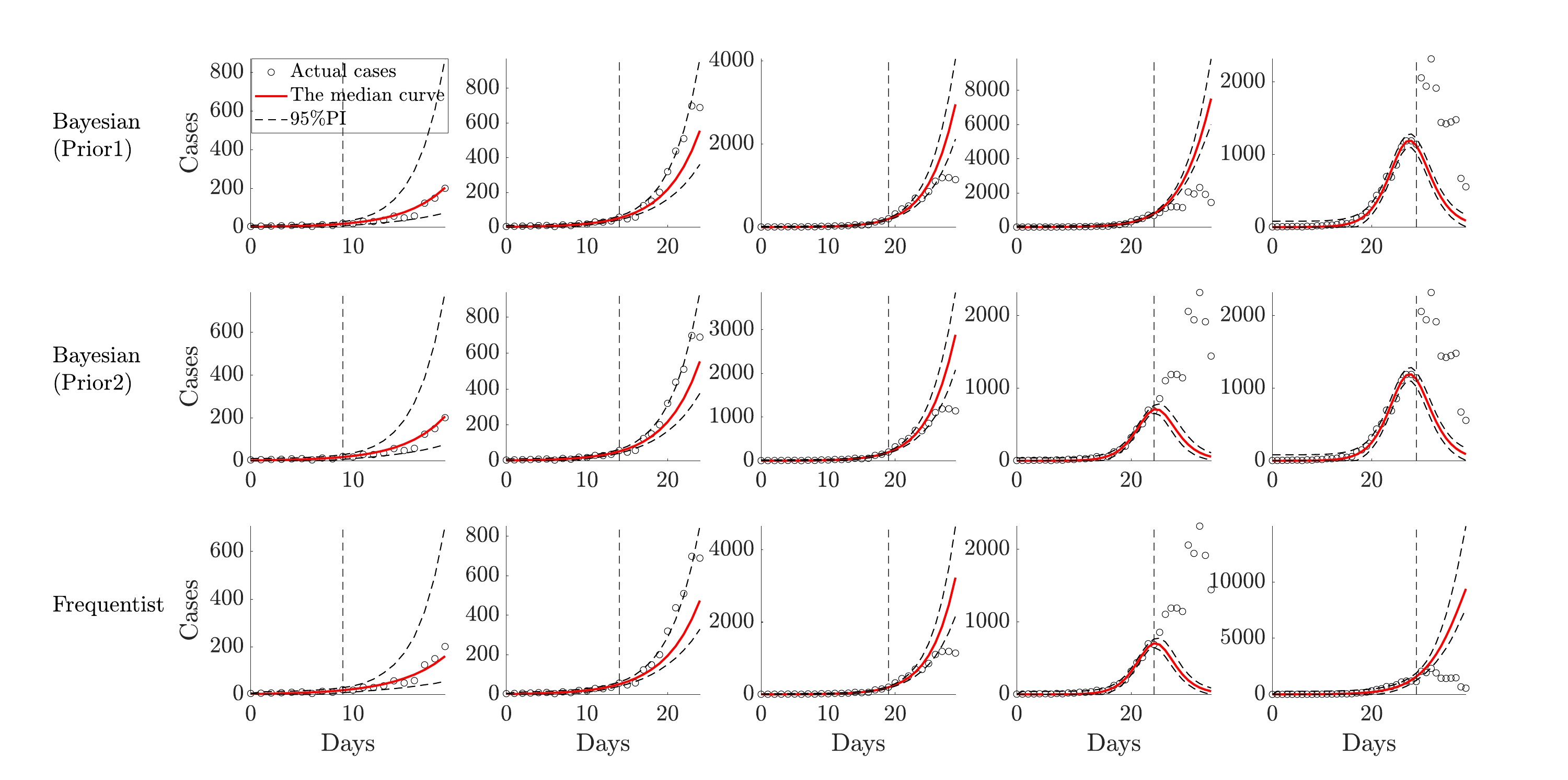}
    \caption{\footnotesize Panel showcasing the fitting of three different methods for the San Francisco 1918 flu data, using several calibration periods: 10, 15, 20, 25, and 30 days, with a forecasting horizon of 10 days. The SEIR model is utilized, fitting the data to the newly infected people $\frac{dC}{dt}$. The population size is 550,000, assuming a normal error structure and the initial condition $(549996,0,4,0,4)$.}
    \label{fig:S-freq-bayes-san-10}
\end{figure}

\begin{table}[H]
\centering
\begin{tabular}{|c|c|c|c|c|}
\hline
\textbf{Calibration} & \textbf{Metrics} & \textbf{Bayesian (Prior1)} & \textbf{Bayesian (Prior2)} & \textbf{Frequentist} \\ \hline
\multirow{4}{*}{10}  & MAE              & \textbf{11.1}              & 11.93                      & 14.86                \\ \cline{2-5} 
                     & RMSE             & \textbf{16.63}             & 17.22                      & 19.02                \\ \cline{2-5} 
                     & WIS              & 12.62                      & 12.13                      & \textbf{11.22}       \\ \cline{2-5} 
                     & 95\%PI           & 100                        & 100                        & 100.00               \\ \hline
\multirow{4}{*}{15}  & MAE              & \textbf{92.5}              & 93.09                      & 119.15               \\ \cline{2-5} 
                     & RMSE             & \textbf{122.53}            & 123.32                     & 157.26               \\ \cline{2-5} 
                     & WIS              & \textbf{57.96}             & 59.34                      & 77.70                \\ \cline{2-5} 
                     & 95\%PI           & \textbf{60}                & \textbf{60}                & 40.00                \\ \hline
\multirow{4}{*}{20}  & MAE              & 436.3                      & \textbf{420.02}            & 499.71               \\ \cline{2-5} 
                     & RMSE             & 706.86                     & \textbf{679.62}            & 815.39               \\ \cline{2-5} 
                     & WIS              & 331.46                     & \textbf{316.9}             & 374.67               \\ \cline{2-5} 
                     & 95\%PI           & 50                         & 50                         & 50.00                \\ \hline
\multirow{4}{*}{25}  & MAE              & 1970.91                    & \textbf{1195.34}           & 1219.06              \\ \cline{2-5} 
                     & RMSE             & 2692.68                    & \textbf{1367.50}           & 1387.82              \\ \cline{2-5} 
                     & WIS              & 1756.89                    & \textbf{1068.99}           & 1187.28              \\ \cline{2-5} 
                     & 95\%PI           & 0                          & \textbf{0}                 & 0.00                 \\ \hline
\multirow{4}{*}{30}  & MAE              & 1086.32                    & \textbf{1084}              & 3608.35              \\ \cline{2-5} 
                     & RMSE             & 1137.21                    & \textbf{1134.93}           & 4628.69              \\ \cline{2-5} 
                     & WIS              & 1049.69                    & \textbf{1045.93}           & 3165.42              \\ \cline{2-5} 
                     & 95\%PI           & \textbf{0}                 & 0                          & \textbf{10.00}       \\ \hline
\end{tabular}
\caption{\footnotesize The performance metrics of three different methods for the San Francisco 1918 flu data, using several calibration periods: 10, 15, 20, 25, and 30 days, with a forecasting horizon of 10 days. The SEIR model is utilized, fitting the data to the newly infected people $\frac{dC}{dt}$. The population size is 550,000, assuming a normal error structure and the initial condition $(549996,0,4,0,4)$.}
\label{tab:S-freq-bayes-san-10}
\end{table}

\begin{figure}[H]
    \centering
    \includegraphics[width=1\textwidth]{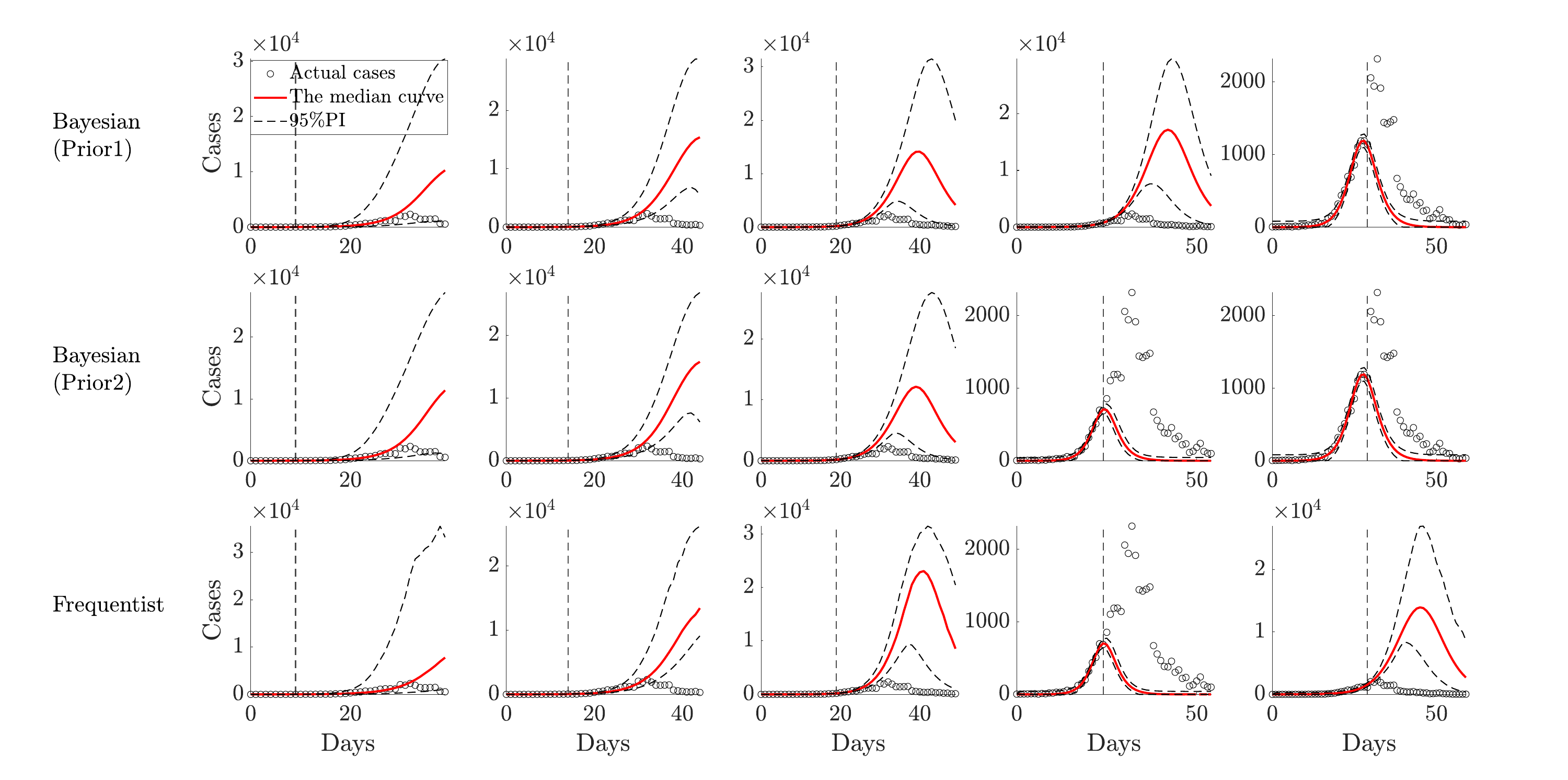}
    \caption{\footnotesize Panel showcasing the fitting of three different methods for the San Francisco 1918 flu data, using several calibration periods: 10, 15, 20, 25, and 30 days, with a forecasting horizon of 30 days. The SEIR model is utilized, fitting the data to the newly infected people $\frac{dC}{dt}$. The population size is 550,000, assuming a normal error structure and the initial condition $(549996,0,4,0,4)$.}
    \label{fig:S-freq-bayes-san-30}
\end{figure}

\begin{table}[H]
\centering
\begin{tabular}{|c|c|c|c|c|}
\hline
\textbf{Calibration} & \textbf{Metrics} & \textbf{Bayesian (Prior1)} & \textbf{Bayesian (Prior2)} & \textbf{Frequentist} \\ \hline
\multirow{4}{*}{10}  & MAE              & 1724.99                    & 1900.63                    & \textbf{1046.27}     \\ \cline{2-5} 
                     & RMSE             & 3344.69                    & 3699.73                    & \textbf{2182.80}     \\ \cline{2-5} 
                     & WIS              & 1126.75                    & 1212.83                    & \textbf{775.45}      \\ \cline{2-5} 
                     & 95\%PI           & 93.33                      & 93.33                      & 93.33                \\ \hline
\multirow{4}{*}{15}  & MAE              & 3828.36                    & 3909.01                    & \textbf{3031.52}     \\ \cline{2-5} 
                     & RMSE             & 6458.55                    & 6602.22                    & \textbf{5237.40}     \\ \cline{2-5} 
                     & WIS              & 2960.42                    & 3100.44                    & \textbf{2587.34}     \\ \cline{2-5} 
                     & 95\%PI           & \textbf{46.67}             & \textbf{46.67}             & 43.33                \\ \hline
\multirow{4}{*}{20}  & MAE              & 5697.21                    & \textbf{4778.58}           & 9406.47              \\ \cline{2-5} 
                     & RMSE             & 7472.54                    & \textbf{6262.83}           & 12504.65             \\ \cline{2-5} 
                     & WIS              & 3875.92                    & \textbf{3312.69}           & 6867.45              \\ \cline{2-5} 
                     & 95\%PI           & 16.67                      & 16.67                      & 16.67                \\ \hline
\multirow{4}{*}{25}  & MAE              & 8126.51                    & \textbf{706.21}            & 715.89               \\ \cline{2-5} 
                     & RMSE             & 9902.49                    & \textbf{942.90}            & 954.47               \\ \cline{2-5} 
                     & WIS              & 5987.67                    & \textbf{634.87}            & 693.63               \\ \cline{2-5} 
                     & 95\%PI           & 0                          & 0                          & 0.00                 \\ \hline
\multirow{4}{*}{30}  & MAE              & 486.28                     & \textbf{485.25}            & 7416.28              \\ \cline{2-5} 
                     & RMSE             & 682.13                     & \textbf{680.74}            & 8613.05              \\ \cline{2-5} 
                     & WIS              & 456                        & \textbf{454.45}            & 6057.73              \\ \cline{2-5} 
                     & 95\%PI           & 16.67                      & 16.67                      & \textbf{3.33}        \\ \hline
\end{tabular}
\caption{\footnotesize The performance metrics of three different methods for the San Francisco 1918 flu data, using several calibration periods: 10, 15, 20, 25, and 30 days, with a forecasting horizon of 30 days. The SEIR model is utilized, fitting the data to the newly infected people $\frac{dC}{dt}$. The population size is 550,000, assuming a normal error structure and the initial condition $(549996,0,4,0,4)$.}
\label{tab:S-freq-bayes-san-30}
\end{table}

\begin{table}[H]
\centering
\begin{tabular}{|l|c|c|c|c|c|}
\hline
\multirow{21}{*}{\textbf{\rotatebox[origin=c]{90}{Estimate (CI)}}} & \textbf{calibration} & \textbf{parameter} & \textbf{Bayesian(Prior1)} & \textbf{Bayesian (Prior2)} & \textbf{Frequentist} \\ \cline{2-6} 
                   & \multirow{4}{*}{10}  & $\beta$            & 1.18 (0.52,1.85)          & 1.06 (0.61,1.64)           & 1.32 (0.73, 1.8)     \\ \cline{3-6} 
                   &                      & $\gamma$           & 0.69 (0.07,0.99)          & 0.59 (0.19,1.01)           & 0.99 (0.18, 1)       \\ \cline{3-6} 
                   &                      & $\kappa$           & 1.14 (0.25,1.95)          & 1.09 (0.37,2.02)           & 2 (0.42, 2)          \\ \cline{3-6} 
                   &                      & $\rho$             & 0.77 (0.21,0.99)          & 0.78 (0.24,0.99)           & 1 (0.42, 1)          \\ \cline{2-6} 
                   & \multirow{4}{*}{15}  & $\beta$            & 1.16 (0.5,1.84)           & 1.03 (0.61,1.59)           & 1.3 (0.78, 1.94)     \\ \cline{3-6} 
                   &                      & $\gamma$           & 0.68 (0.14,0.98)          & 0.59 (0.23,0.98)           & 0.88 (0.42, 1)       \\ \cline{3-6} 
                   &                      & $\kappa$           & 1.09 (0.29,1.96)          & 1.07 (0.38,2)              & 1.82 (0.39, 2)       \\ \cline{3-6} 
                   &                      & $\rho$             & 0.79 (0.39,0.99)          & 0.79 (0.4,0.99)            & 0.98 (0.65, 1)       \\ \cline{2-6} 
                   & \multirow{4}{*}{20}  & $\beta$            & 1.1 (0.39,1.92)           & 1.04 (0.52,1.78)           & 0.71 (0.47, 2)       \\ \cline{3-6} 
                   &                      & $\gamma$           & 0.4 (0.02,0.95)           & 0.46 (0.09,0.9)            & 0.23 (0.01, 0.91)    \\ \cline{3-6} 
                   &                      & $\kappa$           & 0.7 (0.12,1.91)           & 0.86 (0.22,1.85)           & 0.91 (0.07, 1.8)     \\ \cline{3-6} 
                   &                      & $\rho$             & 0.46 (0.16,0.94)          & 0.41 (0.15,0.89)           & 0.7 (0.31, 1)        \\ \cline{2-6} 
                   & \multirow{4}{*}{25}  & $\beta$            & 1.2 (0.54,1.88)           & 1.39 (0.91,1.93)           & 1.77 (1.42, 1.83)    \\ \cline{3-6} 
                   &                      & $\gamma$           & 0.68 (0.17,0.98)          & 0.55 (0.19,0.99)           & 0.92 (0.54, 1)       \\ \cline{3-6} 
                   &                      & $\kappa$           & 1.08 (0.27,1.95)          & 1.26 (0.53,2.14)           & 1.61 (0.88, 2)       \\ \cline{3-6} 
                   &                      & $\rho$             & 0.74 (0.34,0.99)          & 0.01 (0.01,0.02)           & 0.01 (0.01, 0.02)    \\ \cline{2-6} 
                   & \multirow{4}{*}{30}  & $\beta$            & 1.58 (1.27,1.81)          & 1.55 (1.16,1.99)           & 1.35 (0.76, 1.99)    \\ \cline{3-6} 
                   &                      & $\gamma$           & 0.93 (0.64,1)             & 0.9 (0.55,1.25)            & 1 (0.23, 1)          \\ \cline{3-6} 
                   &                      & $\kappa$           & 1.76 (0.99,1.99)          & 1.76 (0.96,2.57)           & 2 (0.13, 2)          \\ \cline{3-6} 
                   &                      & $\rho$             & 0.03 (0.03,0.04)          & 0.03 (0.03,0.04)           & 1 (0.39, 1)          \\ \hline
\end{tabular}
\caption{\footnotesize The parameter estimation of three different methods for the San Francisco 1918 flu data, using several calibration periods: 10, 15, 20, 25, and 30 days. The SEIR model is utilized, fitting the data to the newly infected people $\frac{dC}{dt}$. The population size is 550,000, assuming a normal error structure and the initial condition $(549996,0,4,0,4)$.}
\label{tab:S-paramessannew30}
\end{table}


\section{Flu Cumberland 1918}

\begin{figure}[H]
    \centering
    \includegraphics[width=1\textwidth]{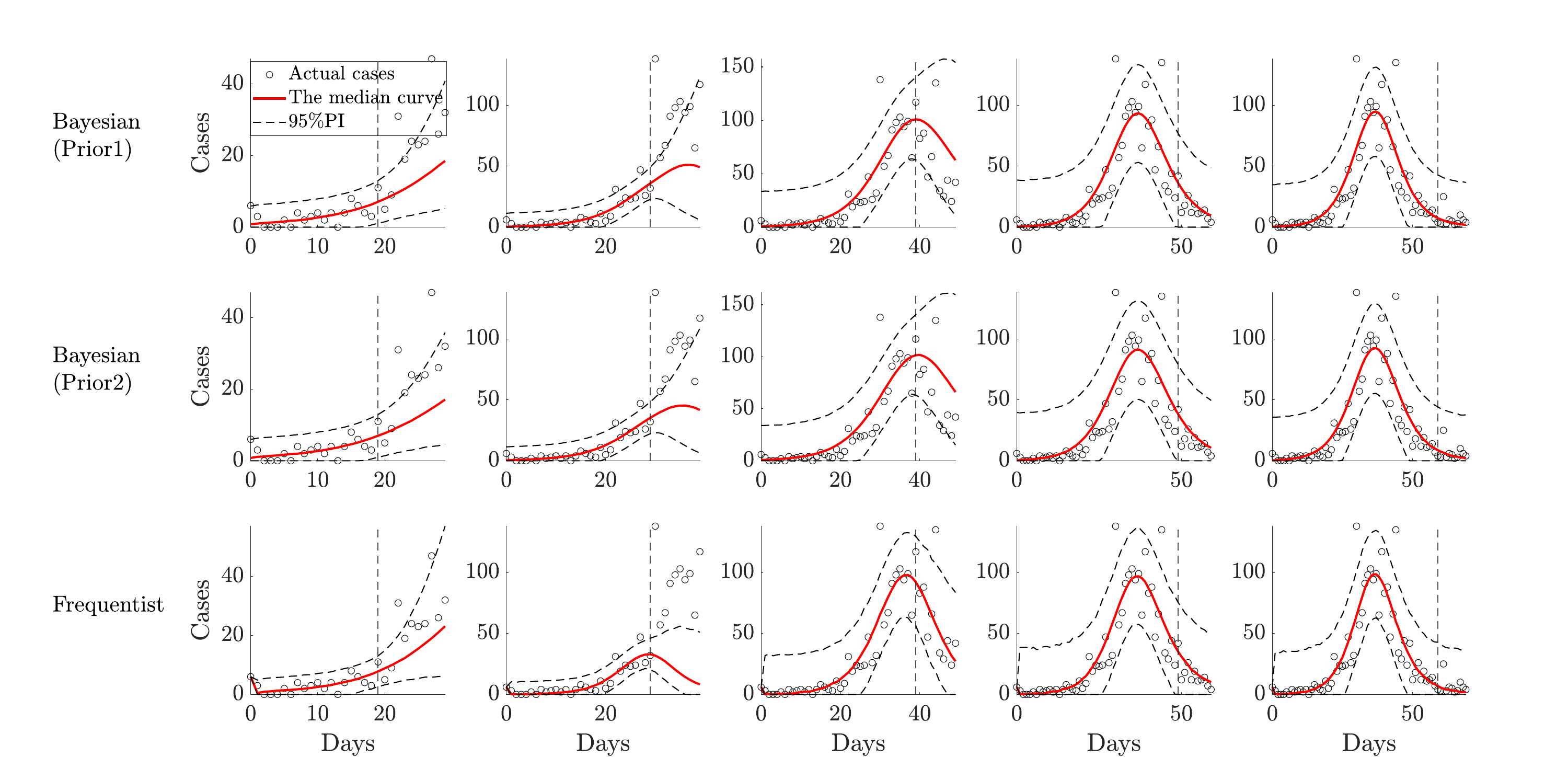}
    \caption{\footnotesize Panel showcasing the fitting of three different methods for the 1918 Cumberland flu epidemic, using several calibration periods: 20, 30, 40, 50, and 60 days, with a forecasting horizon of 10 days. The SEIR model is utilized, fitting the data to the newly infected people $\frac{dC}{dt}$. The population size is 10,000, assuming a normal error structure and the initial condition $(9994, 0, 6, 0, 6)$.}
    \label{fig:S-freq-bayes-cum-10}
\end{figure}

\begin{table}[H]
\centering
\begin{tabular}{|c|c|c|c|c|}
\hline
\textbf{Calibration} & \textbf{Metrics} & \textbf{Bayesian (Prior1)} & \textbf{Bayesian (Prior2)} & \textbf{Frequentist} \\ \hline
\multirow{4}{*}{10}  & MAE              & 37.05                      & \textbf{36.22}             & 36.67                \\ \cline{2-5} 
                     & RMSE             & 54.50                      & \textbf{53.66}             & 54.87                \\ \cline{2-5} 
                     & WIS              & 22.72                      & 22.87                      & \textbf{21.46}       \\ \cline{2-5} 
                     & 95\%PI           & 96.67                      & 96.67                      & \textbf{100.00}      \\ \hline
\multirow{4}{*}{15}  & MAE              & 47.63                      & 48.07                      & \textbf{45.98}       \\ \cline{2-5} 
                     & RMSE             & 61.84                      & 62.33                      & \textbf{60.22}       \\ \cline{2-5} 
                     & WIS              & 34.6                       & 37.48                      & \textbf{32.24}       \\ \cline{2-5} 
                     & 95\%PI           & 56.67                      & 40                         & \textbf{70.00}       \\ \hline
\multirow{4}{*}{20}  & MAE              & 34.76                      & 36.64                      & \textbf{32.45}       \\ \cline{2-5} 
                     & RMSE             & 46.24                      & 49.14                      & \textbf{41.33}       \\ \cline{2-5} 
                     & WIS              & 25.08                      & 28.28                      & \textbf{20.19}       \\ \cline{2-5} 
                     & 95\%PI           & 60                         & 53.33                      & \textbf{86.67}       \\ \hline
\multirow{4}{*}{25}  & MAE              & \textbf{21.52}             & 25.28                      & 34.15                \\ \cline{2-5} 
                     & RMSE             & \textbf{29.63}             & 36.91                      & 43.68                \\ \cline{2-5} 
                     & WIS              & \textbf{16.56}             & 18.1                       & 18.08                \\ \cline{2-5} 
                     & 95\%PI           & 90                         & 83.33                      & \textbf{96.67}       \\ \hline
\multirow{4}{*}{30}  & MAE              & \textbf{24.76}             & 28.47                      & 48.19                \\ \cline{2-5} 
                     & RMSE             & \textbf{36.94}             & 40.57                      & 59.56                \\ \cline{2-5} 
                     & WIS              & 20.85                      & \textbf{20.82}             & 38.40                \\ \cline{2-5} 
                     & 95\%PI           & \textbf{83.33}             & 70                         & 30.00                \\ \hline
\multirow{4}{*}{35}  & MAE              & 65.13                      & 54.08                      & \textbf{16.52}       \\ \cline{2-5} 
                     & RMSE             & 73.14                      & 61.70                      & \textbf{23.79}       \\ \cline{2-5} 
                     & WIS              & 39.06                      & 33.08                      & \textbf{14.60}       \\ \cline{2-5} 
                     & 95\%PI           & 83.33                      & 83.33                      & \textbf{100.00}      \\ \hline
\multirow{4}{*}{40}  & MAE              & 23.35                      & 24.66                      & \textbf{9.64}        \\ \cline{2-5} 
                     & RMSE             & 27.42                      & 29.04                      & \textbf{17.17}       \\ \cline{2-5} 
                     & WIS              & 14.16                      & 15.16                      & \textbf{6.65}        \\ \cline{2-5} 
                     & 95\%PI           & \textbf{96.67}             & 90                         & \textbf{96.67}       \\ \hline
\multirow{4}{*}{45}  & MAE              & \textbf{14.51}             & 14.77                      & 26.14                \\ \cline{2-5} 
                     & RMSE             & \textbf{18.80}             & 19.52                      & 28.08                \\ \cline{2-5} 
                     & WIS              & \textbf{9.38}              & 9.64                       & 13.25                \\ \cline{2-5} 
                     & 95\%PI           & \textbf{100}               & \textbf{100}               & 96.67                \\ \hline
\multirow{4}{*}{50}  & MAE              & \textbf{4.11}              & 4.45                       & 4.15                 \\ \cline{2-5} 
                     & RMSE             & 6.23                       & 6.74                       & \textbf{6.14}        \\ \cline{2-5} 
                     & WIS              & 4.82                       & 4.95                       & \textbf{3.77}        \\ \cline{2-5} 
                     & 95\%PI           & 100                        & 100                        & 100.00               \\ \hline
\multirow{4}{*}{55}  & MAE              & \textbf{2.96}              & 3.06                       & 3.15                 \\ \cline{2-5} 
                     & RMSE             & \textbf{4.54}              & \textbf{4.54}              & 4.80                 \\ \cline{2-5} 
                     & WIS              & 4.31                       & 4.36                       & \textbf{2.90}        \\ \cline{2-5} 
                     & 95\%PI           & 100                        & 100                        & 100.00               \\ \hline
\multirow{4}{*}{60}  & MAE              & 2.81                       & \textbf{2.78}              & 3.08                 \\ \cline{2-5} 
                     & RMSE             & 4.54                       & \textbf{4.39}              & 4.84                 \\ \cline{2-5} 
                     & WIS              & 4.13                       & 4.15                       & \textbf{2.57}        \\ \cline{2-5} 
                     & 95\%PI           & 100                        & 100                        & 100.00               \\ \hline
\end{tabular}
\caption{\footnotesize The performance metrics of the three methods in forecasting for the Cumberland flu 1918 using several calibration periods: 10, 15, 20, 25, 30, 35, 40, 45, 50, 55, and 60 and forecasting horizon of 10. The SEIR model is utilized, fitting the data to the newly infected people $\frac{dC}{dt}$. The population size is 10,000, assuming a normal error structure and the initial condition $(9994, 0, 6, 0, 6)$.}
\label{tab:S-freq-bayes-cum-10}
\end{table}

\begin{figure}[H]
    \centering
    \includegraphics[width=1\textwidth]{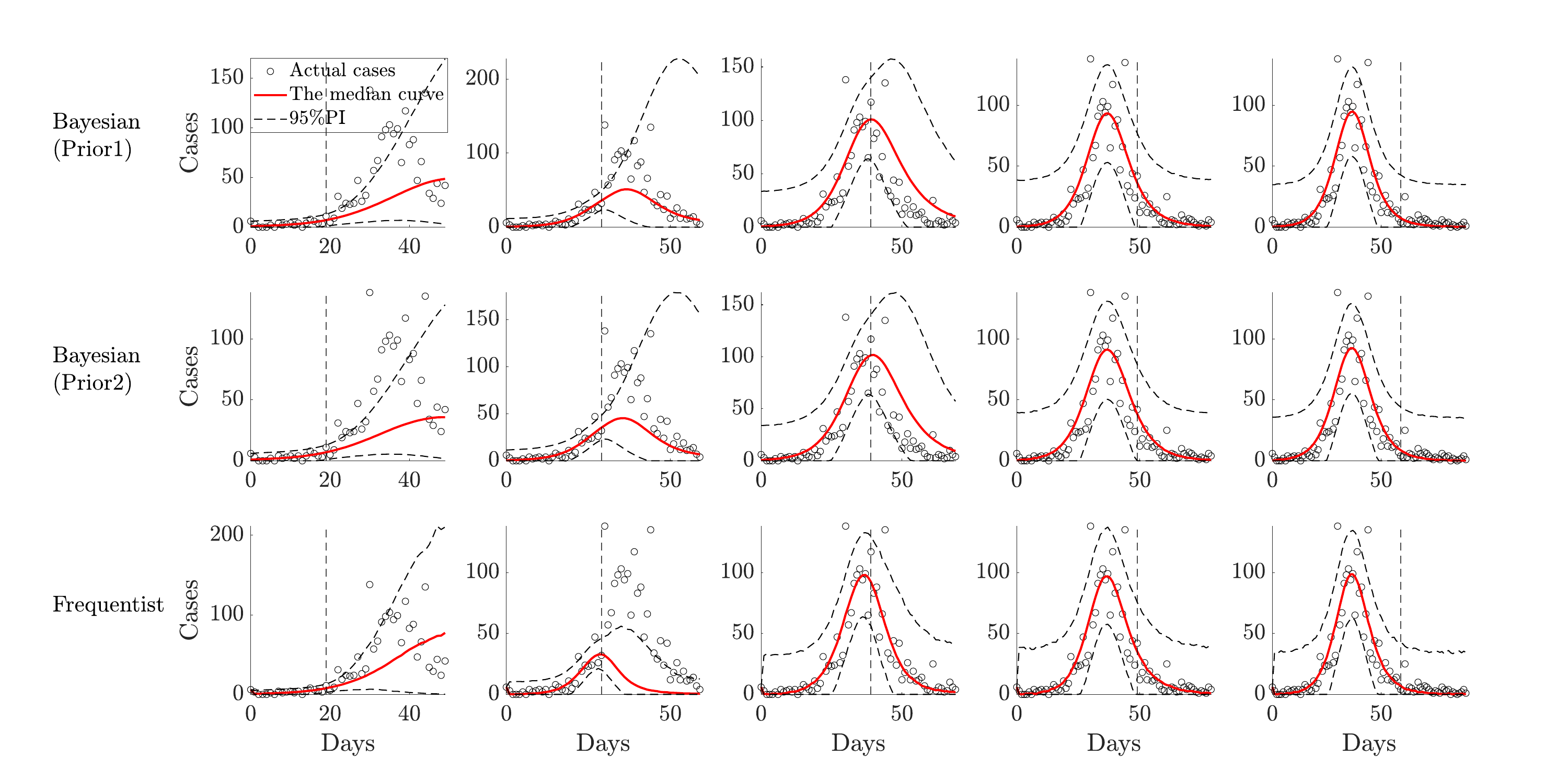}
    \caption{\footnotesize Panel showcasing the fitting of three different methods for the 1918 Cumberland flu epidemic, using several calibration periods: 20, 30, 40, 50, and 60 days, with a forecasting horizon of 30 days. The SEIR model is utilized, fitting the data to the newly infected people $\frac{dC}{dt}$. The population size is 10,000, assuming a normal error structure and the initial condition $(9994, 0, 6, 0, 6)$.
}
    \label{fig:S-freq-bayes-cum-30}
\end{figure}

\begin{table}[H]
\centering
\begin{tabular}{|c|c|c|c|c|}
\hline
\textbf{Calibration} & \textbf{Metrics} & \textbf{Bayesian (Prior1)} & \textbf{Bayesian (Prior2)} & \textbf{Frequentist} \\ \hline
\multirow{4}{*}{10}  & MAE              & 37.05                      & \textbf{36.22}             & 36.67                \\ \cline{2-5} 
                     & RMSE             & 54.50                      & \textbf{53.66}             & 54.87                \\ \cline{2-5} 
                     & WIS              & 22.72                      & 22.87                      & \textbf{21.46}       \\ \cline{2-5} 
                     & 95\%PI           & 96.67                      & 96.67                      & \textbf{100.00}      \\ \hline
\multirow{4}{*}{15}  & MAE              & 47.63                      & 48.07                      & \textbf{45.98}       \\ \cline{2-5} 
                     & RMSE             & 61.84                      & 62.33                      & \textbf{60.22}       \\ \cline{2-5} 
                     & WIS              & 34.6                       & 37.48                      & \textbf{32.24}       \\ \cline{2-5} 
                     & 95\%PI           & 56.67                      & 40                         & \textbf{70.00}       \\ \hline
\multirow{4}{*}{20}  & MAE              & 34.76                      & 36.64                      & \textbf{32.45}       \\ \cline{2-5} 
                     & RMSE             & 46.24                      & 49.14                      & \textbf{41.33}       \\ \cline{2-5} 
                     & WIS              & 25.08                      & 28.28                      & \textbf{20.19}       \\ \cline{2-5} 
                     & 95\%PI           & 60                         & 53.33                      & \textbf{86.67}       \\ \hline
\multirow{4}{*}{25}  & MAE              & \textbf{21.52}             & 25.28                      & 34.15                \\ \cline{2-5} 
                     & RMSE             & \textbf{29.63}             & 36.91                      & 43.68                \\ \cline{2-5} 
                     & WIS              & \textbf{16.56}             & 18.1                       & 18.08                \\ \cline{2-5} 
                     & 95\%PI           & 90                         & 83.33                      & \textbf{96.67}       \\ \hline
\multirow{4}{*}{30}  & MAE              & \textbf{24.76}             & 28.47                      & 48.19                \\ \cline{2-5} 
                     & RMSE             & \textbf{36.94}             & 40.57                      & 59.56                \\ \cline{2-5} 
                     & WIS              & 20.85                      & \textbf{20.82}             & 38.40                \\ \cline{2-5} 
                     & 95\%PI           & \textbf{83.33}             & 70                         & 30.00                \\ \hline
\multirow{4}{*}{35}  & MAE              & 65.13                      & 54.08                      & \textbf{16.52}       \\ \cline{2-5} 
                     & RMSE             & 73.14                      & 61.70                      & \textbf{23.79}       \\ \cline{2-5} 
                     & WIS              & 39.06                      & 33.08                      & \textbf{14.60}       \\ \cline{2-5} 
                     & 95\%PI           & 83.33                      & 83.33                      & \textbf{100.00}      \\ \hline
\multirow{4}{*}{40}  & MAE              & 23.35                      & 24.66                      & \textbf{9.64}        \\ \cline{2-5} 
                     & RMSE             & 27.42                      & 29.04                      & \textbf{17.17}       \\ \cline{2-5} 
                     & WIS              & 14.16                      & 15.16                      & \textbf{6.65}        \\ \cline{2-5} 
                     & 95\%PI           & \textbf{96.67}             & 90                         & \textbf{96.67}       \\ \hline
\multirow{4}{*}{45}  & MAE              & \textbf{14.51}             & 14.77                      & 26.14                \\ \cline{2-5} 
                     & RMSE             & \textbf{18.80}             & 19.52                      & 28.08                \\ \cline{2-5} 
                     & WIS              & \textbf{9.38}              & 9.64                       & 13.25                \\ \cline{2-5} 
                     & 95\%PI           & \textbf{100}               & \textbf{100}               & 96.67                \\ \hline
\multirow{4}{*}{50}  & MAE              & \textbf{4.11}              & 4.45                       & 4.15                 \\ \cline{2-5} 
                     & RMSE             & 6.23                       & 6.74                       & \textbf{6.14}        \\ \cline{2-5} 
                     & WIS              & 4.82                       & 4.95                       & \textbf{3.77}        \\ \cline{2-5} 
                     & 95\%PI           & 100                        & 100                        & 100.00               \\ \hline
\multirow{4}{*}{55}  & MAE              & \textbf{2.96}              & 3.06                       & 3.15                 \\ \cline{2-5} 
                     & RMSE             & \textbf{4.54}              & \textbf{4.54}              & 4.80                 \\ \cline{2-5} 
                     & WIS              & 4.31                       & 4.36                       & \textbf{2.90}        \\ \cline{2-5} 
                     & 95\%PI           & 100                        & 100                        & 100.00               \\ \hline
\multirow{4}{*}{60}  & MAE              & 2.81                       & \textbf{2.78}              & 3.08                 \\ \cline{2-5} 
                     & RMSE             & 4.54                       & \textbf{4.39}              & 4.84                 \\ \cline{2-5} 
                     & WIS              & 4.13                       & 4.15                       & \textbf{2.57}        \\ \cline{2-5} 
                     & 95\%PI           & 100                        & 100                        & 100.00               \\ \hline
\end{tabular}
\caption{\footnotesize The performance metrics of the three methods in forecasting for the Cumberland flu 1918 using several calibration periods: 10, 15, 20, 25, 30, 35, 40, 45, 50, 55, and 60 and forecasting horizon of 30. The SEIR model is utilized, fitting the data to the newly infected people $\frac{dC}{dt}$. The population size is 10,000, assuming a normal error structure and the initial condition $(9994, 0, 6, 0, 6)$.}
\label{tab:S-freq-bayes-cum-30}
\end{table}

\begin{table}[H]
\centering
\begin{tabular}{|c|c|c|c|c|c|}
\hline
\multirow{45}{*}{\textbf{\rotatebox[origin=c]{90}{Estimate (CI)}}} & \textbf{calibration} & \textbf{parameter} & \textbf{Bayesian(Prior1)} & \textbf{Bayesian (Prior2)} & \textbf{Frequentist} \\ \cline{2-6} 
                                         & \multirow{4}{*}{10}  & $\beta$            & 0.68 (0.06,1.76)          & 0.73 (0.22,1.45)           & 1.11 (0, 1.89)       \\ \cline{3-6} 
                                         &                      & $\gamma$           & 0.64 (0.05,0.98)          & 0.62 (0.18,1.06)           & 0.46 (0.01, 1)       \\ \cline{3-6} 
                                         &                      & $\kappa$           & 0.93 (0.06,1.95)          & 0.97 (0.14,1.96)           & 0.98 (0, 2)          \\ \cline{3-6} 
                                         &                      & $\rho$             & 0.54 (0.04,0.97)          & 0.5 (0.03,0.97)            & 0.15 (0, 1)          \\ \cline{2-6} 
                                         & \multirow{4}{*}{15}  & $\beta$            & 0.77 (0.2,1.69)           & 0.74 (0.32,1.32)           & 0.97 (0.29, 1.33)    \\ \cline{3-6} 
                                         &                      & $\gamma$           & 0.65 (0.07,0.98)          & 0.63 (0.22,1.07)           & 0.85 (0.03, 1)       \\ \cline{3-6} 
                                         &                      & $\kappa$           & 0.95 (0.08,1.94)          & 0.98 (0.16,1.98)           & 1.39 (0.12, 2)       \\ \cline{3-6} 
                                         &                      & $\rho$             & 0.62 (0.08,0.98)          & 0.59 (0.09,0.98)           & 0.43 (0.04, 1)       \\ \cline{2-6} 
                                         & \multirow{4}{*}{20}  & $\beta$            & 0.89 (0.21,1.85)          & 0.8 (0.33,1.46)            & 0.38 (0.19, 1.9)     \\ \cline{3-6} 
                                         &                      & $\gamma$           & 0.57 (0.04,0.98)          & 0.57 (0.17,1)              & 0.19 (0.01, 1)       \\ \cline{3-6} 
                                         &                      & $\kappa$           & 0.65 (0.07,1.92)          & 0.85 (0.13,1.89)           & 1.66 (0.02, 2)       \\ \cline{3-6} 
                                         &                      & $\rho$             & 0.48 (0.08,0.97)          & 0.43 (0.06,0.95)           & 0.52 (0.04, 1)       \\ \cline{2-6} 
                                         & \multirow{4}{*}{25}  & $\beta$            & 0.86 (0.22,1.92)          & 0.85 (0.32,1.67)           & 0.47 (0.23, 1.89)    \\ \cline{3-6} 
                                         &                      & $\gamma$           & 0.29 (0.01,0.92)          & 0.43 (0.06,0.86)           & 0.01 (0.01, 0.87)    \\ \cline{3-6} 
                                         &                      & $\kappa$           & 0.34 (0.05,1.85)          & 0.6 (0.09,1.68)            & 0.42 (0.14, 1.62)    \\ \cline{3-6} 
                                         &                      & $\rho$             & 0.23 (0.05,0.87)          & 0.19 (0.05,0.76)           & 0.11 (0.03, 0.66)    \\ \cline{2-6} 
                                         & \multirow{4}{*}{30}  & $\beta$            & 0.95 (0.24,1.93)          & 0.86 (0.38,1.62)           & 0.4 (0.33, 1.23)     \\ \cline{3-6} 
                                         &                      & $\gamma$           & 0.42 (0.02,0.96)          & 0.48 (0.12,0.88)           & 0.03 (0.01, 0.8)     \\ \cline{3-6} 
                                         &                      & $\kappa$           & 0.45 (0.06,1.87)          & 0.7 (0.12,1.73)            & 1.14 (0.33, 1.95)    \\ \cline{3-6} 
                                         &                      & $\rho$             & 0.18 (0.05,0.91)          & 0.17 (0.06,0.81)           & 0.05 (0.04, 0.21)    \\ \cline{2-6} 
                                         & \multirow{4}{*}{35}  & $\beta$            & 0.68 (0.21,1.9)           & 0.78 (0.25,1.73)           & 0.41 (0.27, 1.43)    \\ \cline{3-6} 
                                         &                      & $\gamma$           & 0.19 (0.01,0.89)          & 0.34 (0.03,0.77)           & 0.01 (0.01, 0.67)    \\ \cline{3-6} 
                                         &                      & $\kappa$           & 0.26 (0.04,1.8)           & 0.35 (0.07,1.53)           & 0.28 (0.16, 1)       \\ \cline{3-6} 
                                         &                      & $\rho$             & 0.5 (0.19,0.96)           & 0.49 (0.21,0.96)           & 0.25 (0.13, 1)       \\ \cline{2-6} 
                                         & \multirow{4}{*}{40}  & $\beta$            & 0.81 (0.26,1.93)          & 0.79 (0.33,1.69)           & 0.43 (0.35, 1.83)    \\ \cline{3-6} 
                                         &                      & $\gamma$           & 0.3 (0.01,0.91)           & 0.4 (0.07,0.82)            & 0.01 (0.01, 0.6)     \\ \cline{3-6} 
                                         &                      & $\kappa$           & 0.32 (0.04,1.91)          & 0.48 (0.09,1.59)           & 0.28 (0.04, 1)       \\ \cline{3-6} 
                                         &                      & $\rho$             & 0.31 (0.19,0.83)          & 0.35 (0.21,0.85)           & 0.19 (0.16, 0.43)    \\ \cline{2-6} 
                                         & \multirow{4}{*}{45}  & $\beta$            & 1 (0.29,1.94)             & 0.85 (0.39,1.72)           & 1.24 (0.35, 2)       \\ \cline{3-6} 
                                         &                      & $\gamma$           & 0.34 (0.02,0.93)          & 0.46 (0.11,0.87)           & 0.04 (0.01, 0.87)    \\ \cline{3-6} 
                                         &                      & $\kappa$           & 0.31 (0.04,1.85)          & 0.59 (0.09,1.62)           & 0.05 (0.03, 1.3)     \\ \cline{3-6} 
                                         &                      & $\rho$             & 0.29 (0.21,0.58)          & 0.34 (0.23,0.7)            & 0.33 (0.21, 0.51)    \\ \cline{2-6} 
                                         & \multirow{4}{*}{50}  & $\beta$            & 0.56 (0.25,1.72)          & 0.73 (0.3,1.52)            & 0.71 (0.32, 1.54)    \\ \cline{3-6} 
                                         &                      & $\gamma$           & 0.18 (0.01,0.75)          & 0.33 (0.05,0.76)           & 0.04 (0.01, 0.54)    \\ \cline{3-6} 
                                         &                      & $\kappa$           & 0.44 (0.1,1.85)           & 0.51 (0.14,1.59)           & 0.16 (0.07, 0.72)    \\ \cline{3-6} 
                                         &                      & $\rho$             & 0.23 (0.19,0.33)          & 0.26 (0.2,0.41)            & 0.21 (0.17, 0.28)    \\ \cline{2-6} 
                                         & \multirow{4}{*}{55}  & $\beta$            & 0.47 (0.24,1.57)          & 0.66 (0.28,1.38)           & 0.6 (0.33, 1.3)      \\ \cline{3-6} 
                                         &                      & $\gamma$           & 0.12 (0.01,0.71)          & 0.28 (0.03,0.7)            & 0.04 (0.01, 0.49)    \\ \cline{3-6} 
                                         &                      & $\kappa$           & 0.48 (0.13,1.85)          & 0.5 (0.16,1.55)            & 0.19 (0.1, 0.68)     \\ \cline{3-6} 
                                         &                      & $\rho$             & 0.22 (0.18,0.3)           & 0.24 (0.19,0.37)           & 0.2 (0.17, 0.26)     \\ \cline{2-6} 
                                         & \multirow{4}{*}{60}  & $\beta$            & 0.45 (0.24,1.27)          & 0.62 (0.27,1.32)           & 0.59 (0.34, 1.11)    \\ \cline{3-6} 
                                         &                      & $\gamma$           & 0.11 (0.01,0.57)          & 0.25 (0.03,0.66)           & 0.04 (0.01, 0.53)    \\ \cline{3-6} 
                                         &                      & $\kappa$           & 0.5 (0.14,1.87)           & 0.5 (0.17,1.53)            & 0.2 (0.1, 0.84)      \\ \cline{3-6} 
                                         &                      & $\rho$             & 0.21 (0.18,0.29)          & 0.24 (0.19,0.34)           & 0.2 (0.18, 0.27)     \\ \hline
\end{tabular}
\caption{\footnotesize Parameter estimation of the three methods for the Cumberland flu 1918 using several calibration periods: 10, 15, 20, 25, 30, 35, 40, 45, 50, 55, and 60. The SEIR model is utilized, fitting the data to the newly infected people $\frac{dC}{dt}$. The population size is 10,000, assuming a normal error structure and the initial condition $(9994, 0, 6, 0, 6)$.}
\label{tab:S-paramescumnew30}
\end{table}


\section*{Plague Bombay 1896-97}

\begin{figure}[H]
    \centering
    \includegraphics[width=1\textwidth]{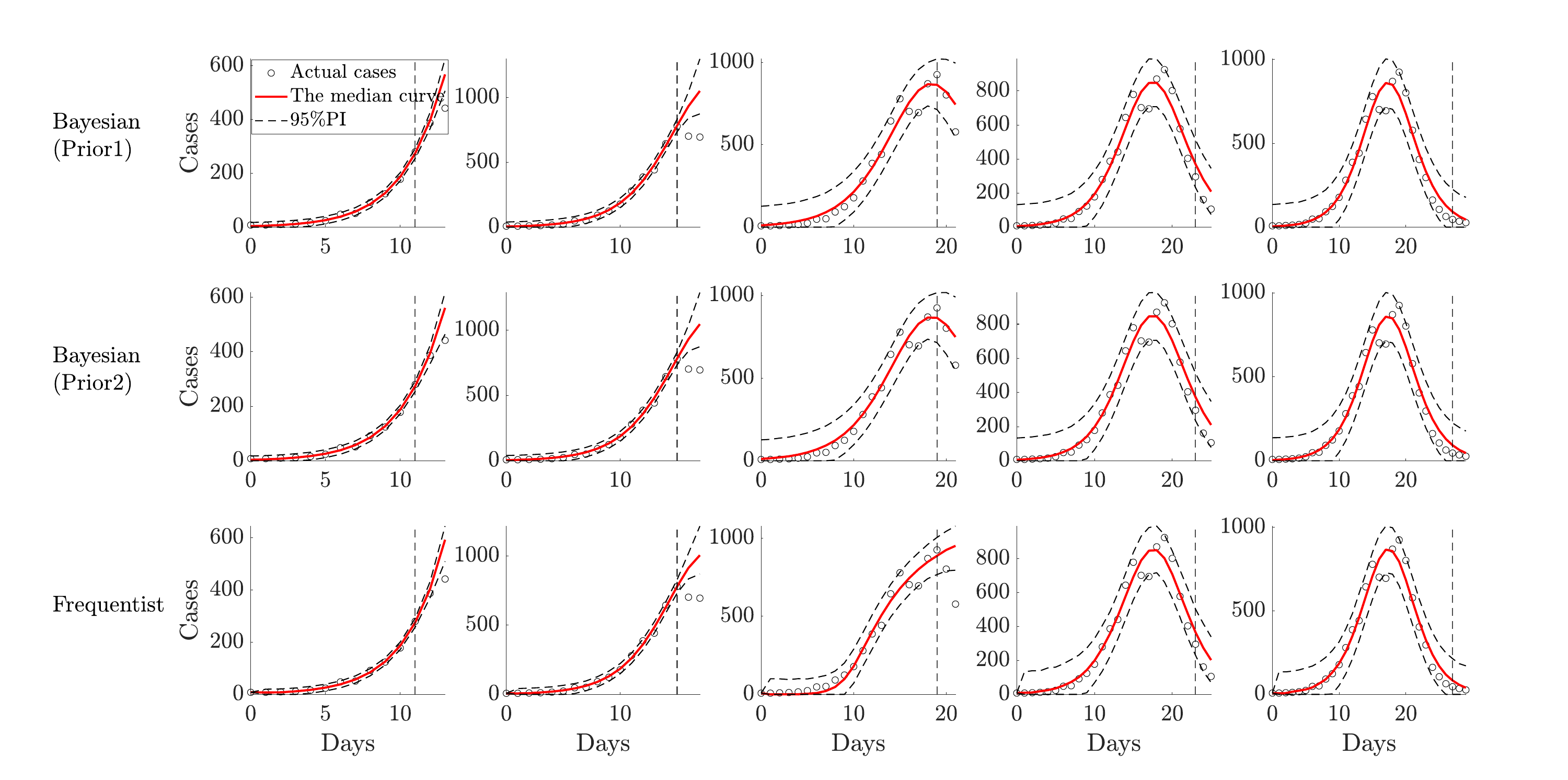}
    \caption{\footnotesize Panel showcasing the fitting of three different methods for the Bombay plague 1896-97 epidemic, using several calibration periods: 12, 16, 20, 24, and 28 weeks, with a forecasting horizon of 2 weeks. The SEIRD model is utilized, fitting the data to the newly infected people $\frac{dD}{dt}$. The population size is 100,000, assuming a normal error structure and the initial condition $(99992,0,8,0,8)$.}
    \label{fig:S-freq-bayes-plague-2}
\end{figure}

\begin{table}[H]
\centering
\begin{tabular}{|c|c|c|c|c|}
\hline
\textbf{Calibration} & \textbf{Metrics} & \textbf{Bayesian (Prior1)} & \textbf{Bayesian (Prior2)} & \textbf{Frequentist} \\ \hline
\multirow{4}{*}{12}  & MAE              & 67.83                      & \textbf{63.41}             & 84.74                \\ \cline{2-5} 
                     & RMSE             & 89.11                      & \textbf{84.40}             & 107.93               \\ \cline{2-5} 
                     & WIS              & 53.44                      & \textbf{44.46}             & 64.68                \\ \cline{2-5} 
                     & 95\%PI           & 50                         & 50                         & 50.00                \\ \hline
\multirow{4}{*}{14}  & MAE              & 210.74                     & \textbf{208.73}            & 230.48               \\ \cline{2-5} 
                     & RMSE             & 221.41                     & \textbf{219.23}            & 242.69               \\ \cline{2-5} 
                     & WIS              & \textbf{164.15}            & 165.16                     & 201.72               \\ \cline{2-5} 
                     & 95\%PI           & 0                          & 0                          & 0.00                 \\ \hline
\multirow{4}{*}{16}  & MAE              & 293.81                     & 289.8                      & \textbf{258.77}      \\ \cline{2-5} 
                     & RMSE             & 300.32                     & 296.15                     & \textbf{263.50}      \\ \cline{2-5} 
                     & WIS              & 243.19                     & 239.92                     & \textbf{220.60}      \\ \cline{2-5} 
                     & 95\%PI           & 0                          & 0                          & 0.00                 \\ \hline
\multirow{4}{*}{18}  & MAE              & 354.01                     & \textbf{350.5}             & 365.90               \\ \cline{2-5} 
                     & RMSE             & 364.36                     & \textbf{360.90}            & 376.68               \\ \cline{2-5} 
                     & WIS              & 317.63                     & \textbf{313.37}            & 327.08               \\ \cline{2-5} 
                     & 95\%PI           & 0                          & 0                          & 0.00                 \\ \hline
\multirow{4}{*}{20}  & MAE              & \textbf{92.15}             & 95.93                      & 245.28               \\ \cline{2-5} 
                     & RMSE             & \textbf{118.04}            & 121.84                     & 275.57               \\ \cline{2-5} 
                     & WIS              & \textbf{60.37}             & 62.05                      & 189.23               \\ \cline{2-5} 
                     & 95\%PI           & \textbf{100}               & \textbf{100}               & 50.00                \\ \hline
\multirow{4}{*}{22}  & MAE              & 144.43                     & 147.55                     & \textbf{127.07}      \\ \cline{2-5} 
                     & RMSE             & 144.44                     & 147.56                     & \textbf{127.09}      \\ \cline{2-5} 
                     & WIS              & 92.63                      & 95.29                      & \textbf{80.29}       \\ \cline{2-5} 
                     & 95\%PI           & 50                         & 0                          & \textbf{100.00}      \\ \hline
\multirow{4}{*}{24}  & MAE              & 110.51                     & 112.05                     & \textbf{103.24}      \\ \cline{2-5} 
                     & RMSE             & 110.78                     & 112.35                     & \textbf{103.60}      \\ \cline{2-5} 
                     & WIS              & 66.33                      & 67.83                      & \textbf{60.94}       \\ \cline{2-5} 
                     & 95\%PI           & 100                        & 100                        & 100.00               \\ \hline
\multirow{4}{*}{26}  & MAE              & 60.06                      & 60.6                       & \textbf{50.89}       \\ \cline{2-5} 
                     & RMSE             & 60.89                      & 61.43                      & \textbf{51.80}       \\ \cline{2-5} 
                     & WIS              & 32.24                      & 32.61                      & \textbf{27.98}       \\ \cline{2-5} 
                     & 95\%PI           & 100                        & 100                        & 100.00               \\ \hline
\multirow{4}{*}{28}  & MAE              & 23.73                      & 24.49                      & \textbf{17.62}       \\ \cline{2-5} 
                     & RMSE             & 24.44                      & 25.13                      & \textbf{18.27}       \\ \cline{2-5} 
                     & WIS              & 17.19                      & 17.34                      & \textbf{14.54}       \\ \cline{2-5} 
                     & 95\%PI           & 100                        & 100                        & 100.00               \\ \hline
\end{tabular}
\caption{\footnotesize The performance metrics of the three different methods for the Bombay plague 1896-97 epidemic, using several calibration periods: 12, 14, 16, 18, 20, 22, 24, 26, and 28 weeks, with a forecasting horizon of 2 weeks. The SEIRD model is utilized, fitting the data to the newly infected people $\frac{dD}{dt}$. The population size is 100,000, assuming a normal error structure and the initial condition $(99992,0,8,0,8)$.}
\label{tab:S-freq-bayes-plague-2}
\end{table}

\begin{figure}[H]
    \centering
    \includegraphics[width=1\textwidth]{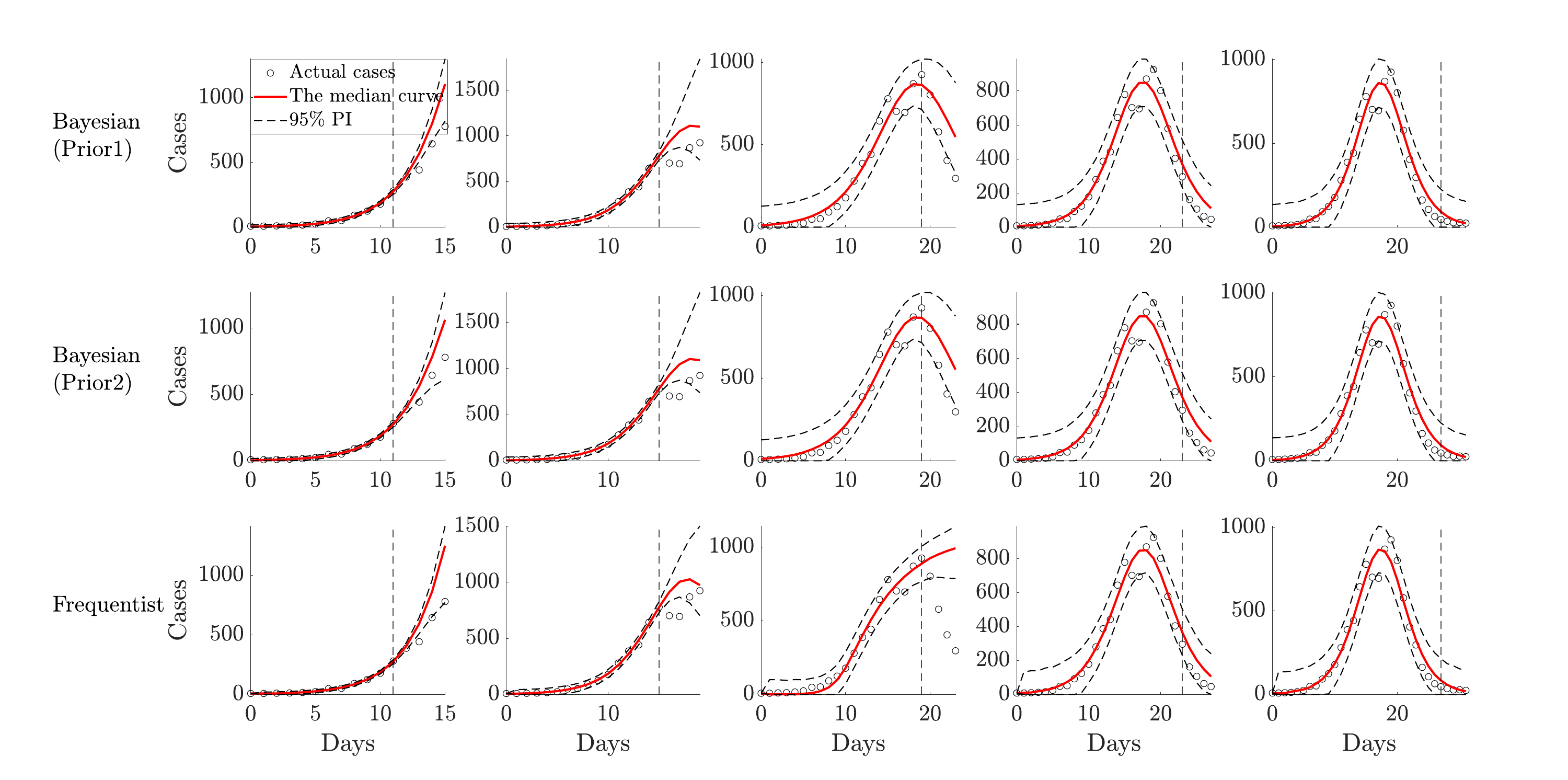}
    \caption{\footnotesize Panel showcasing the fitting of three different methods for the Bombay plague 1896-97 epidemic, using several calibration periods: 12, 16, 20, 24, and 28 weeks, with a forecasting horizon of 4 weeks. The SEIRD model is utilized, fitting the data to the newly infected people $\frac{dD}{dt}$. The population size is 100,000, assuming a normal error structure and the initial condition $(99992,0,8,0,8)$.}
    \label{fig:S-freq-bayes-plague-4}
\end{figure}

\begin{table}[H]
\centering
\begin{tabular}{|c|c|c|c|c|}
\hline
\textbf{Calibration} & \textbf{Metrics} & \textbf{Bayesian (Prior 1)} & \textbf{Bayesian (Prior 2)} & \textbf{Frequentist} \\ \hline
\multirow{4}{*}{12}  & MAE              & 437.07                      & \textbf{378.41}             & 627.56               \\ \cline{2-5} 
                     & RMSE             & 612.71                      & \textbf{525.85}             & 893.22               \\ \cline{2-5} 
                     & WIS              & 297.59                      & \textbf{224.63}             & 461.67               \\ \cline{2-5} 
                     & 95\% PI          & 16.67                       & \textbf{83.33}              & \textbf{33.33}       \\ \hline
\multirow{4}{*}{14}  & MAE              & 384                         & \textbf{380.43}             & 425.63               \\ \cline{2-5} 
                     & RMSE             & 433.79                      & \textbf{430.42}             & 476.08               \\ \cline{2-5} 
                     & WIS              & \textbf{287.68}             & 289.68                      & 377.35               \\ \cline{2-5} 
                     & 95\% PI          & \textbf{33.33}              & \textbf{33.33}              & 0.00                 \\ \hline
\multirow{4}{*}{16}  & MAE              & 258.76                      & 249.5                       & \textbf{152.89}      \\ \cline{2-5} 
                     & RMSE             & 265.93                      & 257.07                      & \textbf{176.79}      \\ \cline{2-5} 
                     & WIS              & 177.03                      & 171.02                      & \textbf{118.52}      \\ \cline{2-5} 
                     & 95\% PI          & 66.67                       & 66.67                       & 66.67                \\ \hline
\multirow{4}{*}{18}  & MAE              & 304.13                      & \textbf{301.78}             & 321.53               \\ \cline{2-5} 
                     & RMSE             & 321.39                      & \textbf{318.84}             & 338.04               \\ \cline{2-5} 
                     & WIS              & 268.38                      & \textbf{265.3}              & 283.64               \\ \cline{2-5} 
                     & 95\% PI          & 0                           & 0                           & 0.00                 \\ \hline
\multirow{4}{*}{20}  & MAE              & \textbf{202.96}             & 207.97                      & 586.37               \\ \cline{2-5} 
                     & RMSE             & \textbf{222.07}             & 227.12                      & 647.59               \\ \cline{2-5} 
                     & WIS              & \textbf{140.1}              & 143.96                      & 494.64               \\ \cline{2-5} 
                     & 95\% PI          & \textbf{33.33}              & 33.33                       & 16.67                \\ \hline
\multirow{4}{*}{22}  & MAE              & 146.25                      & 148.98                      & \textbf{125.87}      \\ \cline{2-5} 
                     & RMSE             & 148.20                      & 150.94                      & \textbf{127.92}      \\ \cline{2-5} 
                     & WIS              & 96.58                       & 98.89                       & \textbf{81.15}       \\ \cline{2-5} 
                     & 95\% PI          & 33.33                       & 16.67                       & \textbf{66.67}       \\ \hline
\multirow{4}{*}{24}  & MAE              & 74.81                       & 76.06                       & \textbf{67.68}       \\ \cline{2-5} 
                     & RMSE             & 81.11                       & 82.44                       & \textbf{74.38}       \\ \cline{2-5} 
                     & WIS              & 43.39                       & 44.38                       & \textbf{39.87}       \\ \cline{2-5} 
                     & 95\% PI          & 100                         & 100                         & 100.00               \\ \hline
\multirow{4}{*}{26}  & MAE              & 30.04                       & 30.41                       & \textbf{24.55}       \\ \cline{2-5} 
                     & RMSE             & 38.74                       & 39.06                       & \textbf{32.31}       \\ \cline{2-5} 
                     & WIS              & 21.55                       & 21.71                       & \textbf{18.23}       \\ \cline{2-5} 
                     & 95\% PI          & 100                         & 100                         & 100.00               \\ \hline
\multirow{4}{*}{28}  & MAE              & 13.48                       & 13.94                       & \textbf{12.41}       \\ \cline{2-5} 
                     & RMSE             & 16.57                       & 17.00                       & \textbf{14.59}       \\ \cline{2-5} 
                     & WIS              & 15.42                       & 15.47                       & \textbf{11.96}       \\ \cline{2-5} 
                     & 95\% PI          & 100                         & 100                         & 100.00               \\ \hline
\end{tabular}
\caption{\footnotesize The performance metrics of the three different methods for the Bombay plague 1896-97 epidemic, using several calibration periods: 12, 14, 16, 18, 20, 22, 24, 26, and 28 weeks, with a forecasting horizon of 4 weeks. The SEIRD model is utilized, fitting the data to the newly infected people $\frac{dD}{dt}$. The population size is 100,000, assuming a normal error structure and the initial condition $(99992,0,8,0,8)$.}
\label{tab:S-freq-bayes-plague-4}
\end{table}

\begin{figure}[H]
    \centering
    \includegraphics[width=1\textwidth]{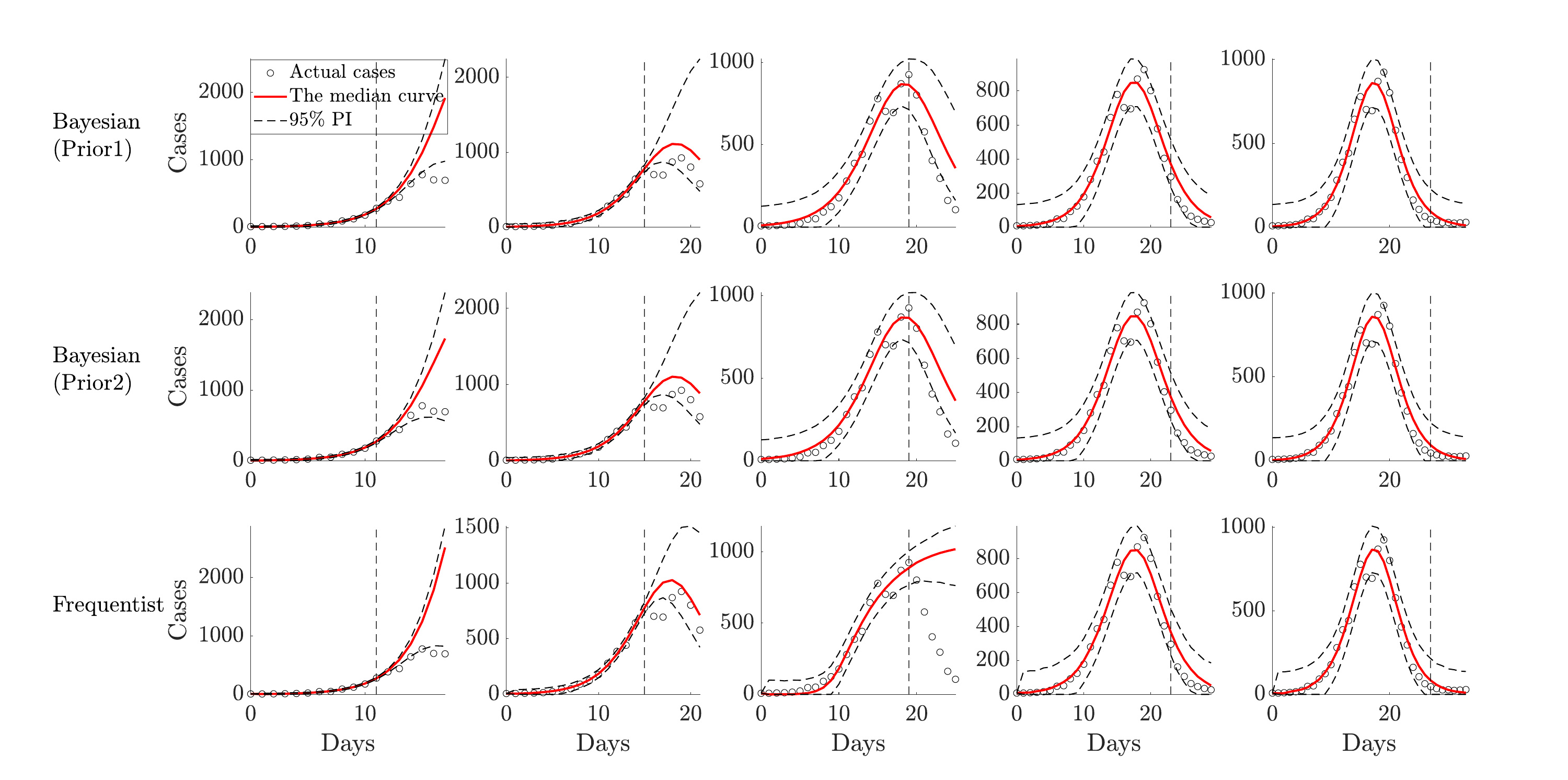}
    \caption{\footnotesize Panel showcasing the fitting of three different methods for the Bombay plague 1896-97 epidemic, using several calibration periods: 12, 16, 20, 24, and 28 weeks, with a forecasting horizon of 6 weeks. The SEIRD model is utilized, fitting the data to the newly infected people $\frac{dD}{dt}$. The population size is 100,000, assuming a normal error structure and the initial condition $(99992,0,8,0,8)$.}
    \label{fig:S-freq-bayes-plague-6}
\end{figure}

\begin{table}[H]
\centering
\begin{tabular}{|c|c|c|c|c|}
\hline
\textbf{Calibration} & \textbf{Metrics} & \textbf{Bayesian (Prior 1)} & \textbf{Bayesian (Prior 2)} & \textbf{Frequentist} \\ \hline
\multirow{4}{*}{12}  & MAE              & 437.07                      & \textbf{378.41}             & 627.56               \\ \cline{2-5} 
                     & RMSE             & 612.71                      & \textbf{525.85}             & 893.22               \\ \cline{2-5} 
                     & WIS              & 297.59                      & \textbf{224.63}             & 461.67               \\ \cline{2-5} 
                     & 95\% PI          & 16.67                       & \textbf{83.33}              & \textbf{33.33}       \\ \hline
\multirow{4}{*}{14}  & MAE              & 384                         & \textbf{380.43}             & 425.63               \\ \cline{2-5} 
                     & RMSE             & 433.79                      & \textbf{430.42}             & 476.08               \\ \cline{2-5} 
                     & WIS              & \textbf{287.68}             & 289.68                      & 377.35               \\ \cline{2-5} 
                     & 95\% PI          & \textbf{33.33}              & \textbf{33.33}              & 0.00                 \\ \hline
\multirow{4}{*}{16}  & MAE              & 258.76                      & 249.5                       & \textbf{152.89}      \\ \cline{2-5} 
                     & RMSE             & 265.93                      & 257.07                      & \textbf{176.79}      \\ \cline{2-5} 
                     & WIS              & 177.03                      & 171.02                      & \textbf{118.52}      \\ \cline{2-5} 
                     & 95\% PI          & 66.67                       & 66.67                       & 66.67                \\ \hline
\multirow{4}{*}{18}  & MAE              & 304.13                      & \textbf{301.78}             & 321.53               \\ \cline{2-5} 
                     & RMSE             & 321.39                      & \textbf{318.84}             & 338.04               \\ \cline{2-5} 
                     & WIS              & 268.38                      & \textbf{265.3}              & 283.64               \\ \cline{2-5} 
                     & 95\% PI          & 0                           & 0                           & 0.00                 \\ \hline
\multirow{4}{*}{20}  & MAE              & \textbf{202.96}             & 207.97                      & 586.37               \\ \cline{2-5} 
                     & RMSE             & \textbf{222.07}             & 227.12                      & 647.59               \\ \cline{2-5} 
                     & WIS              & \textbf{140.1}              & 143.96                      & 494.64               \\ \cline{2-5} 
                     & 95\% PI          & \textbf{33.33}              & 33.33                       & 16.67                \\ \hline
\multirow{4}{*}{22}  & MAE              & 146.25                      & 148.98                      & \textbf{125.87}      \\ \cline{2-5} 
                     & RMSE             & 148.20                      & 150.94                      & \textbf{127.92}      \\ \cline{2-5} 
                     & WIS              & 96.58                       & 98.89                       & \textbf{81.15}       \\ \cline{2-5} 
                     & 95\% PI          & 33.33                       & 16.67                       & \textbf{66.67}       \\ \hline
\multirow{4}{*}{24}  & MAE              & 74.81                       & 76.06                       & \textbf{67.68}       \\ \cline{2-5} 
                     & RMSE             & 81.11                       & 82.44                       & \textbf{74.38}       \\ \cline{2-5} 
                     & WIS              & 43.39                       & 44.38                       & \textbf{39.87}       \\ \cline{2-5} 
                     & 95\% PI          & 100                         & 100                         & 100.00               \\ \hline
\multirow{4}{*}{26}  & MAE              & 30.04                       & 30.41                       & \textbf{24.55}       \\ \cline{2-5} 
                     & RMSE             & 38.74                       & 39.06                       & \textbf{32.31}       \\ \cline{2-5} 
                     & WIS              & 21.55                       & 21.71                       & \textbf{18.23}       \\ \cline{2-5} 
                     & 95\% PI          & 100                         & 100                         & 100.00               \\ \hline
\multirow{4}{*}{28}  & MAE              & 13.48                       & 13.94                       & \textbf{12.41}       \\ \cline{2-5} 
                     & RMSE             & 16.57                       & 17.00                       & \textbf{14.59}       \\ \cline{2-5} 
                     & WIS              & 15.42                       & 15.47                       & \textbf{11.96}       \\ \cline{2-5} 
                     & 95\% PI          & 100                         & 100                         & 100.00               \\ \hline
\end{tabular}
\caption{\footnotesize The performance metrics of the three different methods for the Bombay plague 1896-97 epidemic, using several calibration periods: 12, 14, 16, 18, 20, 22, 24, 26, and 28 weeks, with a forecasting horizon of 6 weeks. The SEIRD model is utilized, fitting the data to the newly infected people $\frac{dD}{dt}$. The population size is 100,000, assuming a normal error structure and the initial condition $(99992,0,8,0,8)$.}
\label{tab:S-freq-bayes-plague-6}
\end{table}

\begin{table}[H]
\centering
\begin{tabular}{|c|c|c|c|c|c|}
\hline
\multirow{37}{*}{\textbf{\rotatebox[origin=c]{90}{Estimate (CI)}}} & \textbf{calibration} & \textbf{parameter} & \textbf{Bayesian(Prior1)} & \textbf{Bayesian (Prior2)} & \textbf{Frequentist} \\ \cline{2-6} 
                                         & \multirow{4}{*}{12}  & $\beta$            & 3.06 (0.89,9.64)          & 5.04 (1,10.19)             & 2.96 (2.02, 9.82)    \\ \cline{3-6} 
                                         &                      & $\gamma$           & 1.57 (0.42,5.53)          & 2.67 (0.51,6.03)           & 1.71 (1.28, 6.38)    \\ \cline{3-6} 
                                         &                      & $\kappa$           & 1.04 (0.14,8.36)          & 0.99 (0.19,7.11)           & 0.96 (0.31, 5.62)    \\ \cline{3-6} 
                                         &                      & $\rho$             & 0.47 (0.19,0.96)          & 0.38 (0.13,0.91)           & 0.83 (0.2, 1)        \\ \cline{2-6} 
                                         & \multirow{4}{*}{14}  & $\beta$            & 5.16 (2.34,9.77)          & 4.78 (2.52,9.31)           & 8.18 (5.74, 8.43)    \\ \cline{3-6} 
                                         &                      & $\gamma$           & 3.84 (1.67,7.33)          & 3.61 (1.84,6.89)           & 6.71 (4.8, 7.03)     \\ \cline{3-6} 
                                         &                      & $\kappa$           & 2.63 (1.39,9.27)          & 2.9 (1.55,7.96)            & 3.71 (2.62, 4.69)    \\ \cline{3-6} 
                                         &                      & $\rho$             & 0.1 (0.06,0.21)           & 0.1 (0.06,0.2)             & 0.11 (0.07, 0.16)    \\ \cline{2-6} 
                                         & \multirow{4}{*}{16}  & $\beta$            & 5.05 (1.94,9.77)          & 4.88 (2.1,9.53)            & 7.84 (4.29, 9.65)    \\ \cline{3-6} 
                                         &                      & $\gamma$           & 3.86 (1.47,7.38)          & 3.74 (1.61,7.2)            & 6.62 (3.69, 7.79)    \\ \cline{3-6} 
                                         &                      & $\kappa$           & 2 (1.2,9.2)               & 2.09 (1.23,7.48)           & 3.21 (2.38, 6.02)    \\ \cline{3-6} 
                                         &                      & $\rho$             & 0.3 (0.18,0.68)           & 0.3 (0.18,0.66)            & 0.28 (0.2, 0.43)     \\ \cline{2-6} 
                                         & \multirow{4}{*}{18}  & $\beta$            & 5.13 (1.79,9.77)          & 4.82 (2.03,9.53)           & 3.58 (2.86, 9.58)    \\ \cline{3-6} 
                                         &                      & $\gamma$           & 3.51 (1.17,6.96)          & 3.44 (1.4,6.62)            & 2.93 (2.16, 7.72)    \\ \cline{3-6} 
                                         &                      & $\kappa$           & 1.93 (0.81,8.95)          & 2.14 (0.96,7.4)            & 7.96 (2.3, 9.22)     \\ \cline{3-6} 
                                         &                      & $\rho$             & 0.13 (0.09,0.19)          & 0.14 (0.09,0.2)            & 0.17 (0.13, 0.23)    \\ \cline{2-6} 
                                         & \multirow{4}{*}{20}  & $\beta$            & 5.96 (2.86,9.8)           & 5.26 (3.07,9.11)           & 9.61 (4.79, 10)      \\ \cline{3-6} 
                                         &                      & $\gamma$           & 5.15 (2.39,8.5)           & 4.58 (2.58,7.86)           & 0.02 (0.01, 0.07)    \\ \cline{3-6} 
                                         &                      & $\kappa$           & 3.69 (2.12,9.37)          & 4.18 (2.34,8.67)           & 0.1 (0.08, 0.21)     \\ \cline{3-6} 
                                         &                      & $\rho$             & 0.44 (0.26,0.78)          & 0.45 (0.27,0.79)           & 0.77 (0.28, 1)       \\ \cline{2-6} 
                                         & \multirow{4}{*}{22}  & $\beta$            & 5.67 (2.64,9.82)          & 5.13 (2.86,9.06)           & 8.87 (5.7, 10)       \\ \cline{3-6} 
                                         &                      & $\gamma$           & 4.78 (2.15,8.3)           & 4.36 (2.35,7.66)           & 7.93 (5.02, 9.21)    \\ \cline{3-6} 
                                         &                      & $\kappa$           & 3.4 (1.85,9.4)            & 3.81 (2.08,8.56)           & 4.96 (3.2, 7.54)     \\ \cline{3-6} 
                                         &                      & $\rho$             & 0.34 (0.23,0.52)          & 0.35 (0.24,0.53)           & 0.35 (0.24, 0.46)    \\ \cline{2-6} 
                                         & \multirow{4}{*}{24}  & $\beta$            & 5.39 (2.29,9.79)          & 4.94 (2.57,9.09)           & 7.85 (5.27, 9.36)    \\ \cline{3-6} 
                                         &                      & $\gamma$           & 4.36 (1.77,7.93)          & 4.05 (2.03,7.35)           & 6.95 (4.59, 8.18)    \\ \cline{3-6} 
                                         &                      & $\kappa$           & 2.9 (1.44,9.14)           & 3.22 (1.64,7.91)           & 4.75 (3.29, 8)       \\ \cline{3-6} 
                                         &                      & $\rho$             & 0.27 (0.18,0.39)          & 0.28 (0.19,0.41)           & 0.31 (0.24, 0.4)     \\ \cline{2-6} 
                                         & \multirow{4}{*}{26}  & $\beta$            & 5.05 (2.11,9.71)          & 4.72 (2.38,9)              & 6.75 (5.07, 9.08)    \\ \cline{3-6} 
                                         &                      & $\gamma$           & 3.98 (1.58,7.64)          & 3.76 (1.82,7.07)           & 5.85 (4.26, 7.92)    \\ \cline{3-6} 
                                         &                      & $\kappa$           & 2.64 (1.24,8.91)          & 2.94 (1.45,7.5)            & 4.48 (2.86, 7.87)    \\ \cline{3-6} 
                                         &                      & $\rho$             & 0.24 (0.16,0.34)          & 0.25 (0.17,0.35)           & 0.28 (0.22, 0.36)    \\ \cline{2-6} 
                                         & \multirow{4}{*}{28}  & $\beta$            & 4.89 (2.11,9.68)          & 4.65 (2.35,8.83)           & 8.7 (6.91, 9.49)     \\ \cline{3-6} 
                                         &                      & $\gamma$           & 3.8 (1.57,7.53)           & 3.68 (1.78,6.87)           & 7.46 (5.78, 8.04)    \\ \cline{3-6} 
                                         &                      & $\kappa$           & 2.54 (1.22,8.61)          & 2.79 (1.39,7.14)           & 3.59 (2.38, 5.15)    \\ \cline{3-6} 
                                         &                      & $\rho$             & 0.23 (0.15,0.32)          & 0.24 (0.16,0.33)           & 0.27 (0.2, 0.32)     \\ \hline
\end{tabular}
\caption{\footnotesize Parameter estimation of the three different methods for the Bombay plague 1896-97 epidemic, using several calibration periods: 12, 14, 16, 18, 20, 22, 24, 26, and 28 weeks. The SEIRD model is utilized, fitting the data to the newly infected people $\frac{dD}{dt}$. The population size is 100,000, assuming a normal error structure and the initial condition $(99992,0,8,0,8)$.}
\label{tab:S-paramesplaguenew6}
\end{table}


\section*{COVID-19 Switzerland}

\begin{figure}[H]
    \centering
    \includegraphics[width=1\textwidth]{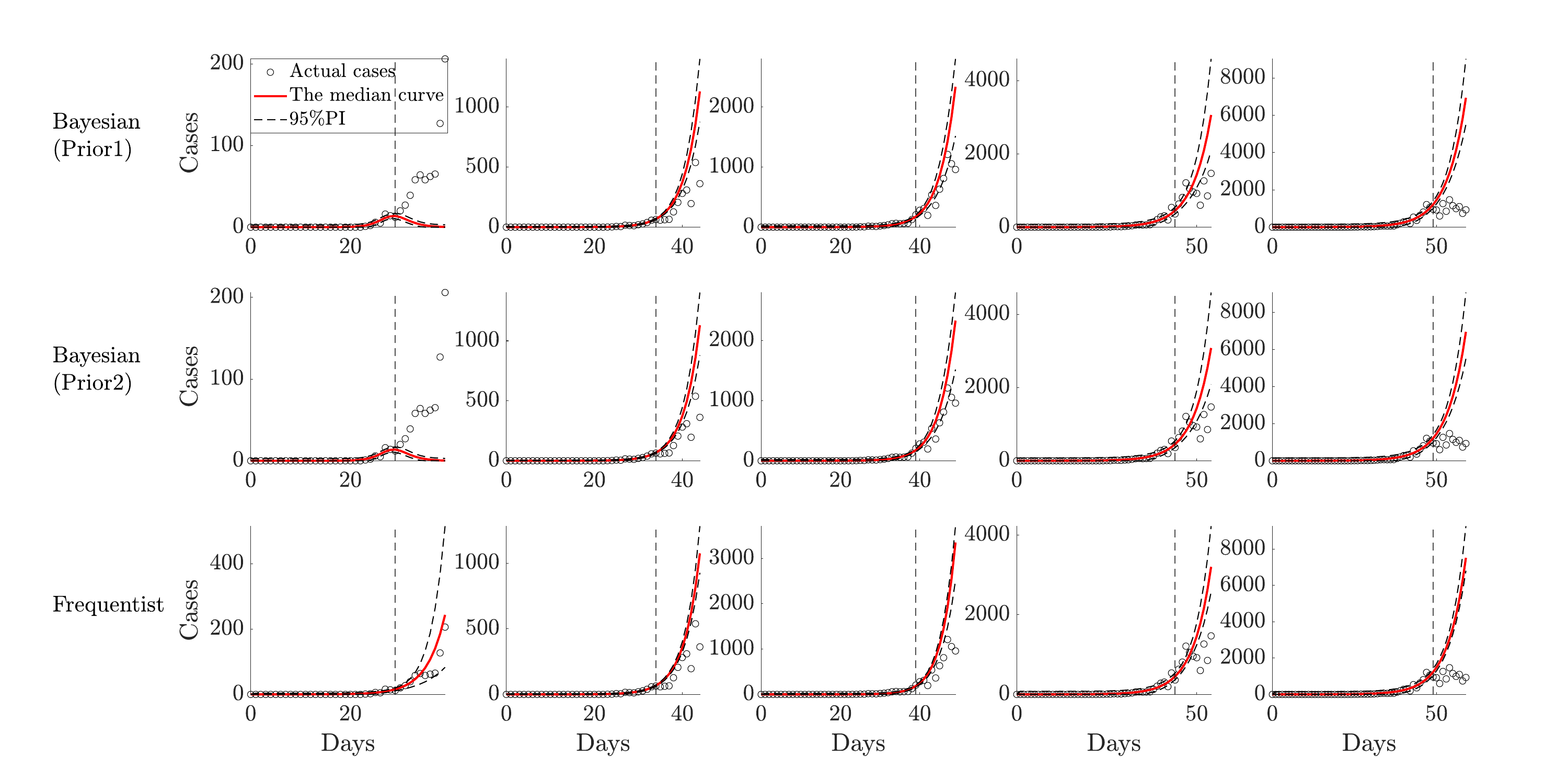}
    \caption{\footnotesize Panel showcasing the fitting of three different methods for the Switzerland COVID-19 data, using several calibration periods: 30, 35, 40, 45, and 50 days, with a forecasting horizon of 10 days. The SEIUR model is utilized, fitting the data to the newly infected people $\frac{dC}{dt}$. The population size is 47,332,614, assuming a normal error structure and the initial condition $(47332613,0,1,0,0,1)$.}
    \label{fig:S-fit-covid-10}
\end{figure}

\begin{table}[H]
\centering
\begin{tabular}{|c|c|c|c|c|}
\hline
\textbf{Calibration} & \textbf{Metrics} & \textbf{Bayesian (Prior 1)} & \textbf{Bayesian (Prior 2)} & \textbf{Frequentist} \\ \hline
\multirow{4}{*}{30}  & MAE              & 68.27                       & 68.26                       & \textbf{25.64}       \\ \cline{2-5} 
                     & RMSE             & 87.77                       & 87.75                       & \textbf{36.25}       \\ \cline{2-5} 
                     & WIS              & 67.08                       & 67.07                       & \textbf{17.67}       \\ \cline{2-5} 
                     & 95\% PI          & 0                           & 0                           & \textbf{100.00}      \\ \hline
\multirow{4}{*}{35}  & MAE              & 216.07                      & 216.32                      & \textbf{196.06}      \\ \cline{2-5} 
                     & RMSE             & 310.91                      & 311.19                      & \textbf{285.50}      \\ \cline{2-5} 
                     & WIS              & 184.79                      & 185.45                      & \textbf{181.55}      \\ \cline{2-5} 
                     & 95\% PI          & 0                           & 0                           & 0.00                 \\ \hline
\multirow{4}{*}{40}  & MAE              & 347.23                      & \textbf{346.19}             & 650.82               \\ \cline{2-5} 
                     & RMSE             & 530.34                      & \textbf{528.80}             & 965.18               \\ \cline{2-5} 
                     & WIS              & \textbf{248.23}             & 248.42                      & 559.86               \\ \cline{2-5} 
                     & 95\% PI          & \textbf{50}                 & \textbf{50}                 & 20.00                \\ \hline
\multirow{4}{*}{45}  & MAE              & \textbf{666.94}             & 672.41                      & 714.64               \\ \cline{2-5} 
                     & RMSE             & \textbf{884.71}             & 893.05                      & 958.10               \\ \cline{2-5} 
                     & WIS              & \textbf{518.03}             & 522.83                      & 616.86               \\ \cline{2-5} 
                     & 95\% PI          & \textbf{30}                 & \textbf{30}                 & 20.00                \\ \hline
\multirow{4}{*}{50}  & MAE              & \textbf{2578.48}            & 2582.69                     & 2799.32              \\ \cline{2-5} 
                     & RMSE             & \textbf{3134.12}            & 3139.82                     & 3414.08              \\ \cline{2-5} 
                     & WIS              & \textbf{2338.28}            & 2338.91                     & 2668.57              \\ \cline{2-5} 
                     & 95\% PI          & \textbf{0}                  & 0                           & 0.00                 \\ \hline
\end{tabular}
\caption{\footnotesize The performance metrics of the three methods for the Switzerland COVID-19 data, using several calibration periods: 30, 35, 40, 45, and 50 days, with a forecasting horizon of 10 days. The SEIUR model is utilized, fitting the data to the newly infected people $\frac{dC}{dt}$. The population size is 47,332,614, assuming a normal error structure and the initial condition $(47332613,0,1,0,0,1)$.}
\label{tab:S-fit-covid-10}
\end{table}

\begin{figure}[H]
    \centering
    \includegraphics[width=1\textwidth]{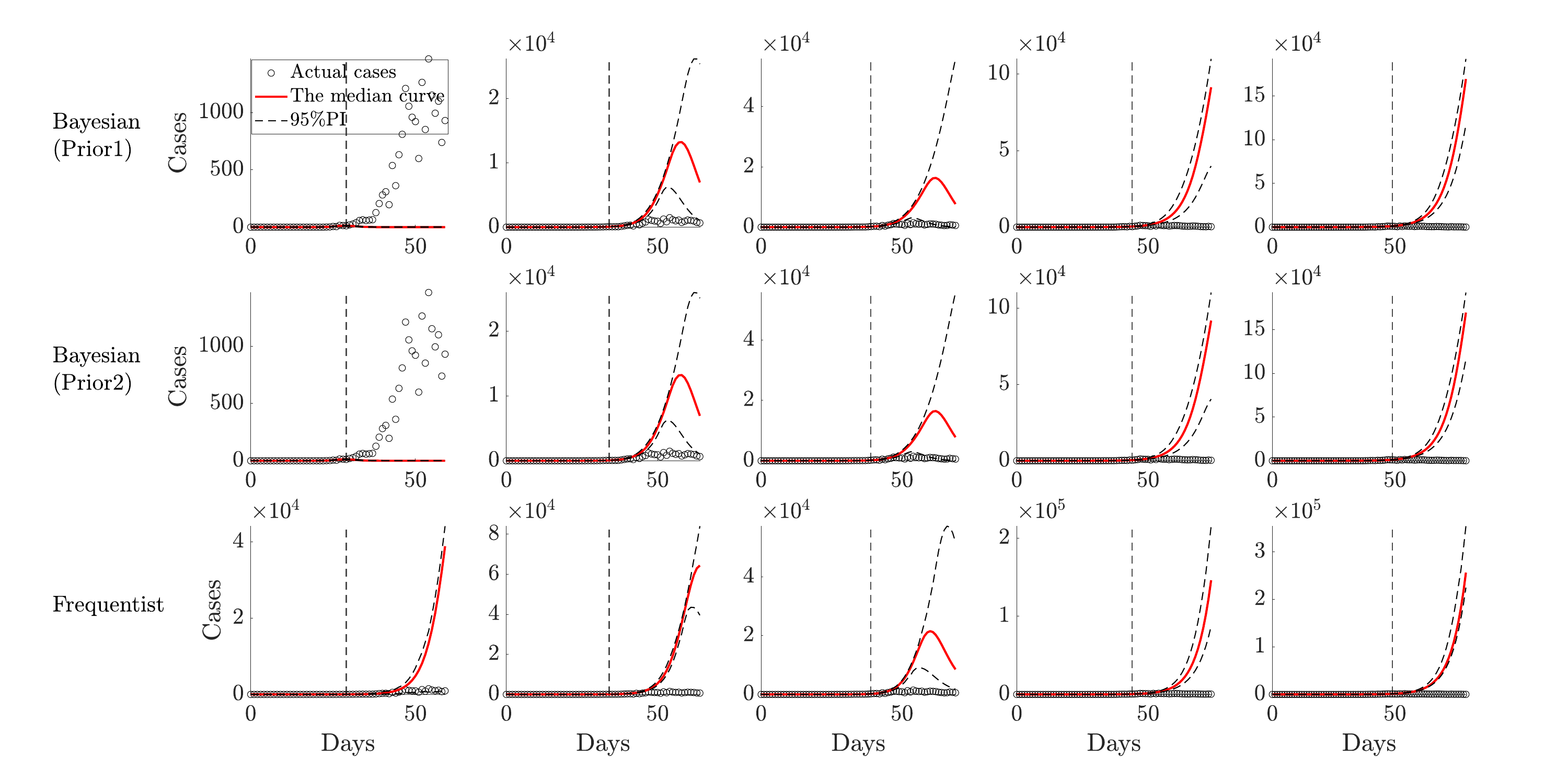}
    \caption{\footnotesize Panel showcasing the fitting of three different methods for the Switzerland COVID-19 data, using several calibration periods: 30, 35, 40, 45, and 50 days, with a forecasting horizon of 30 days. The SEIUR model is utilized, fitting the data to the newly infected people $\frac{dC}{dt}$. The population size is 47,332,614, assuming a normal error structure and the initial condition $(47332613,0,1,0,0,1)$.}
    \label{fig:S-fit-covid-30}
\end{figure}

\begin{table}[H]
\centering
\begin{tabular}{|c|c|c|c|c|}
\hline
\textbf{Calibration} & \textbf{Metrics} & \textbf{Bayesian (Prior 1)} & \textbf{Bayesian (Prior 2)} & \textbf{Frequentist} \\ \hline
\multirow{4}{*}{30}  & MAE              & \textbf{568.37}             & \textbf{568.37}             & 5919.41              \\ \cline{2-5} 
                     & RMSE             & \textbf{727.54}             & \textbf{727.54}             & 11675.53             \\ \cline{2-5} 
                     & WIS              & \textbf{567.33}             & \textbf{567.33}             & 3952.10              \\ \cline{2-5} 
                     & 95\% PI          & 0                           & 0                           & \textbf{90.00}       \\ \hline
\multirow{4}{*}{35}  & MAE              & 4845.63                     & \textbf{4837.24}            & 16185.60             \\ \cline{2-5} 
                     & RMSE             & 6645.08                     & \textbf{6631.79}            & 26796.48             \\ \cline{2-5} 
                     & WIS              & 3632.31                     & \textbf{3622.3}             & 14845.46             \\ \cline{2-5} 
                     & 95\% PI          & 0                           & 0                           & \textbf{0.00}        \\ \hline
\multirow{4}{*}{40}  & MAE              & \textbf{6561.8}             & 6607.45                     & 9057.98              \\ \cline{2-5} 
                     & RMSE             & \textbf{8651.51}            & 8716.16                     & 11699.15             \\ \cline{2-5} 
                     & WIS              & 4108.38                     & \textbf{4104.63}            & 7219.93              \\ \cline{2-5} 
                     & 95\% PI          & \textbf{46.67}              & \textbf{46.67}              & 6.67                 \\ \hline
\multirow{4}{*}{45}  & MAE              & \textbf{20779.68}           & 20947.52                    & 27843.70             \\ \cline{2-5} 
                     & RMSE             & \textbf{33536.18}           & 33769.61                    & 47911.61             \\ \cline{2-5} 
                     & WIS              & \textbf{16420.84}           & 16571.33                    & 24405.43             \\ \cline{2-5} 
                     & 95\% PI          & \textbf{10}                 & \textbf{10}                 & 6.67                 \\ \hline
\multirow{4}{*}{50}  & MAE              & \textbf{39961.57}           & 40023.35                    & 52857.31             \\ \cline{2-5} 
                     & RMSE             & \textbf{62448.95}           & 62525.78                    & 86867.30             \\ \cline{2-5} 
                     & WIS              & 34868.7                     & \textbf{34853.72}           & 50887.84             \\ \cline{2-5} 
                     & 95\% PI          & 0                           & 0                           & 0.00                 \\ \hline
\end{tabular}
\caption{\footnotesize The performance metrics of the three methods for the Switzerland COVID-19 data, using several calibration periods: 30, 35, 40, 45, and 50 days, with a forecasting horizon of 30 days. The SEIUR model is utilized, fitting the data to the newly infected people $\frac{dC}{dt}$. The population size is 47,332,614, assuming a normal error structure and the initial condition $(47332613,0,1,0,0,1)$.}
\label{tab:S-fit-covid-30}
\end{table}

\begin{table}[H]
\centering
\begin{tabular}{|c|c|c|c|c|c|}
\hline
\multirow{11}{*}{\textbf{\rotatebox[origin=c]{90}{Estimate (CI)}}} & \textbf{calibration} & \textbf{parameter} & \textbf{Bayesian(Prior1)} & \textbf{Bayesian (Prior2)} & \textbf{Frequentist} \\ \cline{2-6} 
                                         & \multirow{2}{*}{30}  & $\beta$            & 2.56 (2.52,2.6)           & 2.56 (2.52,2.6)            & 1.25 (0.84, 1.62)    \\ \cline{3-6} 
                                         &                      & $\rho$             & 0 (0,0)                   & 0 (0,0)                    & 0.02 (0, 0.29)       \\ \cline{2-6} 
                                         & \multirow{2}{*}{35}  & $\beta$            & 1.65 (1.59,1.72)          & 1.65 (1.59,1.72)           & 1.26 (1.2, 1.33)     \\ \cline{3-6} 
                                         &                      & $\rho$             & 0.08 (0.05,0.12)          & 0.08 (0.05,0.12)           & 0.02 (0.02, 0.04)    \\ \cline{2-6} 
                                         & \multirow{2}{*}{40}  & $\beta$            & 1.6 (1.5,1.73)            & 1.6 (1.5,1.73)             & 1.37 (1.22, 1.48)    \\ \cline{3-6} 
                                         &                      & $\rho$             & 0.09 (0.04,0.21)          & 0.09 (0.04,0.2)            & 0.01 (0, 0.02)       \\ \cline{2-6} 
                                         & \multirow{2}{*}{45}  & $\beta$            & 1.42 (1.35,1.51)          & 1.42 (1.35,1.51)           & 0.88 (0.81, 0.96)    \\ \cline{3-6} 
                                         &                      & $\rho$             & 0.41 (0.18,0.77)          & 0.4 (0.18,0.77)            & 0.44 (0.21, 1)       \\ \cline{2-6} 
                                         & \multirow{2}{*}{50}  & $\beta$            & 1.38 (1.34,1.43)          & 1.38 (1.34,1.44)           & 0.82 (0.81, 0.9)     \\ \cline{3-6} 
                                         &                      & $\rho$             & 0.62 (0.36,0.93)          & 0.62 (0.36,0.94)           & 1 (0.34, 1)          \\ \hline
\end{tabular}
\caption{\footnotesize The parameter estimation of the three methods for the Switzerland COVID-19 data, using several calibration periods: 30, 35, 40, 45, and 50 days. The SEIUR model is utilized, fitting the data to the newly infected people $\frac{dC}{dt}$. The population size is 47,332,614, assuming a normal error structure and the initial condition $(47332613,0,1,0,0,1)$.}
\label{tab:S-paramescovidnew30}
\end{table}

\begin{table}[H]
\centering
\begin{tabular}{|c|c|c|c|}
\hline
\textbf{Case Study}        & \textbf{Type} & \textbf{Calibration}         & \textbf{Number of Iterations} \\ \hline
Simulated Data 1           & Prior2        & 50,60                        & 40,000                        \\ \hline
Simulated Data 1           & Prior2        & 70,80,90                     & 20,000                        \\ \hline
Simulated Data 1           & Prior1        & 50,60,70,80                  & 20,000                        \\ \hline
Simulated Data 2           & normal        & 130                          & 20,000                        \\ \hline
Simulated Data 2           & normal        & 120                          & 200,000                       \\ \hline
Simulated Data 2           & normal        & 100                          & 400,000                       \\ \hline
Simulated Data 2           & uniform       & 90,100                       & 20,000                        \\ \hline
Simulated Data 2           & uniform       & 110                          & 150,000                       \\ \hline
Simulated Data 2           & uniform       & 120                          & 100,000                       \\ \hline
Simulated Data 2           & uniform       & 130                          & 400,000                       \\ \hline
San Francisco 1918 flu     & Prior2        & 25                           & 150,000                       \\ \hline
Cumberland 1918 flu        & Prior1        & 40                           & 100,000                       \\ \hline
\end{tabular}
\caption{The number of iterations used for conducting the experiments.}
\label{tab:S-niter}
\end{table}

\begin{table}[H]
\centering
\begin{tabular}{|l|c|c|c|c|}
\hline
\textbf{Dataset} & \textbf{Early} & \textbf{Pre-Peak} & \textbf{Peak} & \textbf{Post-Peak} \\
\hline
Simulated Data 1 & 50 days &  & $\sim$60 days & 70--90 days \\
Simulated Data 2 &  & 90 days & $\sim$100 days & 110--130 days \\
SF Flu 1918 & 10--20 days & 25 days & $\sim$30 days & \\
Cumberland Flu 1918 & 10--20 days & 25 days & $\sim$30--45 days & 50--60 days \\
Bombay Plague 1896-97 & 12--14 Fortnights & 16--18 Fortnights & $\sim$20 Fortnights & 22--28 Fortnights \\
Switzerland COVID-19 & 30--40 days & 45 days & $\sim$50 days & \\
\hline
\end{tabular}

\caption{Epidemic phase mapping for each dataset based on calibration periods.}
\label{tab:S-epidemic-phases}
\end{table}

\begin{sidewaystable}
\centering
\begin{tabular}{|l|cc|cc|cc|cc|}
\hline
\multirow{2}{*}{\textbf{Dataset}} & \multicolumn{2}{c|}{\textbf{Early Phase}} & \multicolumn{2}{c|}{\textbf{Pre-peak}} & \multicolumn{2}{c|}{\textbf{Peak}} & \multicolumn{2}{c|}{\textbf{Post-peak}} \\
\cline{2-9}
& \textbf{Short-term} & \textbf{Long-term} & \textbf{Short-term} & \textbf{Long-term} & \textbf{Short-term} & \textbf{Long-term} & \textbf{Short-term} & \textbf{Long-term} \\
\hline
Simulated Data 1 & Frequentist & Prior 1 &  &  & Frequentist & Frequentist & Prior 1 & Prior 1 \\
Simulated Data 2 &  &  & Prior 1 & Prior 2 & Frequentist & Frequentist & Prior 2 & Prior 2 \\
SF Flu 1918 & Prior 1 & Frequentist & Prior 2 & Prior 2 & Prior 2 & Prior 2 &  &  \\
Cumberland Flu 1918 & Frequentist & Frequentist & Prior 1 & Prior 1 & Frequentist & Frequentist & Frequentist & Frequentist \\
Bombay Plague 1896-97 & Prior 2 & Prior 2 & Prior 2/Frequentist & Prior 2/Frequentist & Prior 1 & Prior 1 & Frequentist & Frequentist \\
Switzerland COVID-19 & Frequentist & Prior 2 & Prior 1 & Prior 1 & Prior 1 & Prior 1 & &  \\
\hline
\end{tabular}
\vspace{0.5em}
\caption{Best-performing method (majority of cases) by dataset, phases, and forecast horizon.}
\label{tab:S-summary}
\end{sidewaystable}


\begin{thebibliography}{99}
\bibitem[Cheng et al.(2023)]{cheng2023real}
Cheng, Chieh and Jiang, Wei-Ming and Fan, Byron and Cheng, Yu-Chieh and Hsu, Ya-Ting and Wu, Hsiao-Yu and Chang, Hsiao-Han and Tsou, Hsiao-Hui (2023) \textit{Real-time forecasting of COVID-19 spread according to protective behavior and vaccination: autoregressive integrated moving average models}. BMC Public Health 23(1), 1500.

\bibitem[Lutz et al.(2019)]{lutz2019applying}
Lutz, Chelsea S and Huynh, Mimi P and Schroeder, Monica and Anyatonwu, Sophia and Dahlgren, F Scott and Danyluk, Gregory and Fernandez, Danielle and Greene, Sharon K and Kipshidze, Nodar and Liu, Leann and others (2019) \textit{Applying infectious disease forecasting to public health: a path forward using influenza forecasting examples}. BMC Public Health 19, 1--12.

\bibitem[Shearer et al.(2020)]{shearer2020infectious}
Shearer, Freya M and Moss, Robert and McVernon, Jodie and Ross, Joshua V and McCaw, James M (2020) \textit{Infectious disease pandemic planning and response: Incorporating decision analysis}. PLoS Medicine 17(1), e1003018.

\bibitem[Rahimi et al.(2023)]{rahimi2023review}
Rahimi, Iman and Chen, Fang and Gandomi, Amir H (2023) \textit{A review on COVID-19 forecasting models}. Neural Computing and Applications 35(33), 23671--23681.

\bibitem[Shinde et al.(2020)]{shinde2020forecasting}
Shinde, Gitanjali R and Kalamkar, Asmita B and Mahalle, Parikshit N and Dey, Nilanjan and Chaki, Jyotismita and Hassanien, Aboul Ella (2020) \textit{Forecasting models for coronavirus disease (COVID-19): a survey of the state-of-the-art}. SN Computer Science 1(4), 197.

\bibitem[Bertozzi et al.(2020)]{bertozzi2020challenges}
Bertozzi, Andrea L and Franco, Elisa and Mohler, George and Short, Martin B and Sledge, Daniel (2020) \textit{The challenges of modeling and forecasting the spread of COVID-19}. Proceedings of the National Academy of Sciences 117(29), 16732--16738.

\bibitem[Ioannidis et al.(2022)]{ioannidis2022forecasting}
Ioannidis, John PA and Cripps, Sally and Tanner, Martin A (2022) \textit{Forecasting for COVID-19 has failed}. International Journal of Forecasting 38(2), 423--438.

\bibitem[Chowell et al.(2022a)]{chowell2022ensemble}
Chowell, Gerardo and Dahal, Sushma and Tariq, Amna and Roosa, Kimberlyn and Hyman, James M and Luo, Ruiyan (2022) \textit{An ensemble n-sub-epidemic modeling framework for short-term forecasting epidemic trajectories: Application to the COVID-19 pandemic in the USA}. PLoS Computational Biology 18(10), e1010602.

\bibitem[Chowell et al.(2022b)]{chowell2022sub}
Chowell, Gerardo and Rothenberg, Richard and Roosa, Kimberlyn and Tariq, Amna and Hyman, James M and Luo, Ruiyan (2022) \textit{Sub-epidemic model forecasts during the first wave of the COVID-19 pandemic in the USA and European hotspots}. In: Mathematics of Public Health: Proceedings of the Seminar on the Mathematical Modelling of COVID-19. Springer, 85--137.

\bibitem[McGowan et al.(2019)]{mcgowan2019collaborative}
McGowan, Craig J and Biggerstaff, Matthew and Johansson, Michael and Apfeldorf, Karyn M and Ben-Nun, Michal and Brooks, Logan and Convertino, Matteo and Erraguntla, Madhav and Farrow, David C and Freeze, John and others (2019) \textit{Collaborative efforts to forecast seasonal influenza in the United States, 2015--2016}. Scientific Reports 9(1), 683.

\bibitem[Biggerstaff et al.(2014)]{biggerstaff2014estimates}
Biggerstaff, Matthew and Cauchemez, Simon and Reed, Carrie and Gambhir, Manoj and Finelli, Lyn (2014) \textit{Estimates of the reproduction number for seasonal, pandemic, and zoonotic influenza: a systematic review of the literature}. BMC Infectious Diseases 14(1), 1--20.

\bibitem[Chretien et al.(2015)]{chretien2015mathematical}
Chretien, Jean-Paul and Riley, Steven and George, Dylan B (2015) \textit{Mathematical modeling of the West Africa Ebola epidemic}. eLife 4, e09186.

\bibitem[Chowell et al.(2014)]{chowell2014west}
Chowell, Gerardo and Simonsen, Lone and Viboud, Céline and Kuang, Yang (2014) \textit{Is West Africa approaching a catastrophic phase or is the 2014 Ebola epidemic slowing down? Different models yield different answers for Liberia}. PLoS Currents 6.

\bibitem[Roosa et al.(2020)]{roosa2020multi}
Roosa, Kimberlyn and Tariq, Amna and Yan, Ping and Hyman, James M and Chowell, Gerardo (2020) \textit{Multi-model forecasts of the ongoing Ebola epidemic in the Democratic Republic of Congo, March--October 2019}. Journal of The Royal Society Interface 17(169), 20200447.

\bibitem[Bleichrodt et al.(2023b)]{bleichrodt2023retrospective}
Bleichrodt, Amanda and Luo, Ruiyan and Kirpich, Alexander and Chowell, Gerardo (2023) \textit{Retrospective evaluation of short-term forecast performance of ensemble sub-epidemic frameworks and other time-series models: The 2022-2023 mpox outbreak across multiple geographical scales, July 14th, 2022, through February 26th, 2023}. medRxiv.

\bibitem[Chowell et al.(2024)]{chowell2024growthpredict}
Chowell, Gerardo and Bleichrodt, Amanda and Dahal, Sushma and Tariq, Amna and Roosa, Kimberlyn and Hyman, James M and Luo, Ruiyan (2024) \textit{GrowthPredict: A toolbox and tutorial-based primer for fitting and forecasting growth trajectories using phenomenological growth models}. Scientific Reports 14(1), 1630.

\bibitem[Charniga et al.(2024)]{charniga2024nowcasting}
Charniga, Kelly and Madewell, Zachary J and Masters, Nina B and Asher, Jason and Nakazawa, Yoshinori and Spicknall, Ian H (2024) \textit{Nowcasting and forecasting the 2022 US Mpox outbreak: Support for public health decision making and lessons learned}. Epidemics, 100755.

\bibitem[Gneiting(2008)]{gneiting2008probabilistic}
Gneiting, Tilmann (2008) \textit{Probabilistic forecasting}. Journal of the Royal Statistical Society Series A: Statistics in Society 171(2), 319--321.

\bibitem[Mwambi et al.(2011)]{mwambi2011frequentist}
Mwambi, H and Ramroop, S and White, LJ and Okiro, EA and Nokes, D James and Shkedy, Z and Molenberghs, Geert (2011) \textit{A frequentist approach to estimating the force of infection for a respiratory disease using repeated measurement data from a birth cohort}. Statistical Methods in Medical Research 20(5), 551--570.

\bibitem[Chowell(2017)]{chowell2017fitting}
Chowell, Gerardo (2017) \textit{Fitting dynamic models to epidemic outbreaks with quantified uncertainty: A primer for parameter uncertainty, identifiability, and forecasts}. Infectious Disease Modelling 2(3), 379--398.

\bibitem[Chowell et al.(2020)]{chowell2020real}
Chowell, Gerardo and Luo, Ruiyan and Sun, K and Roosa, Kimberlyn and Tariq, Amna and Viboud, C (2020) \textit{Real-time forecasting of epidemic trajectories using computational dynamic ensembles}. Epidemics 30, 100379.

\bibitem[Pruitt et al.(2024)]{pruitt2024role}
Pruitt, CD and Lovell, AE and Hebborn, C and Nunes, FM (2024) \textit{The role of the likelihood for elastic scattering uncertainty quantification}. arXiv preprint arXiv:2403.00753.

\bibitem[Transtrum and Qiu(2012)]{transtrum2012optimal}
Transtrum, Mark K and Qiu, Peng (2012) \textit{Optimal experiment selection for parameter estimation in biological differential equation models}. BMC Bioinformatics 13, 1--12.

\bibitem[Huang and He(2024)]{huang2024nonlinear}
Huang, Hsin-Hsiung and He, Qing (2024) \textit{Nonlinear regression analysis}. arXiv preprint arXiv:2402.05342.

\bibitem[Chowell et al.(2024)]{chowell2024parameter}
Chowell, Gerardo and Bleichrodt, Amanda and Luo, Ruiyan (2024) \textit{Parameter estimation and forecasting with quantified uncertainty for ordinary differential equation models using QuantDiffForecast: A MATLAB toolbox and tutorial}. Statistics in Medicine, Wiley Online Library.

\bibitem[Grinsztajn et al.(2021)]{grinsztajn2021bayesian}
Grinsztajn, Léo and Semenova, Elizaveta and Margossian, Charles C and Riou, Julien (2021) \textit{Bayesian workflow for disease transmission modeling in Stan}. Statistics in Medicine 40(27), 6209--6234.

\bibitem[Bouman et al.(2024)]{bouman2024bayesian}
Bouman, Judith A and Hauser, Anthony and Grimm, Simon L and Wohlfender, Martin and Bhatt, Samir and Semenova, Elizaveta and Gelman, Andrew and Althaus, Christian L and Riou, Julien (2024) \textit{Bayesian workflow for time-varying transmission in stratified compartmental infectious disease transmission models}. PLoS Computational Biology 20(4), e1011575.

\bibitem[Gelman et al.(2020)]{gelman2020bayesian}
Gelman, Andrew and Vehtari, Aki and Simpson, Daniel and Margossian, Charles C and Carpenter, Bob and Yao, Yuling and Kennedy, Lauren and Gabry, Jonah and Bürkner, Paul-Christian and Modrák, Martin (2020) \textit{Bayesian workflow}. arXiv preprint arXiv:2011.01808.

\bibitem[Belasso et al.(2023)]{belasso2023bayesian}
Belasso, Clyde J and Cai, Zhengchen and Bezgin, Gleb and Pascoal, Tharick and Stevenson, Jenna and Rahmouni, Nesrine and Tissot, Cécile and Lussier, Firoza and Rosa-Neto, Pedro and Soucy, Jean-Paul and others (2023) \textit{Bayesian workflow for the investigation of hierarchical classification models from tau-PET and structural MRI data across the Alzheimer’s disease spectrum}. Frontiers in Aging Neuroscience 15, 1225816.

\bibitem[Annis et al.(2017)]{annis2017bayesian}
Annis, Jeffrey and Miller, Brent J and Palmeri, Thomas J (2017) \textit{Bayesian inference with Stan: A tutorial on adding custom distributions}. Behavior Research Methods 49, 863--886.

\bibitem[Kelter(2020)]{kelter2020bayesian}
Kelter, Riko (2020) \textit{Bayesian survival analysis in STAN for improved measuring of uncertainty in parameter estimates}. Measurement: Interdisciplinary Research and Perspectives 18(2), 101--109.

\bibitem[Sennhenn-Reulen(2018)]{sennhenn2018bayesian}
Sennhenn-Reulen, Holger (2018) \textit{Bayesian regression for a Dirichlet distributed response using Stan}. arXiv preprint arXiv:1808.06399.

\bibitem[Sorensen and Vasishth(2015)]{sorensen2015bayesian}
Sorensen, Tanner and Vasishth, Shravan (2015) \textit{Bayesian linear mixed models using Stan: A tutorial for psychologists, linguists, and cognitive scientists}. arXiv preprint arXiv:1506.06201.

\bibitem[Monnahan et al.(2017)]{monnahan2017faster}
Monnahan, Cole C and Thorson, James T and Branch, Trevor A (2017) \textit{Faster estimation of Bayesian models in ecology using Hamiltonian Monte Carlo}. Methods in Ecology and Evolution 8(3), 339--348.

\bibitem[Bürkner(2017)]{burkner2017brms}
Bürkner, Paul-Christian (2017) \textit{brms: An R package for Bayesian multilevel models using Stan}. Journal of Statistical Software 80, 1--28.

\bibitem[Chowell(2020)]{outbreak_datasets}
Chowell, G. (2020) \textit{Outbreak datasets}. GitHub Repository. Accessed: 5 Aug 2020. Available at: \url{https://github.com/gchowell/outbreak_datasets}.

\bibitem[Andrade and Duggan(2021)]{andrade2021bayesian}
Andrade, Jair and Duggan, Jim (2021) \textit{A Bayesian approach to calibrate system dynamics models using Hamiltonian Monte Carlo}. System Dynamics Review 37(4), 283--309.

\bibitem[Czumbel et al.(2018)]{czumbel2018management}
Czumbel, Ida and Quinten, Chantal and Lopalco, Pierluigi and Semenza, Jan C and ECDC expert panel working group (2018) \textit{Management and control of communicable diseases in schools and other child care settings: systematic review on the incubation period and period of infectiousness}. BMC Infectious Diseases 18, 1--15.

\bibitem[White and Mordechai(2020)]{white2020modeling}
White, Lauren A and Mordechai, Lee (2020) \textit{Modeling the Justinianic Plague: Comparing hypothesized transmission routes}. PloS One 15(4), e0231256.

\bibitem[Grinsztajn et al.(2006)]{grinsztajn2006bayesian}
Grinsztajn, L and Semenova, E and Margossian, C.C and Riou, J (2006) \textit{Bayesian workflow for disease transmission modeling in Stan}. arXiv preprint arXiv:2006.02985.

\bibitem[Roosa et al.(2019)]{roosa2019comparative}
Roosa, Kimberlyn and Luo, Ruiyan and Chowell, Gerardo (2019) \textit{Comparative assessment of parameter estimation methods in the presence of overdispersion: a simulation study}. Math Biosci Eng 16(5), 4299--313.

\bibitem[Gneiting and Raftery(2007)]{gneiting2007strictly}
Gneiting, Tilmann and Raftery, Adrian E (2007) \textit{Strictly proper scoring rules, prediction, and estimation}. Journal of the American Statistical Association 102(477), 359--378.

\bibitem[Kuhn and Johnson(2013)]{kuhn2013applied}
Kuhn, Max and Johnson, Kjell (2013) \textit{Applied Predictive Modeling}. Springer, volume 26.

\bibitem[University of Nicosia(2018)]{universitynicosia2018}
University of Nicosia (2018) \textit{M4 Competition Competitor’s Guide: Prizes and Rules}. University of Nicosia. Accessed: 2023-10-04. Available at: \url{http://www.unic.ac.cy/test/wp-content/uploads/sites/2/2018/09/M4-Competitors-Guide.pdf}.

\bibitem[Bracher et al.(2021)]{bracher2021evaluating}
Bracher, Johannes and Ray, Evan L and Gneiting, Tilmann and Reich, Nicholas G (2021) \textit{Evaluating epidemic forecasts in an interval format}. PLoS Computational Biology 17(2), e1008618.

\bibitem[Cramer et al.(2022)]{cramer2022evaluation}
Cramer, Estee Y and Ray, Evan L and Lopez, Velma K and Bracher, Johannes and Brennen, Andrea and Castro Rivadeneira, Alvaro J and Gerding, Aaron and Gneiting, Tilmann and House, Katie H and Huang, Yuxin and others (2022) \textit{Evaluation of individual and ensemble probabilistic forecasts of COVID-19 mortality in the United States}. Proceedings of the National Academy of Sciences 119(15), e2113561119.

\bibitem[Margossian et al.(2021)]{margossian2021nested}
Margossian, Charles C and Hoffman, Matthew D and Sountsov, Pavel and Riou-Durand, Lionel and Vehtari, Aki and Gelman, Andrew (2021) \textit{Nested \(\hat{R}\): Assessing the convergence of Markov chain Monte Carlo when running many short chains}. arXiv preprint arXiv:2110.13017.

\bibitem[Lopez et al.(2024)]{lopez2024challenges}
Lopez, Velma K and Cramer, Estee Y and Pagano, Robert and Drake, John M and O’Dea, Eamon B and Adee, Madeline and Ayer, Turgay and Chhatwal, Jagpreet and Dalgic, Ozden O and Ladd, Mary A and others (2024) \textit{Challenges of COVID-19 Case Forecasting in the US, 2020--2021}. PLoS Computational Biology 20(5), e1011200.

\bibitem[Reich et al.(2019)]{reich2019collaborative}
Reich, Nicholas G and Brooks, Logan C and Fox, Spencer J and Kandula, Sasikiran and McGowan, Craig J and Moore, Evan and Osthus, Dave and Ray, Evan L and Tushar, Abhinav and Yamana, Teresa K and others (2019) \textit{A collaborative multiyear, multimodel assessment of seasonal influenza forecasting in the United States}. Proceedings of the National Academy of Sciences 116(8), 3146--3154.

\bibitem[Meltzer et al.(2014)]{meltzer2014estimating}
Meltzer, Martin I and Atkins, Charisma Y and Santibanez, Scott and Knust, Barbara and Petersen, Brett W and Ervin, Elizabeth D and Nichol, Stuart T and Damon, Inger K and Washington, Michael L and others (2014) \textit{Estimating the future number of cases in the Ebola epidemic—Liberia and Sierra Leone, 2014--2015}. MMWR Surveillance Summaries 63(Suppl 3), 1--14.

\bibitem[Funk et al.(2019)]{funk2019assessing}
Funk, Sebastian and Camacho, Anton and Kucharski, Adam J and Lowe, Rachel and Eggo, Rosalind M and Edmunds, W John (2019) \textit{Assessing the performance of real-time epidemic forecasts: A case study of Ebola in the Western Area region of Sierra Leone, 2014-15}. PLoS Computational Biology 15(2), e1006785.

\bibitem[Chowell et al.(2017)]{chowell2017perspectives}
Chowell, Gerardo and Viboud, Cécile and Simonsen, Lone and Merler, Stefano and Vespignani, Alessandro (2017) \textit{Perspectives on model forecasts of the 2014--2015 Ebola epidemic in West Africa: lessons and the way forward}. BMC Medicine 15, 1--8.

\bibitem[Bleichrodt et al.(2023a)]{bleichrodt2023real}
Bleichrodt, Amanda and Dahal, Sushma and Maloney, Kevin and Casanova, Lisa and Luo, Ruiyan and Chowell, Gerardo (2023) \textit{Real-time forecasting the trajectory of monkeypox outbreaks at the national and global levels, July--October 2022}. BMC Medicine 21(1), 19.

\bibitem[Banks et al.(2014)]{banks2014modeling}
Banks, Harvey Thomas and Hu, Shuhua and Thompson, W Clayton (2014) \textit{Modeling and inverse problems in the presence of uncertainty}. CRC Press.

\bibitem[Ganyani et al.(2018)]{ganyani2018assessing}
Ganyani, Tapiwa and Faes, Christel and Chowell, Gerardo and Hens, Niel (2018) \textit{Assessing inference of the basic reproduction number in an SIR model incorporating a growth-scaling parameter}. Statistics in Medicine 37(29), 4490--4506.

\bibitem[Girolami(2008)]{girolami2008bayesian}
Girolami, Mark (2008) \textit{Bayesian inference for differential equations}. Theoretical Computer Science 408(1), 4--16.

\bibitem[Kypraios et al.(2017)]{kypraios2017tutorial}
Kypraios, Theodore and Neal, Peter and Prangle, Dennis (2017) \textit{A tutorial introduction to Bayesian inference for stochastic epidemic models using Approximate Bayesian Computation}. Mathematical Biosciences 287, 42--53.

\bibitem[McKinley et al.(2014)]{mckinley2014simulation}
McKinley, Trevelyan J and Ross, Joshua V and Deardon, Rob and Cook, Alex R (2014) \textit{Simulation-based Bayesian inference for epidemic models}. Computational Statistics \& Data Analysis 71, 434--447.

\bibitem[Richardson et al.(2001)]{richardson2001evidence}
Richardson, Martin and Elliman, David and Maguire, Helen and Simpson, John and Nicoll, Angus (2001) \textit{Evidence base of incubation periods, periods of infectiousness and exclusion policies for the control of communicable diseases in schools and preschools}. The Pediatric Infectious Disease Journal 20(4), 380--391.

\bibitem[Book(2003)]{book2003report}
Book, Red (2003) \textit{Report of the Committee on Infectious Diseases}. American Academy of Pediatrics.

\bibitem[Center for Disease Control(1987)]{center1987health}
Center for Disease Control. Bureau of Epidemiology and Center for Disease Control. Quarantine Division and Center for Prevention Services (US). Quarantine Division and National Center for Prevention Services (US). Division of Quarantine and National Center for Infectious Diseases (US). Division of Quarantine (1987) \textit{Health information for international travel}. US Department of Health, Education, and Welfare, Public Health Service.

\bibitem[Cox and Subbarao(2000)]{cox2000global}
Cox, Nancy J and Subbarao, K (2000) \textit{Global epidemiology of influenza: past and present}. Annual Review of Medicine 51(1), 407--421. Annual Reviews 4139 El Camino Way, PO Box 10139, Palo Alto, CA 94303-0139, USA.

\bibitem[Glezen(1996)]{glezen1996emerging}
Glezen, W Paul (1996) \textit{Emerging infections: pandemic influenza}. Epidemiologic Reviews 18(1), 64--76. Citeseer.

\bibitem[Frost(1919)]{frost1919epidemiology}
Frost, Wade H (1919) \textit{The epidemiology of influenza}. Journal of the American Medical Association 73(5), 313--318. American Medical Association.

\bibitem[Chowell et al.(2007)]{chowell2007comparative}
Chowell, Gerardo and Nishiura, Hiroshi and Bettencourt, Luis MA (2007) \textit{Comparative estimation of the reproduction number for pandemic influenza from daily case notification data}. Journal of the Royal Society Interface 4(12), 155--166. The Royal Society London.

\bibitem[Vynnycky et al.(2007)]{vynnycky2007estimates}
Vynnycky, Emilia and Trindall, Amy and Mangtani, Punam (2007) \textit{Estimates of the reproduction numbers of Spanish influenza using morbidity data}. International Journal of Epidemiology 36(4), 881--889. Oxford University Press.

\bibitem[{\O}kland and Mamelund(2019)]{okland2019race}
{\O}kland, Helene and Mamelund, Svenn-Erik (2019) \textit{Race and 1918 influenza pandemic in the United States: A review of the literature}. International Journal of Environmental Research and Public Health 16(14), 2487. MDPI.

\bibitem[Lauer et al.(2020)]{lauer2020incubation}
Lauer, Stephen A and Grantz, Kyra H and Bi, Qifang and Jones, Forrest K and Zheng, Qulu and Meredith, Hannah R and Azman, Andrew S and Reich, Nicholas G and Lessler, Justin (2020) \textit{The incubation period of coronavirus disease 2019 (COVID-19) from publicly reported confirmed cases: estimation and application}. Annals of Internal Medicine 172(9), 577--582. American College of Physicians.

\bibitem[Stringhini et al.(2020)]{stringhini2020seroprevalence}
Stringhini, Silvia and Wisniak, Ania and Piumatti, Giovanni and Azman, Andrew S and Lauer, Stephen A and Baysson, Hélène and De Ridder, David and Petrovic, Dusan and Schrempft, Stephanie and Marcus, Kailing et al. (2020) \textit{Seroprevalence of anti-SARS-CoV-2 IgG antibodies in Geneva, Switzerland (SEROCoV-POP): a population-based study}. The Lancet 396(10247), 313--319. Elsevier.

\bibitem[Kermack and McKendrick(1927)]{kermack1927contribution}
Kermack, William Ogilvy and McKendrick, Anderson G (1927) \textit{A contribution to the mathematical theory of epidemics}. Proceedings of the Royal Society of London. Series A, Containing Papers of a Mathematical and Physical Character 115(772), 700--721. The Royal Society London.

\bibitem[Bacaër(2012)]{bacaer2012model}
Bacaër, Nicolas (2012) \textit{The model of Kermack and McKendrick for the plague epidemic in Bombay and the type reproduction number with seasonality}. Journal of Mathematical Biology 64(3), 403--422. Springer.

\bibitem[Monecke et al.(2009)]{monecke2009modelling}
Monecke, Stefan and Monecke, Hannelore and Monecke, Jochen (2009) \textit{Modelling the black death. A historical case study and implications for the epidemiology of bubonic plague}. International Journal of Medical Microbiology 299(8), 582--593. Elsevier.

\bibitem[Pell et al.(2018)]{pell2018simple}
Pell, Bruce and Phan, Tin and Rutter, Erica M and Chowell, Gerardo and Kuang, Yang (2018) \textit{Simple multi-scale modeling of the transmission dynamics of the 1905 plague epidemic in Bombay}. Mathematical Biosciences 301, 83--92. Elsevier.

\bibitem[Mangiarotti(2015)]{mangiarotti2015low}
Mangiarotti, Sylvain (2015) \textit{Low dimensional chaotic models for the plague epidemic in Bombay (1896--1911)}. Chaos, Solitons \& Fractals 81, 184--196. Elsevier.

\bibitem[Karami et al.(2024)]{karami2024bayesianfitforecast}
Karami, H., Bleichrodt, A., Luo, R., Chowell, G. (2024) \textit{BayesianFitForecast: An R toolbox for parameter estimation and forecasting with quantified uncertainty in ordinary differential equation models}. Under review.

\bibitem[Greenland(2009)]{greenland2009bayesian}
Greenland, Sander (2009) \textit{Bayesian perspectives for epidemiologic research: III. Bias analysis via missing-data methods}. International Journal of Epidemiology 38(6), 1662--1673. Oxford University Press.

\bibitem[Harel et al.(2018)]{harel2018multiple}
Harel, Ofer and Mitchell, Emily M and Perkins, Neil J and Cole, Stephen R and Tchetgen Tchetgen, Eric J and Sun, BaoLuo and Schisterman, Enrique F (2018) \textit{Multiple imputation for incomplete data in epidemiologic studies}. American Journal of Epidemiology 187(3), 576--584. Oxford University Press.

\bibitem[Dunson(2001)]{dunson2001commentary}
Dunson, David B (2001) \textit{Commentary: practical advantages of Bayesian analysis of epidemiologic data}. American Journal of Epidemiology 153(12), 1222--1226. Oxford University Press.

\bibitem[Martin et al.(2020)]{martin2020computing}
Martin, Gael M and Frazier, David T and Robert, Christian P (2020) \textit{Computing Bayes: Bayesian computation from 1763 to the 21st century}. arXiv preprint arXiv:2004.06425.

\bibitem[van de Schoot et al.(2021)]{van2021bayesian}
van de Schoot, Rens and Depaoli, Sarah and King, Ruth and Kramer, Bianca and M{\"a}rtens, Kaspar and Tadesse, Mahlet G and Vannucci, Marina and Gelman, Andrew and Veen, Duco and Willemsen, Joukje et al. (2021) \textit{Bayesian statistics and modelling}. Nature Reviews Methods Primers 1(1), 1. Nature Publishing Group UK London.

\bibitem[Dixon et al.(2022)]{dixon2022comparison}
Dixon, Samuel and Keshavamurthy, Ravikiran and Farber, Daniel H and Stevens, Andrew and Pazdernik, Karl T and Charles, Lauren E (2022) \textit{A comparison of infectious disease forecasting methods across locations, diseases, and time}. Pathogens 11(2), 185. MDPI.

\bibitem[Chowell et al.(2023)]{chowell2023structural}
Chowell, Gerardo and Dahal, Sushma and Liyanage, Yuganthi R and Tariq, Amna and Tuncer, Necibe (2023) \textit{Structural identifiability analysis of epidemic models based on differential equations: a tutorial-based primer}. Journal of Mathematical Biology 87(6), 79. Springer.

\bibitem[Bates and Watts(1988)]{bates1988nonlinear}
Bates, Douglas M and Watts, Donald G (1988) \textit{Nonlinear regression analysis and its applications}. Vol. 2. Wiley, New York.

\bibitem[Cao et al.(2012)]{cao2012penalized}
Cao, Jiguo and Huang, Jianhua Z and Wu, Hulin (2012) \textit{Penalized nonlinear least squares estimation of time-varying parameters in ordinary differential equations}. Journal of Computational and Graphical Statistics 21(1), 42--56. Taylor \& Francis.

\bibitem[Ramsay and Hooker(2017)]{ramsay2017dynamic}
Ramsay, James and Hooker, Giles (2017) \textit{Dynamic data analysis}. Springer, New York, NY.

\bibitem[Seber and Wild(2003)]{seber2003nonlinear}
Seber, George AF and Wild, Christopher John (2003) \textit{Nonlinear regression}. John Wiley \& Sons, Hoboken, New Jersey.

\bibitem[Luo et al.(2024)]{luo2024estimation}
Luo, Ruiyan and Herrera-Reyes, Alejandra D and Kim, Yena and Rogowski, Susan and White, Diana and Smirnova, Alexandra (2024) \textit{Estimation of time-dependent transmission rate for COVID-19 SVIRD model using predictor--corrector algorithm}. In: Mathematical Modeling for Women’s Health: Collaborative Workshop for Women in Mathematical Biology, 213--237. Springer Nature Switzerland, Cham.

\bibitem[Varah(1982)]{varah1982spline}
Varah, James M (1982) \textit{A spline least squares method for numerical parameter estimation in differential equations}. SIAM Journal on Scientific and Statistical Computing 3(1), 28--46. SIAM.

\bibitem[Liang and Wu(2008)]{liang2008parameter}
Liang, Hua and Wu, Hulin (2008) \textit{Parameter estimation for differential equation models using a framework of measurement error in regression models}. Journal of the American Statistical Association 103(484), 1570--1583. Taylor \& Francis.

\bibitem[Ramsay et al.(2007)]{ramsay2007parameter}
Ramsay, Jim O and Hooker, Giles and Campbell, David and Cao, Jiguo (2007) \textit{Parameter estimation for differential equations: a generalized smoothing approach}. Journal of the Royal Statistical Society Series B: Statistical Methodology 69(5), 741--796. Oxford University Press.

\bibitem[Rudi et al.(2022)]{rudi2022parameter}
Rudi, Johann and Bessac, Julie and Lenzi, Amanda (2022) \textit{Parameter estimation with dense and convolutional neural networks applied to the FitzHugh--Nagumo ODE}. In: Mathematical and Scientific Machine Learning, 781--808. PMLR.

\bibitem[Gelman et al.(1996)]{gelman1996physiological}
Gelman, Andrew and Bois, Frederic and Jiang, Jiming (1996) \textit{Physiological pharmacokinetic analysis using population modeling and informative prior distributions}. Journal of the American Statistical Association 91(436), 1400--1412. Taylor \& Francis.

\bibitem[Huang et al.(2006)]{huang2006hierarchical}
Huang, Yangxin and Liu, Dacheng and Wu, Hulin (2006) \textit{Hierarchical Bayesian methods for estimation of parameters in a longitudinal HIV dynamic system}. Biometrics 62(2), 413--423. Oxford University Press.

\bibitem[Huang et al.(2020)]{huang2020bayesian}
Huang, Hanwen and Handel, Andreas and Song, Xiao (2020) \textit{A Bayesian approach to estimate parameters of ordinary differential equation}. Computational Statistics 35, 1481--1499. Springer.


\end{thebibliography}
\end{document}